\documentclass[12pt,a4paper]{article}
\usepackage{amssymb,amsmath,eucal}
\usepackage[dvips]{lscape,graphicx}
\usepackage{epsfig}
\usepackage{cite}
\newcommand{\ct}{\cite}
\newcommand{\lb}{\label}
\newcommand{\bc}{\begin{center}}
\newcommand{\ec}{\end{center}}
\newcommand{\bd}{\begin{displaymath}}
\newcommand{\ed}{\end{displaymath}}
\newcommand{\be}{\begin{equation}}
\newcommand{\ee}{\end{equation}}
\newcommand{\ba}{\begin{array}}
\newcommand{\ea}{\end{array}}
\newcommand{\un}{\underline}
\newcommand{\ov}{\overline}
\newcommand{\bfi}{\begin{figure}}
\newcommand{\efi}{\end{figure}}
\newcommand{\ds}{\displaystyle}
\newcommand{\bee}{\begin{eqnarray}}
\newcommand{\eee}{\end{eqnarray}}

\newcommand{\plaqr}{$\biggl(\quad\raisebox{-2pt}{\mbox{\framebox(0,12){\phantom{a}}
\hspace{-6.9mm} \dashbox{3}(12,12)[b]{\phantom{a}}}}\biggr)$}
\newcommand{\plaql}{$\biggl(\raisebox{-2pt}{\mbox{\framebox(0,12){\phantom{a}}\hspace{-1.4mm}
\dashbox{3}(12,12)[b]{\phantom{a}}}}\biggr)$}
\newcommand{\plaqt}{$\biggl(\raisebox{-2pt}{\mbox{\raisebox{12pt}{\framebox(11,0){\phantom{a}}}
\hspace{-7.0mm} \dashbox{3}(12,12)[b]{\phantom{a}}}}\biggr)$}
\newcommand{\plaqb}{$\biggl(\raisebox{-2pt}{\mbox{\framebox(11,0)
{\phantom{a}}\hspace{-5.5mm}
\dashbox{2}(12,12)[b]{\phantom{a}}}}\biggr)$}

\newcommand{\link}{\ba{l}\begin{picture}(22,4)
    \put(0,2.5){\circle*{4}}
    \put(20,2.5){\circle*{4}}
    \put(0,2.5){\line(1,0){20}}
\end{picture}\ea}

\begin{document}

\title{\Large\bf Phase transition in gauge theories, monopoles and
the Multiple Point Principle}

\author{\bf C.R.~Das${}^{1}$,
\bf L.V.~Laperashvili${}^{1,}$ ${}^{2}$\\[5mm]
\itshape{${}^{1}$ The Institute of Mathematical Sciences,}\\
{\it CIT Campus, Taramani, Chennai 600113, India}\\[5mm]
\itshape{${}^{2}$
Institute of Theoretical and Experimental Physics,}\\
{\it B. Cheremushkinskaya 25, 117218 Moscow, Russia}}

\date{}

\maketitle

\thispagestyle{empty}

\vspace{1cm}



\vspace{8cm}

\vspace{0.5cm} \footnoterule{\noindent\large\bf ${}^{1}$ E--mail:
crdas@imsc.res.in\\ ${}^{2}$ E--mail: laper@itep.ru}

\clearpage\newpage

\thispagestyle{empty}

\begin{abstract}

This review is devoted to the Multiple Point Principle (MPP),
according to which several vacuum states with the same energy
density exist in Nature. The MPP is implemented to the Standard
Model (SM), Family replicated gauge group model (FRGGM) and phase
transitions in gauge theories with/without monopoles. Using
renormalization group equations for the SM, the effective
potential in the two--loop approximation is investigated, and the
existence of its postulated second minimum at the fundamental
scale is confirmed. Phase transitions in the lattice gauge
theories are reviewed. The lattice results for critical coupling
constants are compared with those of the Higgs monopole model, in
which the lattice artifact monopoles are replaced by the
point-like Higgs scalar particles with magnetic charge.
Considering our $(3+1)$--dimensional space--time as, in some way,
discrete or imagining it as a lattice with a parameter
$a=\lambda_P$, where $\lambda_P$ is the Planck length, we have
investigated the additional contributions of monopoles to the
$\beta$--functions of renormalization group equations for running fine
structure constants $\alpha_i(\mu)$ $(i=1,\, 2,\, 3$ correspond to the
$U(1)$, $SU(2)$ and $SU(3)$ gauge groups of the SM) in the FRGGM
extended beyond the SM at high energies. It is shown that
monopoles have $N_{fam}$ times smaller magnetic charge in the
FRGGM than in the SM ($N_{fam}$ is a number of families in the
FRGGM). We have estimated also the enlargement of a number of
fermions in the FRGGM leading to the suppression of the asymptotic
freedom in the non--Abelian theory. We have reviewed that, in
contrast to the case of the Anti--grand unified theory (AGUT),
there exists a possibility of unification of all gauge
interactions (including gravity) near the Planck scale due to
monopoles. The possibility of the $\left[SU(5)\right]^3$ or
$\left[SO(10)\right]^3$
unification at the GUT--scale $\sim 10^{18}$ GeV is briefly
considered.

\end{abstract}

\clearpage\newpage

\pagenumbering{arabic}

\tableofcontents

\clearpage\newpage

\section{Introduction: cosmological constant and the Multiple Point Principle}
\label{sec1}

The contemporary low-energy physics of the electroweak and strong
interactions is described by the Standard Model (SM) which unifies
the Glashow--Salam--Weinberg electroweak theory with QCD --- the
theory of strong interactions.

The gauge symmetry group in the SM is :
\be
  SMG = SU(3)_{c}\times SU(2)_{L}\times U(1)_{Y},
  \lb{1i}
\ee which describes the present elementary particle physics up to
the scale $\approx 100$ GeV.

The vast majority of the available experimental information is
already explained by the SM. All accelerator physics is in
agreement with the SM, except for neutrino oscillations. Presently
only this neutrino physics, together with astrophysics and
cosmology, gives us any phenomenological evidence for going beyond
the SM.

One of the main goals of physics today is to find the fundamental
theory beyond the SM. In first approximation we might ignore the
indications of new physics and consider the possibility that the
SM essentially represents physics well up to the Planck scale.
Developing the ideas of Ref.~\ct{1}, the authors of Ref.~\ct{2}
suggested a scenario, using only the pure SM, in which an
exponentially huge ratio between the fundamental (Planck) and
electroweak scales results:
\be
     \frac{\mu_{fund}}{\mu_{ew}} \sim e^{40}. \lb{A}
\ee This exponentially huge scale ratio occurs due to the required
degeneracy of the three vacuum states (phases) discussed in
Refs.~\ct{2,3,4,5}.

In such a scenario it is reasonable to assume the existence of a
simple and elegant postulate which helps us to explain the SM
parameters: couplings, masses and mixing angles. In model \ct{1,2}
such a postulate is based on a phenomenologically required result
in cosmology \ct{6}: the cosmological constant is zero, or
approximately zero, meaning that the vacuum energy density is very
small. {\it A priori} it is quite possible for a quantum field
theory to have several minima of its effective potential as a
function of its scalar fields. Postulating zero cosmological
constant, we are confronted with a question: is the energy
density, or cosmological constant, equal to zero (or approximately
zero) for all possible vacua or it is zero only for that vacuum in
which we live ?

This assumption would not be more complicated if we postulate that
all the vacua which might exist in Nature, as minima of the
effective potential, should have zero, or approximately zero
cosmological constant. This postulate corresponds to what we call
the Multiple Point Principle (MPP) \ct{7,8,9}.

MPP postulates: {\it there are many vacua with the same energy
density or cosmological constant, and all cosmological constants
are zero or approximately zero.}

There are circa 20 parameters in the SM characterizing the
couplings and masses of the fundamental particles, whose values
can only be understood in speculative models extending the SM. It
was shown in Ref.~\ct{10} that the Family Replicated Gauge Group
Model (FRGGM), suggested in Refs.~\ct{11,12} as an extension of the
SM (see also reviews \ct{13,14}), fits the SM fermion masses and
mixing angles and describes all neutrino experimental data order
of magnitude-wise using only 5 free parameters --- five vacuum
expectation values of the Higgs fields which break the FRGG
symmetry to the SM. This approach based on the FRGG--model was
previously called Anti--Grand Unified Theory (AGUT) and developed
as a realistic alternative to SUSY Grand Unified Theories (GUTs)
\ct{11,12,13,14,15,15a,16,16a,17,18,19}.
In Refs.~\ct{20,21,21aa,21a,21b,21c,21cc,21d,21dd}
the MPP was applied to the investigation
of phase transitions in the regularized gauge theories. A tiny
order of magnitude of the cosmological constant was explained in a
model involving supersymmetry breaking in N=1 supergravity and MPP
\ct{22,23}. The investigation of hierarchy problem in the SM extended
by MPP and two Higgs doublets was developed in Ref.~\ct{24} (see also
\ct{25}). In the recent investigation \ct{26} the MPP was applied to the
flipped $SU(5)\times U(1)$ gauge theory.

The present paper is a review of the MPP implementation to phase
transitions in different gauge theories.

\section{The renormalization group equation for the effective
potential}
\label{sec2}

In the theory of a single scalar field interacting with a gauge
field, the effective potential $V_{eff}(\phi_c)$ is a function of
the classical field $\phi_c$ given by \be
       V_{eff} = - \sum_0^\infty
       \frac{1}{n!}\Gamma^{(n)}(0)\phi_c^n,     \lb{1}
\ee where $\Gamma^{(n)}(0)$ is the one--particle irreducible (1PI)
n--point Green's function calculated at zero external momenta. The
renormalization group equation (RGE) for the effective potential
means that the potential cannot depend on a change in the
arbitrary renormalization scale parameter M:
\be
         \frac{dV_{eff}}{dM} = 0.           \lb{2}
\ee The effects of changing it are absorbed into changes in the
coupling constants, masses and fields, giving so--called running
quantities.

Considering the renormalization group (RG) improvement of the
effective potential \ct{28,29} and choosing the evolution variable as
\be
          t = \log\left(\frac{\mu}{M}\right) = \log\left(\frac{\phi}{M}\right),
  \lb{3}
\ee where $\mu$ is the energy scale, we have the Callan--Symanzik
\ct{30,31} RGE for the full $V_{eff}\left(\phi_c\right)$ with $\phi\equiv
\phi_c$ : \be
     \left(M\frac{\partial}{\partial M} + \beta_{m^2}\frac{\partial}
     {\partial m^2} + \beta_{\lambda}
     \frac{\partial}{\partial \lambda} + \beta_g \frac{\partial}{\partial g}
      + \gamma \phi\frac{\partial}{\partial \phi}\right)
     V_{eff}(\phi) = 0,                            \lb{4}
\ee where M is a renormalization mass scale parameter,
$\beta_{m^2}$, $\beta_{\lambda}$, $\beta_g$ are beta functions for
the scalar mass squared $m^2$, scalar field self--interaction
$\lambda$ and gauge couplings $g$, respectively. Also $\gamma$ is
the anomalous dimension, and the gauge coupling constants: $g_i =
\left(g',\, g,\, g_3\right)$ correspond to the $U(1)_Y$, $SU(2)_L$ and
$SU(3)_c$
groups of the SM. Here the couplings depend on the renormalization
scale M: $\lambda = \lambda(M)$, $m^2 = m^2(M)$ and $g_i =
g_i(M)$. In general, we also consider the top quark Yukawa
coupling $h\stackrel{def}{=}g_t$ and neglect the Yukawa couplings
of all lighter fermions.

It is convenient to introduce a more compact notation for the
parameters of theory. Define: \be
          \lambda_p = \left(m^2,\, \lambda,\, g\right),    \lb{5}
\ee so that the RGE  can be abbreviated as
\be
   \left(M\frac{\partial}{\partial M} + \beta_p\frac{\partial}{\partial
\lambda_p} +
     \gamma  \phi \frac{\partial}{\partial \phi}\right)V_{eff}
     = 0.                                                     \lb{6}
\ee The general solution of the above-mentioned RGE has the
following form \ct{28}: \be
  V_{eff} = - \frac{m^2(\phi)}{2}\left[G(t)\phi\right]^2
            + \frac{\lambda(\phi)}{8}\left[G(t)\phi\right]^4 + C,  \lb{6a}
\ee where \be
            G(t) = \exp\left(-\int_0^t \gamma\left(t'\right)dt'\right).
 \lb{6b}
\ee We shall also use the notation $\lambda(t) = \lambda(\phi)$,
$m^2(t) = m^2(\phi)$, $g_i^2(t) = g_i^2(\phi)$, which should not
lead to any misunderstanding.

\subsection{The second minimum of the Standard Model (SM) effective potential}
\label{subsec2.1}

In this Section our goal is to show the possible existence of a
second (non--standard) minimum of the effective potential in the
pure SM at the fundamental scale \ct{1,2}:
\be
            \phi_{min2} \gg v = \phi_{min1}.      \lb{51b}
\ee
The tree--level Higgs potential with the standard ``weak scale
minimum" at $\phi_{min1} =v$ is given by:
\be
       V\left(\mbox{tree--level}\right) = \frac{\lambda}{8}\left(\phi^2 -
v^2\right)^2 + C.
                                               \lb{52}
\ee In accord with cosmological results, we take the cosmological
constants $C$ for both vacua equal to zero (or approximately
zero): $C=0$ (or $C\approx 0$). The following requirements must be
satisfied in order that the SM effective potential should have two
degenerate minima:
\bee
        V_{eff}\left(\phi_{min1}^2\right) &=&
V_{eff}\left(\phi_{min2}^2\right) = 0,
\lb{53}
\\
        V'_{eff}\left(\phi_{min1}^2\right) &=&
V'_{eff}\left(\phi_{min2}^2\right) = 0,
\lb{54}
\eee where \be
         V'(\phi^2) = \frac{\partial V}{\partial \phi^2}.
                                             \lb{55}
\ee These degeneracy conditions first considered in Ref.~\ct{1}
correspond to the MPP expectation. {\it The first minimum is the
standard ``Weak scale minimum", and the second one is the
non--standard ``Fundamental scale minimum" (if it exists).} An
illustrative schematic picture of $V_{eff}$ is presented in
Fig.~\ref{f1}.

Here we consider the SM theory with zero temperature ($T=0$). As
was shown in Ref.~\ct{1}, the above MPP--requirements lead to the
condition that our electroweak vacuum is barely stable at $T=0$.

With good accuracy, the predictions of Ref.~\ct{1} for the top
quark and Higgs masses from the MPP requirement of a second
degenerate vacuum, together with the identification of its
position with the Planck scale $\phi_{min2} = M_{\rm Planck}$, were as
follows: \be M_t = 173 \pm 5\, {\mbox{GeV}}, \quad M_H = 135 \pm
9\, {\mbox{GeV}}. \lb{56} \ee Later, in Ref.~\ct{32}, an
alternative metastability requirement for the electroweak (first)
vacuum was considered, which gave a Higgs mass prediction of $122
\pm 11$ GeV, close to the LEP lower bound of 115 GeV (Particle
Data Group \ct{33}).

Following Ref.~\ct{1}, let us now investigate the conditions,
Eqs.~(\ref{53}, \ref{54}), for the existence of a second degenerate
vacuum at the fundamental scale: $ \phi_{min2}\sim {\mu}_{fund.}$.

For large values of the Higgs field: $\phi^2 \gg m^2$ the effective
potential $V_{eff}$ is very well approximated by the quartic term
in Eq.~(\ref{6a}) and the degeneracy condition (\ref{53}) gives:
\be
       \lambda\left(\phi_{min2}\right) = 0. \lb{61}
\ee The condition (\ref{54}) for a turning value then gives: \be
 \lambda'\left(\phi_{min2}\right) = 0    \lb{62}
\ee which can be expressed in the form: \be
      \beta_{\lambda}\left(\phi_{min2},\, \lambda=0\right) = 0.  \lb{63}
\ee In Ref.~\ct{2} the scale $\phi_{min2}$ depending on the
experimental data uncertainties was calculated when the degeneracy
conditions (\ref{61}--\ref{63}) were taken into account. For
central values of experimentally observable quantities the result
(\ref{A}) was obtained and gave an exponentially huge ratio
between the fundamental and electroweak scales.

\section{Phase transitions. Triple point of water analogy}
\label{sec3}

In general, it is quite possible that there exist a lot of vacua
in Nature. If several vacua are degenerate then the phase diagram
of theory contains {\it a special point} in which the
corresponding phases meet together (see Fig.~\ref{f2}). This special
point is the Multiple Critical Point (MCP). The phase diagram of
any gauge theory is presented by a space which has axes given by
bare coupling constants (and maybe by bare masses).

Here it is useful to remind you a triple point of water analogy.

It is well known in the thermal physics that in the range of fixed
extensive quantities: volume, energy and a number of moles the
degenerate phases of water (namely, ice, water and vapour
presented in Fig.~\ref{f3}) exist on the phase diagram $(P,\, T)$ of
Fig.~\ref{f4}
at the fine-tuned values of the intensive variables --- pressure
$P$ and temperature $T$: \be T_c\approx 0.01\, ^{\rm o}\mbox{C}
\quad{\mbox{and}} \quad
P_c\approx
  4.58\, {\mbox{mm Hg}},
                            \lb{64}
\ee
giving the critical (triple) point $O$ shown in Fig.~\ref{f4}. This
is a triple point of water analogy.

The idea of the Multiple Point Principle has its origin from the
lattice investigations of gauge theories. In particular, Monte
Carlo simulations of $U(1)$--, $SU(2)$-- and $SU(3)$--gauge theories on
lattice indicate the existence of the triple critical point.

\section{Lattice theories}
\label{sec4}

\subsection{Mathematical structure of lattice gauge theories}
\label{subsec4.1}

A lattice contains sites, links and plaquettes. Link variables
defined on the edges of the lattice are fundamental variables of
the lattice theory. These variables are simultaneously the
elements of the gauge group $G$, describing a symmetry of the
corresponding lattice gauge theory: \be
     {\cal U}(x\,\link y) \in G.
                            \lb{18}
\ee It is easy to understand the sense of this variable turning to
the differential geometry of the continuum space--time in which
our gauge fields exist. Such a space geometrically is equivalent
to curvilinear space and an operator, which compares fields at
different points, is an operator of the parallel transport between
the points $x,\, y$: \be
  {\cal U}(x, y) = Pe^{ig\int_{C_{xy}} A_{\mu}(x)dx^{\mu}}, \lb{19}
\ee where $P$ is the path ordering operator and $C_{xy}$ is a
curve from point $x$ till point $y$. Moreover, the operator:
\be
    W = {\rm Tr}\left(Pe^{ig\oint_C A_{\mu}(x)dx^{\mu}}\right)  \lb{20}
\ee is the well--known Wilson--loop. In the case of scalar field
$\phi(x)$, interacting with gauge field $A_{\mu}$, we have an
additional gauge invariant observable: \be
     \phi^+(y)\left[Pe^{ig\int_{C_{xy}} A_{\mu}(x)dx^{\mu}}\right]\phi(x).
\lb{21}
\ee The link variable (\ref{18}) is a lattice version of
Eq.~(\ref{19}):
\be
      {\cal U}(x\,\link y) = e^{i\Theta_{\mu}(n)} \equiv {\cal U}_{\mu}(n).
                                \lb{22}
\ee This link variable connects the point $n$ and the point $n +
a_{\mu}$, where the index $\mu$ indicates the direction of a link
in the hypercubic lattice with parameter $a$. Considering the
infinitesimal increment of the operator (\ref{22}) in the
continuum limit, we have: \be
             \Theta_{\mu}(n) = a{\hat A}_{\mu}(x),    \lb{23}
\ee where the quantity \be
      {\hat A}_{\mu}(x) = gA^j_{\mu}(x) t^j      \lb{24}
\ee contains the generator $t^j$ of the group $G$ if $G=SU(N)$.

For $G=SU(3)$ we have $t^j = \lambda^j/2$, where $\lambda^j$ are
the well--known Gell--Mann matrices.

For $G=U(1)$: \be
      {\hat A}_{\mu}(x) = gA_{\mu}(x).      \lb{25}
\ee Plaquette variables are not independent because they are
products of link variables:
\be {\cal U}_p\equiv {\cal
U}\left(\square\right){\stackrel{def}{=}}
      {\cal U}\mbox{\plaqb} {\cal U}\mbox{\plaqr} {\cal U}\mbox{\plaqt}
      {\cal U}\mbox{\plaql}.
                                   \lb{26}
\ee

\subsection{Lattice actions}
\label{subsec4.2}

The lattice action $S[\cal U]$ is invariant under the gauge
transformations on a lattice \be
       {\cal U}(x\,\link y) \longrightarrow \Lambda(x)
       {\cal U}(x\,\link y)\Lambda^{-1}(y),
\ee where $\Lambda(x)\in G$.

The simplest action $S[\cal U]$ is given by the expression:
\be
     S[{\cal U}] = \sum_q \frac{\beta_q}{{\mbox{dim}}\, q}\sum_p
                  Re\left( {\rm Tr}\left({\cal U}_p^{(q)}\right)\right).
                                                   \lb{28}
\ee Here $q$ is the index of the representation of the group $G$,
${\mbox{dim}}\, q$ is the dimension of this representation, and
$\beta_q = 1/g^2_q$, where $g_q$ is the coupling constant of gauge
fields corresponding to the representation $q$.

The path integral
\be
   Z = \int D{\cal U}(\,\link )e^{ - S[{\cal U}(\,\link )]},          \lb{29}
\ee which is an analogue of the partition function, describes the
lattice gauge theory in the Euclidean four--dimensional space.

It is necessary to construct the lattice field theory such that,
for $a \to 0$, i.e. in the continuum limit, it leads to a
regularized smooth gauge theory of fields $A_{\mu}^j(x)$, where
$j$ is the symmetry subscript. In the opposite case, a passage to
the continuum limit is not unique \ct{12p}.

Let us consider the simplest case of the group $G = U(1)$, using
the only representation of this group in Eq.~(\ref{28}):
\be
           S\left[{\cal U}_p\right] = \beta \sum_p Re\left({\cal U}_p\right).
       \lb{30}
\ee Here the quantity ${\cal U}_p$ is given by Eq.~(\ref{26}) in
which the link variables ${\cal U}(\,\link )$ are complex numbers
with their moduli equal to unity, i.e.,
 \be
     {\cal U}(x\,\link y) = \left\{z | z\in {\bf C},\, |z| = 1\right\}.   \lb{31}
\ee In the lattice model, the Lorentz gauge condition has the form
\be
           \prod_{x\,\link y}{\cal U}(x\,\link y) = 1.      \lb{32}      
\ee Introducing the notation
\be
             z = e^{i\Theta} ,     \lb{33}
\ee we can write:
\be
            {\cal U}_p = e^{i\Theta_p}.           \lb{34}
\ee The variables ${\cal U}_p$ are not independent, they satisfy
the identity: \be
   \prod_{\Box\in ({\mbox{lattice cube}})}{\cal U}\left(\Box\right) = I,
\lb{35}
\ee called the Bianchi identity. In Eq.~(\ref{35}) the product is
taken over all plaquettes belonging to the cell (cube) of the
hypercubic lattice.

According to Eqs.~(\ref{30}) and (\ref{34}), the simplest lattice
$U(1)$ action has the form: \be
        S\left[{\cal U}_p\right] = \beta \sum_p\cos\Theta_p.    \lb{35a}
\ee For the compact lattice QED: $\beta=1/e_0^2$, where $e_0$ is
the bare electric charge.

The lattice $SU(N)$ gauge theories were first introduced by
K.~Wilson \ct{1s} for studying the problem of confinement. He
suggested the following simplest action:
\be
      S = - \frac{\beta}{N}\sum_p Re\left({\rm Tr}\left({\cal
U}_p\right)\right),          \lb{36}
\ee where the sum runs over all plaquettes of a hypercubic lattice
and ${\cal U}_p\equiv U(\Box)$ belongs to the adjoint
representation of $SU(N)$.

Monte Carlo simulations of these simple Wilson lattice theories in
four dimensions showed a (or an almost) second--order deconfining
phase transition for $U(1)$ \ct{2s,3s}, a crossover behaviour
for $SU(2)$ and $SU(3)$ \ct{4s,5s}, and a first--order phase
transition for $SU(N)$ with $N\ge 4$ \ct{6s}.

Bhanot and Creutz \ct{7s,8s} have generalized the simple
Wilson theory, introducing two parameters in the $SU(N)$ action:
\be
   S = \sum_p\left[-\frac{\beta_f}{N}Re\left({\rm Tr}\left({\cal
U}_p\right)\right) -
               \frac{\beta_A}{N^2-1}Re\left({\rm Tr}_A\left({\cal
U}_p\right)\right)\right],   \lb{37} 
\ee where $\beta_f,\, Tr$ and $\beta_A,\, Tr_A$ are respectively the
lattice constants and traces in the fundamental and adjoint
representations of $SU(N)$.

The phase diagrams, obtained for the generalized lattice $SU(2)$
and $SU(3)$ theories (\ref{37}) by Monte Carlo methods in
Refs.~\ct{7s,8s} (see also \ct{8ss}) are shown in Figs.~\ref{f5a},
\ref{f5b}.
They indicate the existence of a triple point which is a boundary
point of three first--order phase transitions: the ``Coulomb--like"
and $SU(N)/Z_N$ and $Z_N$ confinement phases meet together at this
point. From the triple point emanate three phase border lines
which separate the corresponding phases. The $Z_N$ phase
transition is a discrete transition, occurring when lattice
plaquettes jump from the identity to nearby elements in the group.
The $SU(N)/Z_N$ phase transition is due to a condensation of
monopoles (a consequence of the non--trivial $\Pi_1$ of the group).

The phase diagram of the lattice gauge theory described by the
action with mixed $SU(2)$--$SO(3)$ symmetries \ct{9ss} is presented
in Fig.~\ref{f6}. Here I is a range where the densities of $Z_2$--vortices
($E$) and $Z_2$--monopoles ($M$) accept the values $E \sim M \sim
0.5$. The range II corresponds to $E \sim 0.5,\, M \sim 0$ and in
the range III we have $E \sim M \sim 0$. The closed $Z_2$--vortex
and $Z_2$--monopole (in the three--dimensional lattice) are shown in
Fig.~\ref{f6a} and Fig.~\ref{f6b}, respectively.

Monte Carlo simulations of the $U(1)$ gauge theory, described by
the two-parameter lattice action \ct{9s,10s}:
\be
     S = \sum_p\left[\beta^{lat} \cos \Theta_p + \gamma^{lat}
\cos2\Theta_p\right], \quad
     {\mbox {where}}  \quad {\cal U}_p = e^{i\Theta_p},
                            \lb{38}
\ee also indicate the existence of a triple point on the
corresponding phase diagram: ``Coulomb--like", totally confining
and $Z_2$ confining phases come together at this triple point (see
Fig.~\ref{f7}).

In general, we have a number of phases meeting at the MCP. For
example, Fig.~\ref{f8} demonstrates the meeting of the five phases in the
case of the gauge theory with the $U(1)\times SU(2)$ symmetry
considered in Ref.~\ct{8}.

Recently N.~Arkani--Hamed \ct{10ss} referred to the modern
cosmological theory which assumes the existence of a lot of
degenerate vacua in the Universe.

In Ref.~\ct{10sx} the coexistence of different quantum vacua of our
Universe is explained by the MPP. It was shown that these vacua are
regulated by the baryonic charge and all the coexisting vacua exhibit
the baryonic asymmetry. The present baryonic asymmetry of the Universe
is discussed.

Lattice theories are given in reviews \ct{12p}. The next efforts
of the lattice simulations of the $SU(N)$ gauge theories are
presented in the review \ct{10as}, etc.

\subsection{Lattice artifact monopoles}
\label{subsec4.3}

Lattice monopoles are responsible for the confinement in lattice
gauge theories what is confirmed by many numerical and theoretical
investigations (see reviews \ct{11s} and papers \ct{12s}).

In the compact lattice gauge theory the monopoles are not physical
objects: they are lattice artifacts driven to infinite mass in the
continuum limit. Weak coupling (``Coulomb") phase terminates
because of the appearance for non--trivial topological
configurations which are able to change the vacuum. These
topological excitations are closed monopole loops (or universe
lines of monopole--anti--monopole pairs). When these monopole loops
are long and numerous, they are responsible for the confinement.
But when they are dilute and small, the Coulomb (``free photon")
phase appears. Banks et al. \ct{[22]} have shown that in the
Villain form of the $U(1)$ lattice gauge theory \ct{[23]} it is
easy to exhibit explicitly the contribution of the topological
excitations.  The Villain lattice action is:
\be
  S_V =  \frac{\beta}{2}\sum_p {\left(\theta_p - 2\pi k\right)}^2,  \quad k\in
Z.
                                                \lb{39}
\ee In such a model the partition function $Z$ may be written in a
factorized form: \be
                Z = Z_C Z_M,                     \lb{40}
\ee where $Z_C$ is a part describing the photons: \be
        Z_C \sim \int\limits_{-\infty}^{\infty}d\Theta
      \exp \left[-\frac{\beta}{2} \sum_{\Box}{\Theta}^2\left(\Box\right)\right],
 \lb{41}
\ee and $Z_M$ is the partition function of a gas of the monopoles
\ct{[24],[25]}:
\be
  Z_M \sim \sum_{m\in Z} \exp \left[-2{\pi}^2\beta \sum_{x,\,
y}m(x)v(x-y)m(y)\right].
                                         \lb{42}
\ee
In Eq.~(\ref{42}) $v(x-y)$ is a lattice version of $1/r$--potential
and $m(x)$ is the charge of the monopole, sitting in an elementary
cube $c$ of the dual lattice, which can be simply expressed in
terms of the integer variables $n_p$ :
\be
             m =\sum_{p\in \partial c}n_p,
                                            \lb{43}    
\ee where $n_p$ is a number of Dirac strings passing through the
plaquettes of the cube $c$.

The Gaussian part $Z_C$ provides the usual Coulomb potential,
while the monopole part $Z_M$ leads, at large separations, to a
linearly confining potential.

It is more complicated to exhibit the contribution of monopoles
even in the $U(1)$ lattice gauge theory described by the simple
Wilson action (\ref{35a}).
 Let us consider the Wilson loop as a
rectangle of length $T$ in the 1--direction (time) and width $R$ in
the 2--direction (space--like distance), then we can extract the
potential $V(R)$ between two static charges of opposite signs: \be
         V(R) = - \lim_{T\to\infty}\frac{1}{T}\log(\langle W\rangle),   \lb{44}
\ee and obtain:
\bee
   V(R) &=& - \frac{\alpha (\beta)}{R} \quad -  \quad
        \mbox{in ``Coulomb" phase},                     \lb{45}
\\
  V(R) &=& \sigma R -\frac{\alpha (\beta)}{R} +
O\left(\frac{1}{R^3}\right) +
\mbox{const}\mbox{ -- in confinement phase}.
                                                 \lb{46}
\eee

\subsection{The behaviour of electric fine structure constant $\alpha$
near the phase transition point. ``Freezing" of $\alpha$}
\label{subsec4.4}

The lattice investigators were not able to obtain the lattice
triple point values of $\alpha_{i,\, crit}$ for $i=2,\, 3$ by Monte Carlo
simulations method. Only the critical value of the electric fine
structure constant $\alpha(\beta)$ was obtained in Ref.~\ct{10s} in the
compact QED described by the Wilson and Villain actions
(\ref{35a}) and (\ref{39}) respectively:
\be
\alpha_{crit}^{lat} =
0.20\pm 0.015 \quad {\mbox{and}}  \quad {\tilde \alpha}_{crit}^{lat}
= 1.25\pm 0.10  \quad{\mbox{at}} \quad
\beta_T\equiv\beta_{crit}\approx{1.011}.
                                                     \lb{47}
\ee Here
\be
             \alpha = \frac{e^2}{4\pi} \quad {\mbox{and}} \quad
             \tilde \alpha = \frac{g^2}{4\pi},
                                                     \lb{47a}
\ee
where $g$ is a magnetic charge of monopoles.

Using the Dirac relation for elementary charges (see below
Subsection 11.1), we have: \be
   eg = 2\pi,  \quad{\mbox{or}} \quad \alpha \tilde \alpha =
   \frac{1}{4}.
                                             \lb{54y}
\ee The behaviour of $\alpha(\beta)$ in the vicinity of the phase
transition point $\beta_T$ (given by Ref.~\ct{10s}) is shown in
Fig.~\ref{f9a} for the Wilson and Villain lattice actions. Fig.~\ref{f9b}
demonstrates the comparison of the function $\alpha(\beta)$
obtained by Monte Carlo method for the Wilson lattice action and
by theoretical calculation of the same quantity. The theoretical
(dashed) curve was calculated by so--called ``Parisi improvement
formula" \ct{13p}: \be
    \alpha (\beta )=\left[4\pi \beta W_p\right]^{-1}.     \lb{48}
\ee Here $W_p=\langle\cos \Theta_p \rangle$ is a mean value of the plaquette
energy. The corresponding values of $W_p$ are taken from
Ref.~\ct{9s}.

The theoretical value of $\alpha_{crit}$ is less than the
``experimental" (Monte Carlo) value (\ref{47}):
\be
      \alpha_{crit}\left(\mbox{lattice theory}\right)\approx{0.12}.
                                                    \lb{49}
\ee This discrepancy between the theoretical and ``experimental"
results is described by monopole contributions: the fine structure
constant $\alpha$ is renormalised by an amount proportional to the
susceptibility of the monopole gas \ct{[24]}: \be
          K = \frac{{\alpha}_{crit}\left(\mbox{Monte Carlo}\right)}{{\alpha}_
              {crit}\left(\mbox{lattice theory}\right)}\\
              \approx{\frac{0.20}{0.12}}\approx{1.66}.   \lb{50}
\ee Such an enhancement of the  critical fine structure constant
is due to vacuum monopole loops \ct{[25]}.

According to Fig.~\ref{f9c}: \be
                 \alpha_{crit.,\, theor.}^{-1}\approx 8.5.    \lb{50a}
\ee This result does not coincide with the lattice result
(\ref{47}) which gives the following value:
\be
                 \alpha_{crit.,\, theor.}^{-1}\approx 5.    \lb{50b}
\ee The deviation of theoretical calculations of $\alpha(\beta )$
from the lattice ones, which is shown in Figs.~\ref{f9b}, \ref{f9c}, has the
following explanation: ``Parisi improvement formula" (\ref{48}) is
valid in Coulomb phase where the mass of artifact monopoles is
infinitely large and photon is massless. But in the vicinity of
the phase transition (critical) point the monopole mass $m\to 0$
and photon acquires the non--zero mass $m_0\neq 0$ in the
confinement range. This phenomenon leads to the ``freezing" of
$\alpha$: the effective electric fine structure constant is almost
unchanged in the confinement phase and approaches to its maximal
value $\alpha=\alpha_{max}$. The authors of Ref.~\ct{14p} predicted
that in the confinement phase, where we have the formation of
strings, the fine structure constant $\alpha$ cannot be infinitely
large, but has the maximal value: \be \alpha_{max} =
         \frac{\pi}{12}\approx 0.26, \lb{50c}
\ee due to the Casimir effect for strings. The authors of
Ref.~\ct{14pa} developed this viewpoint in the spinor QED: the
vacuum polarization induced by thin ``strings"--vortices of the
magnetic flux leads to the suggestion of an analogue of the
``spaghetti vacuum" \ct{14pb} as a possible mechanism for avoiding
the divergences in the perturbative QED. According to
Ref.~\ct{14pa}, the non--perturbative sector of QED arrests the
growth of the effective $\alpha$ to infinity and confirms the
existence of $\alpha_{max}$.

We see that Fig.~\ref{f9a} demonstrates the tendency to freezing of
$\alpha$ in the compact QED.

The analogous ``freezing" of $\alpha_s$ was considered in QCD in
Ref.~\ct{15p}.

\section{The Higgs Monopole Model and phase transition in the
regularized $U(1)$ gauge theory}
\label{sec5}

The simplest effective dynamics describing the confinement
mechanism in the pure gauge lattice $U(1)$ theory is the dual
Abelian Higgs model of scalar monopoles \ct{13s} (see also
Refs.~\ct{11s} and \ct{12s}).

In the previous papers \ct{8} and \ct{18} the calculations of the
$U(1)$ phase transition (critical) coupling constant were connected
with the existence of artifact monopoles in the lattice gauge
theory and also in the Wilson loop action model \ct{18}.

In Ref.~\ct{18} the authors have put forward the speculations of
Refs.~\ct{8} and \ct{13} suggesting that the modifications of the
form of the lattice action might not change too much the phase
transition value of the effective continuum coupling constant. The
purpose was to investigate this approximate stability of the
critical coupling with respect to a somewhat new regularization
being used instead of the lattice, rather than just modifying the
lattice in various ways. In \ct{18} the Wilson loop action was
considered in the approximation of circular loops of radii $R\ge
a$. It was shown that the phase transition coupling constant is
indeed approximately independent of the regularization method:
${\alpha}_{crit}\approx{0.204}$, in correspondence with the Monte
Carlo simulation result on lattice:
${\alpha}_{crit}\approx{0.20\pm 0.015}$ (see Eq.~(\ref{47})).

But in Refs.~\ct{20,21,21aa,21a,21b,21c,21cc,21d}
instead of using the lattice or
Wilson loop cut--off we have considered the Higgs Monopole Model
(HMM) approximating the lattice artifact monopoles as fundamental
point--like particles described by the Higgs scalar fields.
Considering the renormalization group improvement of the effective
Coleman--Weinberg potential \ct{28,29}, written in
Ref.~\ct{21} for the dual sector of scalar electrodynamics in the
two--loop approximation, we have calculated the $U(1)$ critical
values of the magnetic fine structure constant:
\be{\tilde\alpha}_{crit} = \frac{g^2_{crit}}{4\pi}\approx 1.20\ee and
electric fine structure constant \be\alpha_{crit} =
\frac{\pi}{g^2_{crit}}\approx 0.208  \quad \left(\mbox{by the Dirac
relation}\right).\ee These values coincide with the lattice result
(\ref{47}). The next Subsections follow the review of the
HMM calculations of the $U(1)$ critical couplings obtained in
Refs.~\ct{21,21aa,21a}.

\subsection{The Coleman--Weinberg effective potential for the HMM}
\label{subsec5.1}

As it was mentioned above, the dual Abelian Higgs model of scalar
monopoles (shortly HMM) describes the dynamics of the confinement
in lattice theories. This model, first suggested in Ref.~\ct{13s},
considers the following Lagrangian:
$$
    L = - \frac{1}{4g^2} F_{\mu\nu}^2(B) + \frac{1}{2} \left|
\left(\partial_{\mu} -
           iB_{\mu} \right)\Phi\right|^2 - U(\Phi),$$
where
\be
 U(\Phi) = \frac{1}{2}\mu^2 {|\Phi|}^2 +
 \frac{\lambda}{4}{|\Phi|}^4   \lb{5y}
\ee is the Higgs potential of scalar monopoles with magnetic
charge $g$, and $B_{\mu}$ is the dual gauge (photon) field
interacting with the scalar monopole field $\Phi$.  In this theory
the parameter $\mu^2$ is negative. In Eq.~(\ref{5y}) the complex
scalar field $\Phi$ contains the Higgs ($\phi$) and Goldstone
($\chi$) boson fields: \be
          \Phi = \phi + i\chi.             \lb{7y}
\ee The effective potential in the Higgs model of scalar
electrodynamics was first calculated by Coleman and Weinberg
\ct{28} in the one--loop approximation. The general method of its
calculation is given in the review \ct{29}. Using this method, we
can construct the effective potential for HMM. In this case the
total field system of the gauge  $\left(B_{\mu}\right)$ and magnetically
charged ($\Phi$) fields is described by the partition function
which has the following form in Euclidean space:
\be
      Z = \int [DB][D\Phi] \left[D\Phi^{+}\right]e^{-S},     \lb{8y}
\ee where the action $S = \int d^4x L(x) + S_{gf}$ contains the
Lagrangian (\ref{5y}) written in Euclidean space and gauge fixing
action $S_{gf}$.

Let us consider now a shift: \be
 \Phi (x) = \Phi_b + {\hat \Phi}(x)                \lb{9y}
\ee with $\Phi_b$ as a background field and calculate the
following expression for the partition function in the one-loop
approximation:
\bee
  Z &=& \int [DB] \left[D\hat \Phi\right] \left[D{\hat \Phi}^{+}\right]
   \exp \left\{ - S \left(B,\Phi_b\right)
- \int d^4x  \left[\left.\frac{\delta S(\Phi)}{\delta
\Phi(x)} \right|_{\Phi=
   \Phi_b}{\hat \Phi}(x) + h.c. \right]\right\}\nonumber\\
    &=&\exp \left\{ - F \left(\Phi_b,\, g^2,\, \mu^2,\,
\lambda\right)\right\}.      \lb{10y}
\eee Using the representation (\ref{7y}), we obtain the effective
potential: \be
  V_{eff} = F \left(\phi_b,\, g^2,\, \mu^2,\, \lambda\right)        \lb{11y}
\ee given by the function $F$ of Eq.~(\ref{10y}) for the real
constant background field $ \Phi_b = \phi_b = \mbox{const}$. In
this case the one--loop effective potential for monopoles
coincides with the expression of the effective potential
calculated by the authors of Ref.~\ct{28} for scalar
electrodynamics and extended to the massive theory (see review
\ct{29}):
\bee
 V_{eff}(\phi_b^2) &=& \frac{\mu^2}{2} {\phi_b}^2 +
                 \frac{\lambda}{4} {\phi_b}^4
  + \frac{1}{64\pi^2}\left[ 3g^4
{\phi_b}^4\log\left(\frac{\phi_b^2}{M^2}\right)
+  { \left(\mu^2 +
3\lambda {\phi_b}^2\right)}^2\log\left(\frac{\mu^2 +
3\lambda\phi_b^2}{M^2}\right)\right. \nonumber\\
&&\left.+ { \left(\mu^2 +\lambda
\phi_b^2\right)}^2\log\left(\frac{\mu^2
 + \lambda \phi_b^2}{M^2}\right)\right] + C,
                           \lb{12y}
\eee where $M$ is the cut--off scale and $C$ is a constant not
depending on $\phi_b^2$.

The effective potential (\ref{11y}) has several minima. Their
position depends on $g^2,\, \mu^2$ and $\lambda$. If the first local
minimum occurs at $\phi_b=0$ and $V_{eff}(0)=0$, it corresponds to
the so--called ``symmetrical phase", which is the Coulomb--like phase
in our description. Then it is easy to determine the constant C in
Eq.~(\ref{12y}): \be
        C = - \frac{\mu^4}{16{\pi^2}}\log\left( \frac {\mu}M\right),  \lb{13y}
\ee and we have the effective potential for HMM described by the
following expression:
\be V_{eff}(\phi^2_b) =
\frac{\mu^2_{run}}{2}\phi_b^2 + \frac{\lambda_{run}}{4}\phi_b^4
     + \frac{\mu^4}{64\pi^2}\log\left(\frac{ \left(\mu^2 + 3\lambda
\phi_b^2\right) \left(\mu^2 +
       \lambda \phi_b^2\right)}{\mu^4}\right).
                                                \lb{14y}
\ee Here $\lambda_{run}$ is the running scalar field
self--interaction constant given by the expression standing in
front of $\phi_b^4$ in Eq.~(\ref{12y}):
\be
  \lambda_{run}(\phi_b^2)
   = \lambda + \frac{1}{16\pi^2}  \left[ 3g^4\log\left(
\frac{\phi_b^2}{M^2}\right)
   + 9{\lambda}^2\log\left(\frac{\mu^2 + 3\lambda \phi_b^2}{M^2}\right)+
     {\lambda}^2\log\left(\frac{\mu^2 + \lambda\phi_b^2}{M^2}\right)\right].
\lb{15y}
\ee The running squared mass of the Higgs scalar monopoles also
follows from Eq.~(\ref{12y}):
\be
   \mu^2_{run}(\phi_b^2)
   = \mu^2 + \frac{\lambda\mu^2}{16\pi^2} \left[ 3\log\left(\frac{\mu^2 +
   3\lambda \phi_b^2}{M^2}\right) + \log\left(\frac{\mu^2 +
\lambda\phi_b^2}{M^2}\right)\right].
                                  \lb{16y}
\ee As it was shown in Ref.~\ct{28}, the effective potential can be
improved by consideration of the renormalization group equation
(RGE).

\subsection{Renormalization group equations in the HMM}
\label{subsec5.2}

The RGE for the effective potential are given by
Eqs.~(\ref{4}--\ref{6b}). A set of ordinary differential equations
(RGE) corresponds to Eq.~(\ref{4}): \bee
    \frac{d\lambda_{run}}{dt} &=& \beta_{\lambda}\left(g_{run}(t),\,
                    \lambda_{run}(t)\right),      \lb{22y}
\\
    \frac{d\mu^2_{run}}{dt} &=& \mu^2_{run}(t)\beta_{(\mu^2)}
             \left(g_{run}(t),\, \lambda_{run}(t)\right),
                                                  \lb{23y}
\\
    \frac{dg^2_{run}}{dt} &=&
\beta_g\left(g_{run}(t),\, \lambda_{run}(t)\right).
                                               \lb{24y}
\eee So far as the mathematical structure of HMM is equivalent to
the Higgs scalar electrodynamics, we can use all results of the
last theory in our calculations, replacing the electric charge $e$
and photon field $A_{\mu}$ by magnetic charge $g$ and dual gauge
field $B_{\mu}$.

Let us write now the one--loop potential (\ref{14y}) as \be
         V_{eff} = V_0 + V_1,          \lb{25y}
\ee where
$$
   V_0 = \frac{\mu^2}2 \phi^2 + \frac{\lambda}4 \phi^4,
   $$
\bee
  V_1 &=& \frac{1}{64\pi^2}\left[ 3g^4
{\phi}^4\log\left(\frac{\phi^2}{M^2}\right)
+ {\left(\mu^2 + 3\lambda {\phi}^2\right)}^2\log\left(\frac{\mu^2 +
3\lambda\phi^2}{M^2}\right)\right.
\nonumber\\
    &&+\left. { \left(\mu^2 +\lambda
\phi^2\right)}^2\log\left(\frac{\mu^2
     + \lambda \phi^2}{M^2}\right) -
2\mu^4\log\left(\frac{\mu^2}{M^2}\right)\right].
                                                    \lb{27y}
\eee We can plug this $V_{eff}$ into RGE (\ref{4}) and obtain the
following equation (see \ct{29}):
\be
    \left( \beta_{\lambda}\frac{\partial}{\partial \lambda} +
    \beta_{(\mu^2)}{\mu^2}\frac{\partial}{\partial \mu^2} -
    \gamma \phi^2 \frac{\partial}{\partial \phi^2}\right) V_0 =
     - M^2\frac{\partial V_1}{\partial M^2}.
                                           \lb{28y}
\ee Equating $\phi^2$ and $\phi^4$ coefficients, we obtain the
expressions of $\beta_{\lambda}$ and $\beta_{(\mu^2)}$ in the
one--loop approximation: \bee
    \beta_{\lambda}^{(1)}
&=& 2\gamma \lambda_{run}
           + \frac{5\lambda_{run}^2}{8\pi^2} +
                \frac{3g_{run}^4}{16\pi^2},         \lb{29y}
\\
    \beta_{(\mu^2)}^{(1)}
&=& \gamma + \frac{\lambda_{run}}{4\pi^2}. \lb{30y} \eee The
one--loop result for $\gamma$ is given in Ref.~\ct{28} for scalar
field with electric charge $e$, but it is easy to rewrite this
$\gamma$--expression for monopoles with charge $g=g_{run}$: \be
          \gamma^{(1)} = - \frac{3g_{run}^2}{16\pi^2}.   \lb{30ay}
\ee Finally we have: \bee \frac{d\lambda_{run}}{dt}&\approx&
\beta_{\lambda}^{(1)} = \frac 1{16\pi^2}  \left( 3g^4_{run} +10
\lambda^2_{run} - 6\lambda_{run}g^2_{run}\right),
                                     \lb{31y}
\\ \frac{d\mu^2_{run}}{dt}&\approx& \beta_{(\mu^2)}^{(1)} =
\frac{\mu^2_{run}}{16\pi^2} \left( 4\lambda_{run} - 3g^2_{run} \right).
                                                \lb{32y}
\eee The expression of $\beta_g$--function in the one--loop
approximation also is given by the results of Ref.~\ct{28}: \be
    \frac{dg^2_{run}}{dt}\approx
     \beta_g^{(1)} = \frac{g^4_{run}}{48\pi^2}.  \lb{33y}
\ee The RG $\beta$--functions for different renormalizable gauge
theories with semisimple group have been calculated in the
two--loop approximation \ct{22y,23y,24y,25y,26y,27y} and even beyond
\ct{28y}. But in this paper we made use the results of
Refs.~\ct{22y} and \ct{25y} for calculation of $\beta$--functions
and anomalous dimension in the two--loop approximation, applied to
the HMM with scalar monopole fields. The higher approximations
essentially depend on the renormalization scheme \ct{28y}. Thus,
on the level of two--loop approximation we have for all
$\beta$--functions: \be
  \beta = \beta^{(1)} + \beta^{(2)},           \lb{34y}
\ee where
\be
  \beta_{\lambda}^{(2)} = \frac{1}{ \left(16\pi^2\right)^2} \left( - 25\lambda^3
+
   \frac{15}{2}g^2{\lambda}^2 - \frac{229}{12}g^4\lambda -
\frac{59}{6}g^6\right).
                                                    \lb{35y}
\ee and \be \beta_{(\mu^2)}^{(2)} =
\frac{1}{ \left(16\pi^2\right)^2} \left(\frac{31}{12}g^4 + 3\lambda^2\right).
                                             \lb{36y}
\ee The gauge coupling $\beta_g^{(2)}$--function is given by
Ref.~\ct{22y}: \be
     \beta_g^{(2)} = \frac{g^6}{ \left(16\pi^2\right)^2}.  \lb{37y}
\ee Anomalous dimension follows from calculations made in
Ref.~\ct{25y}: \be
    \gamma^{(2)} = \frac{1}{ \left(16\pi^2\right)^2}\frac{31}{12}g^4.
                                                   \lb{38y}
\ee In Eqs.~(\ref{34y}--\ref{38y}) and below, for simplicity, we
have used the following notations: $\lambda\equiv \lambda_{run}$,
$g\equiv g_{run}$ and $\mu\equiv \mu_{run}$.

\subsection{The phase diagram in the HMM}
\label{subsec5.3}

Let us apply the effective potential calculation as a technique
for the getting phase diagram information for the condensation of
monopoles in HMM. As it was mentioned in the Subsection \ref{subsec5.1}, the
effective potential (\ref{11y}) can have several minima. Their
positions depend on $g^2$, $\mu^2$ and $\lambda$: \be
             \phi_0 = \phi_{min1} = f \left(g^2,\, \mu,\, \lambda\right).
\lb{38yy}
\ee

The first local minimum at $\phi_0 = 0$ and $V_{eff}(0)
= 0$ corresponds to the ``symmetric", or Coulomb--like phase,
presented in Fig.~\ref{f10}. In the case when the effective potential has
the second local minimum at $\phi_0 = \phi_{min2} \neq 0$ with
$V_{eff}^{min} \left(\phi_{min2}^2\right) < 0$, we have the confinement
phase (see Fig.~\ref{f11}). The phase transition between the
Coulomb--like and confinement phases is given by degeneracy of the
first local minimum (at $\phi_0 = 0$) with the second minimum (at
$\phi_0 = \phi_{min2}$). These degenerate minima are shown in
Fig.~\ref{f12} by the solid curve 1. They correspond to the different
vacua arising in the present model. The dashed curve 2 in Fig.~\ref{f12}
describes the appearance of two minima corresponding to the
confinement phases (see details in Subsection \ref{subsec5.5}).

The conditions of the existence of degenerate vacua are given by
the following equations: \be
           V_{eff}(0) = V_{eff} \left(\phi_0^2\right) = 0,     \lb{39y}
\ee \be
    \left.\frac{\partial V_{eff}}{\partial \phi}\right|_{\phi=0} =
    \left.\frac{\partial V_{eff}}{\partial \phi}\right|_{\phi=\phi_0} = 0,
 \quad{\mbox{or}} \quad V'_{eff} \left(\phi_0^2\right)\equiv
\left.\frac{\partial
V_{eff}}{\partial \phi^2}\right|_{\phi=\phi_0} = 0,
                                                    \lb{40y}
\ee and inequalities \be
     \left.\frac{\partial^2 V_{eff}}{\partial \phi^2}\right|_{\phi=0} > 0,
\quad
     \left.\frac{\partial^2 V_{eff}}{\partial \phi^2}\right|_{\phi=\phi_0} >
0.
                                               \lb{41y}
\ee The first equation (\ref{39y}), applied to Eq.~(\ref{6a}),
gives: \be
    \mu^2_{run} = - \frac{1}{2} \lambda_{run} \left(t_0\right) \phi_0^2 G^2
\left(t_0\right),
 \quad{\mbox{where}} \quad t_0 = \log \left(\frac{\phi_0^2}{M^2}\right).
                                    \lb{42y}
\ee Calculating the first derivative of $V_{eff}$ given by
Eq.~(\ref{40y}), we obtain the following expression: \bee
      V'_{eff} \left(\phi^2\right) &=& \frac{V_{eff}
\left(\phi^2\right)}{\phi^2} \left(1 +
 2\frac{d\log (G)}{dt}\right) + \frac 12 \frac{d\mu^2_{run}}{dt} G^2(t)
\nonumber\\
      &&+ \frac 14 \left(\lambda_{run}(t) + \frac{d\lambda_{run}}{dt}
+
      2\lambda_{run}\frac{d\log (G)}{dt}\right)G^4(t)\phi^2.
                                                \lb{45y}
\eee From Eq.~(\ref{6b})  we have: \be
          \frac{d\log (G)}{dt} = - \frac{1}{2}\gamma .  \lb{46y}
\ee It is easy to find the joint solution of equations \be
      V_{eff} \left(\phi_0^2\right) = V'_{eff} \left(\phi_0^2\right) = 0.
\lb{47y}
\ee Using RGE (\ref{22y}), (\ref{23y}) and Eqs.~(\ref{42y}--\ref{46y}),
we obtain: \be
 V'_{eff} \left(\phi_0^2\right) =\frac{1}{4} \left( -
\lambda_{run}\beta_{(\mu^2)} +
\lambda_{run} + \beta_{\lambda} - \gamma
\lambda_{run}\right)G^4 \left(t_0\right)\phi_0^2 = 0,
                                                    \lb{48y}
\ee or \be
    \beta_{\lambda} + \lambda_{run} \left(1 - \gamma - \beta_{(\mu^2)}\right) =
0.
                                            \lb{49y}
\ee Putting into Eq.~(\ref{49y}) the functions
$\beta_{\lambda}^{(1)},\, \beta_{(\mu^2)}^{(1)}$ and $\gamma^{(1)}$
given by Eqs.~(\ref{29y}--\ref{30ay}) and (\ref{33y}), we obtain
in the one--loop approximation the following equation for the
phase transition border: \be
     g^4_{PT} = - 2\lambda_{run} \left(\frac{8\pi^2}3 + \lambda_{run}\right).
                                                 \lb{50y}
\ee The curve (\ref{50y}) is represented on the phase diagram
$ \left(\lambda_{run};\, g^2_{run}\right)$ of Fig.~\ref{f13} by the curve
``1" which
describes the border between the ``Coulomb--like" phase with
$V_{eff} \ge 0$ and the confinement one with $V_{eff}^{min} < 0$.
This border corresponds to the one--loop approximation.

Using Eqs.~(\ref{29y}--\ref{30ay}) and (\ref{33y}--\ref{38y}), we
are able to construct the phase transition border in the two--loop
approximation. Substituting these equations into Eq.~(\ref{49y}),
we obtain the following phase transition border curve equation in
the two--loop approximation: \be
 3y^2 - 16\pi^2 + 6x^2 + \frac{1}{16\pi^2} \left(28x^3 + \frac{15}{2}x^2y +
  \frac{97}{4}xy^2 - \frac{59}{6}y^3\right) = 0,            \lb{51y}
\ee where $x = - \lambda_{PT}$ and $y = g^2_{PT}$ are the phase
transition values of $ - \lambda_{run}$ and $g^2_{run}$. Choosing
the physical branch corresponding to $g^2 \ge 0$ and $g^2\to 0$
when $\lambda \to 0$, we have received the curve 2 on the phase
diagram $ \left(\lambda_{run};\, g^2_{run}\right)$ shown in Fig.~\ref{f13}. This
curve
corresponds to the two--loop approximation and can be compared with
the  curve 1 of Fig.~\ref{f13}, which describes the same phase border
calculated in the one--loop approximation. It is easy to see that
the accuracy of the one--loop approximation is not excellent and can
commit errors of order 30\%.

According to the phase diagram drawn in Fig.~\ref{f13}, the confinement
phase begins at $g^2 = g^2_{max}$ and exists under the phase
transition border line in the region $g^2 \le g^2_{max}$, where
$e^2$ is large: $e^2\ge  \left(2\pi/g_{max}\right)^2$ due to the Dirac
relation (see Eq.~(\ref{54y})). Therefore, we have:
\bee
g^2_{crit} &=& g^2_{max1}\approx 18.61  \quad - \quad \mbox{in the
one--loop approximation},
\nonumber\\
   g^2_{crit} &=& g^2_{max2}\approx
  15.11  \quad - \quad \mbox{in the two--loop approximation}.  \lb{52y}
\eee We see the deviation of results of order 20\%. The results
(\ref{52y}) give:
\bee
   \tilde \alpha_{crit} &=& \frac {g^2_{crit}}{4\pi}\approx 1.48
 \quad - \quad \mbox{in the one--loop approximation},
\nonumber\\
   \tilde \alpha_{crit} &=& \frac {g^2_{crit}}{4\pi}\approx 1.20
   \quad - \quad \mbox{in the two--loop approximation}.
                                                  \lb{53y}
\eee Using the Dirac relation (\ref{54y}): $\alpha \tilde \alpha =
1/4$, we obtain the following values for the critical electric
fine structure constant:
\bee
        \alpha_{crit} &=& \frac{1}{4{\tilde \alpha}_{crit}}\approx 0.17
 \quad - \quad \mbox{in the one--loop approximation},
\nonumber\\
        \alpha_{crit} &=& \frac{1}{4{\tilde \alpha}_{crit}}\approx
        0.208
   \quad - \quad \mbox{in the two--loop approximation}.
                                               \lb{55y}
\eee The last result coincides with the lattice values (\ref{47})
obtained for the compact QED by Monte Carlo method \ct{10s}.

Writing Eq.~(\ref{24y}) with $\beta_g$--function given by
Eqs.~(\ref{33y}, \ref{34y}) and (\ref{37y}), we have the following
RGE for the monopole charge in the two--loop approximation:
\be
              \frac{dg^2_{run}}{dt}\approx \frac{g^4_{run}}{48\pi^2}
     + \frac{g^6_{run}}{\left(16\pi^2\right)^2},          \lb{55yy}
\ee or
\be
    \frac{d\log (\tilde \alpha)}{dt}\approx \frac{\tilde
    \alpha}{12\pi}\left(1 + 3\frac{\tilde \alpha}{4\pi}\right).    \lb{56yy}
\ee The values (\ref{52y}) for $g^2_{crit} = g^2_{max1,\, 2}$
indicate that the contribution of two loops described by the
second term of Eq.~(\ref{55yy}), or Eq.~(\ref{56yy}), is about 30\%,
confirming the validity of perturbation theory.

In general, we are able to estimate the validity of the two--loop
approximation for all $\beta$--functions and $\gamma$, calculating
the corresponding ratios of the two--loop contributions to the
one--loop contributions at the maxima of curves 1 and 2: \be
\ba{|l|l|}
\hline %
&\\[-0.2cm]%
\lambda_{crit} = \lambda_{run}^{max1}
\approx{-13.16}&\lambda_{crit} =
\lambda_{run}^{max2}\approx{-7.13}\\[0.5cm]
g^2_{crit} = g^2_{max1}\approx{18.61}& g^2_{crit} = g^2_{max2}
\approx{15.11}\\[0.5cm]
\frac{\ds\gamma^{(2)}}{\ds\gamma^{(1)}}\approx{-0.0080}&\frac{\ds
\gamma^{(2)}}{\ds\gamma^{(1)}}\approx{-0.0065}\\[0.5cm]
\frac{\ds\beta_{\mu^2}^{(2)}}{\ds\beta_{\mu^2}^{(1)}}\approx{-0.0826}
&\frac{\ds\beta_{\mu^2}^{(2)}}{\ds\beta_{\mu^2}^{(1)}}
\approx{-0.0637}\\[0.8cm]
\frac{\ds\beta_{\lambda}^{(2)}}{\ds\beta_{\lambda}^{(1)}}\approx{0.1564}
&\frac{\ds\beta_{\lambda}^{(2)}}
{\ds\beta_{\lambda}^{(1)}}\approx{0.0412}\\[0.8cm]
\frac{\ds\beta_g^{(2)}}{\ds\beta_g^{(1)}}\approx{0.3536}&\frac{\ds
\beta_g^{(2)}}{\ds\beta_g^{(1)}}\approx{0.2871}\\[0.5cm]
\hline
\ea
                                         \lb{57y}
\ee Here we see that all ratios are sufficiently small, i.e. all
two--loop contributions are small in comparison with one--loop
contributions, confirming the validity of perturbation theory in
the two--loop approximation, considered in this model. The accuracy
of deviation is worse ($\sim 30$\%) for $\beta_g$--function. But
it is necessary to emphasize that calculating the border curves 1
and 2 of Fig.~\ref{f13}, we have not used RGE (\ref{37y}) for monopole
charge: $\beta_g$--function is absent in Eq.~(\ref{49y}).
Therefore, the calculation of $g^2_{crit}$ according to
Eq.~(\ref{51y}) does not depend on the approximation of
$\beta_g$--function. The above--mentioned $\beta_g$--function
appears only in the second order derivative of $V_{eff}$ which is
related with the monopole mass $m$ (see Subsection \ref{subsec5.5}).

Eqs.~(\ref{47}) and (\ref{55y}) give the result (\ref{50b}):
\be
          \alpha_{crit}^{-1}\approx 5.            \lb{56y}
\ee which is important for the phase transition at the Planck
scale predicted by the MPP.

\subsection{Approximate universality of the critical coupling
constants}
\label{subsec5.4}

The review of all existing results for $\alpha_{crit}$ and
${\tilde \alpha}_{crit}$ gives:

\begin{enumerate}
\item[1.]

\be \alpha_{crit}^{lat} = 0.20\pm 0.015 \quad {\mbox{and}}  \quad
{\tilde \alpha}_{crit}^{lat} = 1.25\pm 0.10  \lb{1A}\ee --- in the
compact QED with the Wilson lattice action \ct{10s};

\item[2.]

\be \alpha_{crit}^{lat}\approx 0.204, \quad  {\tilde
\alpha}_{crit}^{lat}\approx 1.25  \lb{2A} \ee --- in the model with
the Wilson loop action \ct{18};

\item[3.]

\be
    \alpha_{crit}\approx 0.18,  \quad {\tilde \alpha}_{crit}\approx
    1.36     \lb{3A}
\ee --- in the compact QED with the Villain lattice action
\ct{[23]};

\item[4.]

\be
  \alpha_{crit} = \alpha_A \approx 0.208,  \quad
  {\tilde \alpha}_{crit} ={\tilde \alpha}_A \approx 1.20  \lb{A4}
\ee --- in the HMM \ct{21,21a}.
\end{enumerate}

It is necessary to emphasize, that the functions $\alpha(\beta)$
in Fig.~\ref{f9a}, describing the effective electric fine structure
constant in the vicinity of the phase transition point
$\beta_{crit}\approx 1$, are different for the Wilson and Villain
lattice actions in the $U(1)$ lattice gauge theory, but the critical
values of $\alpha(\beta)$ coincide for both theories \ct{10s}.

Hereby we see an additional arguments for the previously hoped
\ct{8,18} ``approximate universality" of the first order
phase transition critical coupling constants: for example, at the
phase transition point the fine structure constant $\alpha$ is
approximately the same one for various parameters and different
regularization schemes.

The most significant conclusion of MPP, which predicts the values
of gauge couplings arranging just the MCP, where all phases of the
given theory meet, is possibly that the calculations of
Refs.~\ct{8,18} suggest the validity of the approximate
universality of the critical couplings \ct{15a,17}. It was shown in
Refs.~\ct{20,21,21aa,21a,21b,21c,21cc,21d}
that one can crudely calculate the phase
transition couplings without using any specific lattice, rather
only approximating the lattice artifact monopoles as fundamental
(point--like) magnetically charged particles condensing. Thus, the
details of the lattice --- hypercubic or random, with
multi--plaquette terms or without them, etc., --- also the details
of the regularization --- lattice or Wilson loops, lattice or the
Higgs monopole model --- do not matter for values of the phase
transition couplings so much. Critical couplings depend only on
groups with any regularization. Such an approximate universality
is, of course, absolutely needed if there is any sense in relating
lattice phase transition couplings to the experimental couplings
found in Nature. Otherwise, such a comparison would only make
sense if we could guess the true lattice in the right model, what
sounds too ambitious.

\subsection{Triple point of the HMM phase diagram }
\label{subsec5.5}

In this Section we demonstrate the existence of the triple point
on the phase diagram of HMM \ct{21}.

Considering the second derivative of the effective potential: \be
                V''_{eff}\left(\phi_0^2\right)
    \equiv \frac{\partial^2 V_{eff}}{\partial {\left(\phi^2\right)}^2},
                                                 \lb{58y}
\ee we can calculate it for the RG improved effective potential
(\ref{6a}):
\bee
{V''}_{eff}\left(\phi^2\right) &=& \frac
{{V'}_{eff}\left(\phi^2\right)}{\phi^2} + \left(
- \frac 12
 \mu^2_{run}
+ \frac 12 \frac{d^2\mu^2_{run}}{dt^2} + 2\frac{d\mu^2_{run}}{dt}
 \frac{d\log (G)}{dt}+ \mu^2_{run}\frac{d^2\log (G)}{dt^2}\right.
\nonumber\\
&& \left.  +
   2\mu^2_{run}{\left(\frac{d\log (G)}{dt}\right)}^2\right)\frac {G^2}{\phi^2} +
\left(
   \frac 12 \frac{d\lambda_{run}}{dt}+ \frac 14 \frac {d^2\lambda_{run}}
  {dt^2} + 2\frac{d\lambda_{run}}{dt}\frac{d\log (G)}{dt}\right.
\nonumber\\
&&\left.
+ 2\lambda_{run}\frac{d\log (G)}{dt}
 + \lambda_{run}\frac{d^2\log (G)}{dt^2} +
     4\lambda_{run}{\left(\frac{d\log (G)}{dt}\right)}^2\right) G^4(t).
                                                    \lb{59y}
\eee Let us consider now the case when this second derivative
changes its sign giving a maximum of $V_{eff}$ instead of the
minimum at $\phi^2 = \phi_0^2$. Such a possibility is shown in
Fig.~\ref{f12} by the dashed curve 2. Now the two additional minima at
$\phi^2 = \phi_1^2$ and $\phi^2 = \phi_2^2$ appear in our theory.
They correspond to the two different confinement phases for the
confinement of electrically  charged particles if they exist in
the system. When these two minima are degenerate, we have the
following requirements: \be
       V_{eff}\left(\phi_1^2\right) = V_{eff}\left(\phi_2^2\right) < 0 \quad
{\mbox{and}} \quad
        {V'}_{eff}\left(\phi_1^2\right) = {V'}_{eff}\left(\phi_2^2\right) = 0,
\lb{61y}
\ee which describe the border between the confinement phases
``Conf. 1" and ``Conf. 2" presented in Fig.~\ref{f14}. This border is
given
as a curve ``3" at the phase diagram $\left(\lambda_{run};\,
g^4_{run}\right)$
shown in Fig.~\ref{f14}. The curve ``3" meets the curve ``1" at the
triple
point A. According to the illustration of Fig.~\ref{f14}, the triple
point A is given by the following requirements: \be
    V_{eff}\left(\phi_0^2\right) = V'_{eff}\left(\phi_0^2\right) =
V''_{eff}\left(\phi_0^2\right) = 0.
                                         \lb{62y}
\ee In contrast to the requirements:
\be
       V_{eff}\left(\phi_0^2\right) = V'_{eff}\left(\phi_0^2\right) = 0,
\lb{63y}
\ee describing the curve ``1", let us consider the joint solution
of the following equations:
\be
         V_{eff}\left(\phi_0^2\right) = V''_{eff}\left(\phi_0^2\right) = 0 .
\lb{64y}
\ee For simplicity, we have considered the one--loop
approximation. Using Eqs.~(\ref{42y}, \ref{59y}) and
(\ref{31y}--\ref{33y}), it is easy to obtain the solution of
Eq.~(\ref{64y}) in the one--loop approximation:
\be
            {\cal F}\left(\lambda_{run},\, g^2_{run}\right) = 0,
                                                            \lb{65y}
\ee where
\be
{\cal F}\left(\lambda_{run}, g^2_{run}\right) = 5g_{run}^6 +
  24\pi^2g_{run}^4 + 12\lambda_{run}g_{run}^4 - 9\lambda_{run}^2g_{run}^2
+ 36\lambda_{run}^3 + 80\pi^2\lambda_{run}^2 +
64\pi^4\lambda_{run}.
                                                \lb{66y}
\ee The dashed curve ``2" of Fig.~\ref{f14} represents the solution of
Eq.~(\ref{65y}) which is equivalent to Eqs.~(\ref{64y}). The curve
``2" is going very close to the maximum of the curve ``1". Assuming
that the position of the triple point A coincides with this
maximum let us consider the border between the phase ``Conf. 1",
having the first minimum at nonzero $\phi_1$  with
$V_{eff}^{min}\left(\phi_1^2\right) = c_1 < 0$, and the phase ``Conf. 2"
which
reveals two minima with the second minimum being the deeper one
and having $V_{eff}^{min}\left(\phi_2^2\right)=c_2 < 0$. This border
(described by the curve ``3" of Fig.~\ref{f14}) was calculated in the
vicinity of the triple point A by means of Eq.~(\ref{61y}) with
$\phi_1$ and $\phi_2$ represented as $ \phi_{1,\, 2} = \phi_0 \pm
\epsilon$ with $\epsilon \ll \phi_0$. The result of such
calculations gives the following expression for the curve ``3": \be
  g^4_{PT} = \frac {5}{2} \left( 5\lambda_{run} + 8 \pi^2\right) \lambda_{run} +
8\pi^4.
                                                 \lb{67y}
\ee The curve ``3" meets the curve ``1" at the triple point A.

The piece of the curve ``1" to the left of the point A describes
the border between the ``Coulomb--like" phase and phase ``Conf. 1".
In the vicinity of the triple point A the second derivative
$V_{eff}''\left(\phi_0^2\right)$ changes its sign leading to the existence of
the maximum at $\phi^2=\phi_0^2$, in correspondence with the
dashed curve ``2" of Fig.~\ref{f14}.  By this reason, the curve ``1"
of
Fig.~\ref{f14} does not already describe a phase transition border up to
the next point B when the curve ``2" again intersects the curve ``1"
at $\lambda_{(B)}\approx - 12.24$. This intersection (again giving
$V''_{eff}\left(\phi_0^2\right) > 0$) occurs surprisingly quickly.

The right piece of the curve ``1" along to the right of the point B
shown in Fig.~\ref{f14} separates the ``Coulomb" phase and the phase
``Conf. 2". But between the points A and B the phase transition
border is going slightly upper the curve ``1". This deviation is
very small and cannot be distinguished on Fig.~\ref{f14}.

It is necessary to note that only $V''_{eff}\left(\phi^2\right)$ contains the
derivative $dg^2_{run}/dt$. The joint solution of equations
(\ref{62y}) leads to the joint solution of Eqs.~(\ref{50y}) and
(\ref{65y}). This solution was obtained numerically and gave the
following triple point values of $\lambda_{run}$ and $g^2_{run}$:
\be
    \lambda_{(A)}\approx{ - 13.41}, \quad
              g^2_{(A)}\approx{18.61}.            \lb{68y}
\ee The solution (\ref{68y}) demonstrates that the triple point A
exists in the very neighbourhood of the maximum of the curve
(\ref{50y}). The position of this maximum is given by the
following analytical expressions, together with their approximate
values: \bee
     \lambda_{(A)}&\approx& - \frac{4\pi^2}3\approx -13.2,
                                                       \lb{69y}
\\
     g^2_{(A)} &=&\left.
g^2_{crit}\right|_{\mbox{ for }\lambda_{run}=\lambda_{(A)}}
                 \approx \frac{4\sqrt{2}}3{\pi^2}\approx 18.6.
                                                        \lb{70y}
\eee Finally, we can conclude that the phase diagram shown in
Fig.~\ref{f14} gives such a description: there exist three phases in the
dual sector of the Higgs scalar electrodynamics --- the
Coulomb--like phase and confinement phases ``Conf. 1" and ``Conf. 2".

The border ``1", which is described by the curve (\ref{50y}),
separates the Coulomb--like phase (with $V_{eff} \ge 0$) and
confinement phases (with $V_{eff}^{min}(\phi_0^2) < 0$). The curve
``1" corresponds to the joint solution of the equations
$V_{eff}\left(\phi_0^2\right)=V'_{eff}\left(\phi_0^2\right)=0$.

The dashed curve ``2" represents the solution of the equations
$V_{eff}\left(\phi_0^2\right)=V''_{eff}\left(\phi_0^2\right)=0$.

The phase border ``3" of Fig.~\ref{f14} separates the two confinement
phases. The following requirements take place for this border:
$$
          V_{eff}\left(\phi_{1,\, 2}^2\right) < 0, \quad
         V_{eff}\left(\phi_1^2\right) = V_{eff}\left(\phi_2^2\right), \quad
         V'_{eff}\left(\phi_1^2\right) = V'_{eff}\left(\phi_2^2\right) = 0,
$$
\be
         V''_{eff}\left(\phi_1^2\right) > 0, \quad
V''_{eff}\left(\phi_2^2\right) > 0.
                                               \lb{71y}
\ee The triple point A is a boundary point of all three phase
transitions shown in the phase diagram of Fig.~\ref{f14}. For $g^2 <
g^2_{({A})}$ the field system, described by our model, exists in
the confinement phase where all electric charges have to be
confined.

Taking into account that monopole mass $m$ is given by the
following expression: \be
  V''_{eff}\left(\phi_0^2\right) =\left.
\frac 1{4\phi_{0A}^2}\frac {d^2V_{eff}}{d\phi^2}\right|_{\phi=\phi_0}
            = \frac {m^2}{4\phi_{0A}^2},                \lb{72y}
\ee we see that monopoles acquire zero mass in the vicinity of the
triple point A: \be
  V''_{eff}\left(\phi_{0A}^2\right) = \frac {m^2_{(A)}}{4\phi_{0A}^2} = 0.
                               \lb{74y}
\ee This result is in agreement with the result of the compact QED
\ct{29y}: $m^2\to 0$ in the vicinity of the critical point.

\section{``ANO--strings", or the vortex description of the
confinement phases}
\label{sec6}

As it was shown in the previous Subsection, two regions between
the curves ``1", ``3" and ``3", ``1", given by the phase
diagram of
Fig.~\ref{f14}, correspond to the existence of the two confinement
phases, different in the sense that the phase ``Conf. 1" is produced
by the second minimum, but the phase ``Conf. 2" corresponds to the
third minimum of the effective potential. It is obvious that in
this case both phases have nonzero monopole condensate in the
minima of the effective potential, when
$V_{eff}^{min}\left(\phi_{1,\, 2}\neq 0\right) < 0$. By this reason, the
Abrikosov--Nielsen--Olesen (ANO) electric vortices (see
Refs.~\ct{30y,31y}) may exist in these both phases, which are
equivalent in the sense of the ``string" formation.  If electric
charges are present in a model (they are absent in HMM), then
these charges are placed at the ends of the vortices--``strings"
and therefore are confined. But only closed ``strings" exist in the
confinement phases of HMM. The properties of the ``ANO--strings" in
the $U(1)$ gauge theory were investigated in Ref.~\ct{21}.

In the London's limit ($\lambda \to \infty$) the dual Abelian
Higgs model developed in Refs.~\ct{30y,31y} and described by
the Lagrangian (\ref{5y}), gives the formation of monopole
condensate with amplitude $\phi_0$, which repels and suppresses
the electromagnetic field $F_{\mu\nu}$ almost everywhere, except
the region around the vortex lines. In this limit, we have the
following London equation: \be
      \mbox{rot}\; {\vec j}^m = {\delta}^{-2}{\vec E},      \lb{112x}
\ee where ${\vec j}^m$ is the microscopic current of monopoles,
${\vec E}$ is the electric field strength and $\delta$ is the
penetration depth. It is clear that ${\delta}^{-1}$ is the photon
mass $m_V$, generated by the Higgs mechanism. The closed equation
for $\vec E$ follows from the Maxwell equations and
Eq.~(\ref{112x}) just in the London's limit.

In our case $\delta$ is defined by the following relation: \be
            {\delta}^{-2} \equiv m_V^2 = g^2 \phi_0^2.    \lb{113x}
\ee On the other hand, the field $\phi$ has its own correlation
length $\xi$, connected to the mass of the field $\phi$ (``the
Higgs mass"): \be
        \xi = {m_S}^{-1},  \quad m_S^2 = \lambda \phi_0^2.   \lb{114x}
\ee The London's limit for our ``dual superconductor of the second
type" corresponds to the following relations: \be
      \delta \gg \xi, \quad m_V \ll m_S, \quad g \ll \lambda,
\lb{115x}
\ee and ``the string tension" --- the vortex energy per unit length
\ct{30y} --- is (for the minimal electric vortex flux $2\pi$): \be
       \sigma = \frac {2\pi}{g^2\delta^2}\ln \left(\frac{\delta}{\xi}\right)
              = 2\pi \phi_0^2 \ln \left(\frac{m_S}{m_V}\right),  \quad {\mbox
        {where}} \quad \frac{\delta}{\xi} = \frac {m_S}{m_V} \gg 1.
\lb{116x}
\ee We see that the ANO--theory in the London's limit implies the
photon mass generation: $m_V = 1/\delta$, which is much less than
the Higgs mass $m_S = 1/\xi$.

Let us wonder now, whether our ``strings" are thin or thick.

The vortex may be considered as thin, if the distance $L$ between
the electric charges sitting at its ends, i.e. the string length,
is much larger than the penetration length $\delta$: \be
            L \gg \delta \gg \xi.         \lb{117x}
\ee It is obvious that only rotating ``strings" can exist as stable
states. In the framework of classical calculations, it is not
difficult to obtain the mass $M$ and angular momentum $J$ of the
rotating ``string": \be
             J = \frac{1}{2\pi \sigma} M^2,  \quad
             M = \frac {\pi}{2} \sigma L.               \lb{118x}
\ee The following relation follows from Eq.~(\ref{118x}): \be
        L = 2 \sqrt{\frac{2J}{\pi \sigma}},             \lb{119x}
\ee or \be
       L = \frac{2g\delta}{\pi}\sqrt{\frac{J}{\ln
\left(\frac{m_S}{m_V}\right)}}.
                             \lb{120x}
\ee For $J=1$ we have: \be
       \frac{L}{\delta} = \frac{2g}{\pi \sqrt{\ln \left(\frac
{m_S}{m_V}\right)}},
                                                            \lb{121x}
\ee what means that for $ m_S \gg m_V $ the length of this ``string"
is small and does not obey the requirement (\ref{117x}).
It is easy to see from Eq.~(\ref{120x}) that in the London's limit
the ``strings" are very thin $\left(L/\delta \gg 1\right)$ only for the
enormously large angular momenta $J \gg 1$.

The phase diagram of Fig.~\ref{f14} shows the existence of the
confinement phase for $\alpha \ge \alpha_{(A)}$. This means that
the formation of (closed) vortices begins at the triple point
$\alpha = \alpha_{(A)}$: for $\alpha > \alpha_{(A)}$, i.e. $\tilde
\alpha < \tilde \alpha_{(A)}$, we have nonzero $\phi_0$ leading to
the creation of vortices.

In Section \ref{subsec4.4} we have shown that the lattice investigations lead
to the ``freezing" of the electric fine structure constant at the
value $\alpha = \alpha_{max}$ and mentioned that the authors of
Ref.~\ct{14p} predicted: $\alpha_{max}={\pi}/{12} \approx 0.26$.

Let us estimate now the region of values of the magnetic charge
$g$ in the confinement phase considered in this paper:
\bee
               g_{min} &\le& g \le g_{max},
\nonumber\\
           g_{max} &=& g_{(A)}\approx{\sqrt {15.1}}\approx 3.9,
\nonumber\\
           g_{min} &=& \sqrt {\frac {\pi}{\alpha_{max}}}\approx 3.5.
                                             \lb{122x}
\eee Then for $m_S = 10m_V$ (considered as an example) we have from
Eq.~(\ref{121x}) the following estimate of the ``string" length when
$J=1$: \be
            1.5 \stackrel{<}{\sim}\frac{L}{\delta}\stackrel{<}{\sim} 1.8.
                              \lb{123x}
\ee We see that in the $U(1)$ gauge theory the low-lying states of
``strings" correspond to the short and thick vortices.

In general, the way of receiving of the Nambu--Goto strings from
the dual Abelian Higgs model of scalar monopoles was demonstrated
in Ref.~\ct{32y}.

\section{Phase transition couplings in the regularized $SU(N)$ gauge theories}
\label{sec7}

It was shown in a lot of investigations (see for example,
\ct{11s,12s} and references there) that the confinement in
the $SU(N)$ lattice gauge theories effectively comes to the same
$U(1)$ formalism. The reason is the Abelian dominance in the
monopole vacuum: monopoles of the Yang--Mills theory are the
solutions of the $U(1)$--subgroups, arbitrary embedded into the
$SU(N)$ group. After a partial gauge fixing (Abelian projection by
't Hooft \ct{24p}) $SU(N)$ gauge theory is reduced to the Abelian
$U(1)^{N-1}$ theory with $N-1$ different types of Abelian
monopoles. Choosing the Abelian gauge for dual gluons, it is
possible to describe the confinement in the lattice $SU(N)$ gauge
theories by the analogous dual Abelian Higgs model of scalar
monopoles.

\subsection{The ``abelization" of monopole vacuum in the non--Abelian
theories}
\label{subsec7.1}

A lattice imitates the non--perturbative vacuum of zero
temperature $SU(2)$ and $SU(3)$ gluodynamics as a condensate of
monopoles which emerge as leading non--perturbative fluctuations
of the non--Abelian $SU(N)$ gauge theories in the gauge of the
Abelian projections by G.~'t Hooft \ct{24p} (see also the review
\ct{25p} and Refs.~\ct{25pa,25pb,25pc}). It is possible to find
such a gauge, in which monopole degrees of freedom, hidden in the
given field configuration, become explicit.

Let us consider the $SU(N)$ gluodynamics. For any composite operator
$\bf X\in$ the adjoint representation of $SU(N)$ group ($\bf X$ may
be ${\left(F_{\mu\nu}\right)}_{ij}$, where $ i,\, j=1,\, 2,\, ...,\, N$)
we can find such
a gauge: \be {\bf    X \to X' = VXV^{-1}, } \lb{1h} \ee
where the unitary matrix $\bf V$ transforms $\bf X$ to diagonal
$\bf X'$: \be {\bf    X \to X' = VXV^{-1} = {\mbox{diag}}\left(
\lambda_1, \lambda_2,..., \lambda_N\right).           } \lb{2h} \ee We
can choose the ordering of $\lambda_i$: \be
            \lambda_1 \le \lambda_2 \le...\le \lambda_N.         \lb{3h}
\ee The matrix $\bf X'$ belongs to the Cartain, or Maximal Abelian
subgroup of the $SU(N)$ group: \be
             U(1)^{N-1}\in SU(N).
                                        \lb{4h}
\ee Let us consider the field $A_{\mu}$ in the diagonal gauge: \be
          {\bar A}_{\mu} =
V \left( A_{\mu} + \frac{i}{g}\partial_{\mu}\right)V^{-1}.
                                        \lb{5h}
\ee This field transforms according to the subgroup $U(1)^{N-1}$:
its diagonal elements
$$
{\left(a_{\mu}\right)}_i \equiv {\left({\bar A}_{\mu}\right)}_{ii}
$$
transform as Abelian gauge fields (photons): \be
{\left(a_{\mu}\right)}_i \to {\left(a'_{\mu}\right)}_i =
{\left(a_{\mu}\right)}_i +
              \frac{1}{g}\partial_{\mu} \alpha_i,
                                       \lb{6h}
\ee but its non--diagonal elements
$$
{\left(c_{\mu}\right)}_{ij} \equiv {\left({\bar A}_{\mu}\right)}_{ij} \quad
{\mbox{with}} \quad i\neq j
$$
transform as charged fields: \be  {\left(c'_{\mu}\right)}_{ij} =
\exp\left[i\left(\alpha_i - \alpha_j\right)\right]{\left(c_{\mu}\right)}_{ij},
    \lb{7h}
\ee where  $i,\, j = 1,\, 2,\, ...,\, N$.

 According to G.'t Hooft \ct{24p}, if some $\lambda_i$ coincide, then
the singularities, having the properties of monopoles, appear in
the ``Abelian part" of the non--Abelian gauge fields. Indeed, let
us consider the strength tensor of the ``Abelian gluons":
\bee
{\left(f_{\mu\nu}\right)}_i &=& \partial_{\mu}{\left(a_{\nu}\right)}_i
-
\partial_{\nu}{\left(a_{\mu}\right)}_i
\nonumber\\
 &=& V F_{\mu\nu}V^{-1} + ig\left[
V \left( A_{\mu} + \frac{i}{g}\partial_{\mu}\right)V^{-1}, V\left( A_{\nu} +
\frac{i}{g}\partial_{\nu}\right)V^{-1}\right].
                                        \lb{8h}
\eee The monopole current is: \be      {\left(K_{\mu}\right)}_i =
\frac{1}{8\pi}\epsilon_{\mu\nu\rho\sigma}
\partial_{\nu}{\left(f_{\rho\sigma}\right)}_i,
                                        \lb{8ha}
\ee and it is conserved: \be
\partial_{\mu}{\left(K_{\mu}\right)}_i = 0.
                                    \lb{9h}
\ee $ F_{\mu\nu}$ had no singularities.
Therefore, all singularities can come from the commutator, which
is written in Eq.~(\ref{8h}).

The magnetic charge $m_i(\Omega)$ in 3d--volume $\Omega $ is: \be
  m_i(\Omega)= \int_{\Omega} d^3 \sigma_{\mu}{\left(K_{\mu}\right)}_i =
   \frac{1}{8\pi}\int_{\partial{\Omega}} d^2 \sigma_{\mu\nu}
    {\left(f_{\mu\nu}\right)}_i.
                                    \lb{10h}
\ee If $\lambda_1=\lambda_2$ (coincide) at the point $
x^{(1)}$ in 3d--volume $\Omega$, then we have a singularity on the
curve in 4d--space, which is a world--line of the magnetic
monopole, and $x = x^{(1)}$ is a singular point of the gauge
transformed fields ${\bar A}_{\mu}$ and $ {\left(a_{\mu}\right)}_i$.

As it was shown by 't Hooft \ct{24p}: \be  {\left(f_{\mu\nu}\right)}_i\sim
O\left(\left|x - x^{(1)}\right|^{-2}\right)              \lb{11h} \ee only in
the
vicinity of $x^{(1)}$, where it behaves as a magnetic field of the
point--like monopole.

Finally,  we have the following conclusions:

\begin{enumerate}
\item[1.] The initial potentials $A_{\mu}$ and strength tensor $
F_{\mu\nu}$ had no singularities.

\item[2.] At large distances ${\left(f_{\mu\nu}\right)}_i$ doesn't have a
behaviour
$$
 {\left(f_{\mu\nu}\right)}_i \sim O\left(\left|x - x^{(1)}\right|^{-2}\right)
$$
and the monopoles exist only near $x = x^{(1)}$.

\item[3.] Fields ${\bar A}_{\mu}$ and ${\left(a_{\mu}\right)}_i$ are not
classical solutions: they are a result of the quantum fluctuations
of gluon fields.

\item[4.] Any distribution of gluon fields in the vacuum can undergo the
Abelian projection.
\end{enumerate}

We have seen that in the $SU(N)$ gauge theories quantum fluctuations
(non--perturbative effects) of gluon fields reveal an Abelian
vacuum monopoles and suppress the non--diagonal components of the
strength tensor ${\left(F_{\mu\nu}\right)}_{ij}$. As it will be shown below,
this phenomenon gives very important consequences for the Planck
scale physics.

Using the idea of the monopole vacuum ``abelization" of the $SU(N)$
lattice gauge theories, a method of theoretical estimate of the
$SU(N)$ critical couplings was developed in Ref.~\ct{21aa}.

\subsection{Monopoles strength group dependence}
\label{subsec7.2}

Lattice non--Abelian gauge theories also have lattice artifact
monopoles. It was supposed in Ref.~\ct{21aa} that only those lattice
artifact monopoles are important for the phase transition
calculations which have the smallest monopole charges.

Let us consider the lattice gauge theory with the gauge group
$SU(N)/Z_N$ as a main example. That is to say, we consider the
adjoint representation action and do not distinguish link
variables forming the same one multiplied by any element of the
center of the group. The group $SU(N)/Z_N $ is not simply
connected and has the first homotopic group $\Pi_1(SU(N)/Z_N)$
equal to $Z_N$. The lattice artifact monopole with the smallest
magnetic charge may be described as a three--cube (or rather a
chain of three--cubes describing the time track) from which
radiates magnetic field corresponding to the $U(1)$ subgroup of
the gauge group $SU(N)/Z_N$ with the shortest length insight of
this group, but still homotopically non--trivial. In fact, this
$U(1)$ subgroup is obtained by the exponentiating generator: \be
\frac{``\lambda_8"}{2} = \frac {1}{\sqrt{2N(N-1)}}
\left(\ba{*{4}{c}}
N-1 & 0 & \cdots & 0\\
0 & -1 & \cdots & 0 \\
\vdots & \vdots &\ddots & \vdots\\
0 & 0 &  \cdots & -1 \\
\ea
\right). \lb{1z} \ee This specific form is one gauge choice; any
similarity transformation of this generator would describe
physically the same monopole. If one has somehow already chosen
the gauge monopoles with different but similarity transformation
related generators, they would be physically different. Thus,
after gauge choice, there are monopoles corresponding to different
directions of the Lie algebra generators in the form ${\cal
U}\frac{``\lambda_8"}{2}{\cal U}^{+}$.

Now, when we  want to apply the effective potential calculation as
a technique for the getting phase diagram information for the
condensation of the lattice artifact monopoles in the non--Abelian
lattice gauge theory, we have to correct the Abelian case
calculation for the fact that after gauge choice we have a lot of
different monopoles. If a couple of monopoles happens to have
their generators just in the same directions in the Lie algebra,
they will interact with each other as Abelian monopoles (in first
approximation). In general, the interaction of two monopoles by
exchange of a photon will be modified by the following factor: \be
  \frac{ {\rm Tr} \left({\cal U}_1\frac{``\lambda_8"}{2} {\cal U}_1^{+}
{\cal U}_2\frac{``\lambda_8"}{2} {\cal U}_2^{+}\right)}
{{\rm Tr}{\left(\frac{``\lambda_8"}2\right)}^2}.          \lb{2z} \ee We
shall
assume that we can correct these values of monopole orientations
in the Lie algebra in a statistical way. That is to say, we want
to determine an effective coupling constant ${\tilde g}_{eff}$
describing the monopole charge as if there is only one Lie algebra
orientation--wise type of monopole. It should be estimated
statistically in terms of the magnetic charge ${\tilde
g}_{\rm genuine}$ valid to describe the interaction between monopoles
with generators oriented along the same $U(1)$ subgroup. A very
crude intuitive estimate of the relation between these two
monopole charge concepts ${\tilde g}_{\rm genuine}$ and ${\tilde
g}_{eff}$ consists in playing that the generators are randomly
oriented in the whole $N^2 - 1$ dimensional Lie algebra. When even
the sign of the Lie algebra generator associated with the monopole
is random --- as we assumed in this crude argument --- the
interaction between two monopoles with just one photon exchanged
averages out to zero. Therefore, we can get a non--zero result only
in the case of exchange by two photons or more. That is, however,
good enough for our effective potential calculation since only
${\tilde g}^4$ (but not the second power) occurs in the Coleman--Weinberg
effective potential in the one--loop approximation (see
\ct{28,29}). Taking into account this fact that we can
average imagining monopoles with generators along a basis vector
in the Lie algebra, the chance of interaction by double photon
exchange between two different monopoles is just $1/(N^2 -
1)$, because there are $N^2 - 1$ basis vectors in the basis of the
Lie algebra. Thus, this crude approximation gives: \be
    {\tilde g}^4_{eff} = \frac{1}{N^2 - 1}{\tilde g}^4_{\rm genuine}.
                                     \lb{3z}
\ee Note that considering the two photons exchange which is forced
by our statistical description, we must concern the fourth power
of the monopole charge $\tilde g$.

The relation (\ref{3z}) was not derived correctly, but its
validity can be confirmed if we use a more correct statistical
argument. The problem with our crude estimate is that the
generators making monopole charge to be minimal must go along the
shortest type of $U(1)$ subgroups with non--trivial homotopy.

\subsubsection*{Correct averaging}

The $\lambda_8$--like generators ${\cal
U}\frac{``\lambda_8"}{2}{\cal U}^+$  maybe written as
\be
{\cal U} \frac{``\lambda_8"}{2} {\cal U}^{+} = - \sqrt{\frac
{1}{2N(N-1)}} {\bf 1\!\!\! 1} + \sqrt{\frac{N}{2(N-1)}} {\cal
P}, \lb{4z}
\ee
where $\cal P$ is a projection metrics into one--dimensional state
in the \un{N} representation. It is easy to see that averaging
according to the Haar measure distribution of $\cal U$, we get the
average of $\cal P$ projection on ``quark" states with a
distribution corresponding to the rotationally invariant one on
the unit sphere in the N--dimensional \un{N}--Hilbert space.

If we denote the Hilbert vector describing the state on which
$\cal P$ shall project as \be\left(
\ba{c}
\psi_1\\
\psi_2\\
\vdots\\
\psi_N
\ea
\right),
                                       \lb{5z}
\ee then the probability distribution on the unit sphere becomes:
\be P \left(\left(
\ba{c}
\psi_1\\
\psi_2\\
\vdots\\
\psi_N
\ea
\right)\right) {\prod}_{i=1}^N d{\psi}_i\propto \delta\left({\sum}_{i=1}^N
{\left|{\psi}_i\right|}^2 - 1\right) \prod_{i=1}^N
d\left({\left|\psi_i\right|}^2\right).
\lb{6z} \ee Since, of course, we must have ${\left|{\psi}_i\right|}^2 \ge 0$
for all $i=1,\, 2,\, ...,\, N$, the $\delta$--function is easily seen to
select a flat distribution on a $(N - 1)$--dimensional equilateral
simplex. The average of the two photon exchange interaction given
by the correction factor (\ref{2z}) squared (numerically):
\be
\frac{{\rm Tr}\left({\cal U}_1 \frac{``\lambda_8"}{2} {\cal U}_1^{+}{\cal
U}_2 {\frac{``\lambda_8"}{2}}{\cal
U}_2^{+}\right)^2}{{\rm
Tr}\left({\left(\frac{``\lambda_8"}{2}\right)}^2\right) ^2}
                                 \lb{7z}
\ee
can obviously be replaced by the expression where we take as
random only one of the ``random"  $\lambda_8$--like generators,
while the other one is just taken as $\frac{``\lambda_8"}{2}$,
i.e. we can take say ${\cal U}_2 = {\bf 1\!\!\!1}$ without
changing the average.

Considering the two photon exchange diagram, we can write the
correction factor (obtained by the averaging) for the fourth power
of magnetic charge: \be \frac{{\tilde g}^4_{eff}}{{\tilde
g}^4_{{\rm genuine}}} = {\rm average}\left\{\frac{{\rm
Tr}{\left(\frac{``\lambda_8
"}{2} {{\cal U}_{1}} {\frac{``\lambda_8"}{2}}{{\cal
U}_{1}}^{+}\right)}^2} {{\rm
Tr}{\left({\left(\frac{``\lambda_8"}{2}\right)}^2\right)}^2}\right\}.
\lb{8z} \ee Substituting the expression (\ref{4z}) in
Eq.~(\ref{8z}), we have: \be \frac{{\tilde g}^4_{eff}}{{\tilde
g}^4_{{\rm genuine}}} = {\rm average}\left\{\frac{
{\rm Tr}{\left(\frac{``\lambda_8"}{2} \left(-\sqrt{\frac{1} {2N(N-1)}}{\bf
1\!\!\!1}
 + \sqrt{\frac {N}{2(N-1)}}{\cal P}\right)\right)}^2}
{{\rm
Tr}{\left({\left(\frac{``\lambda_8"}{2}\right)}^2\right)}^2}\right\}.
\ee Since $\frac{``\lambda_8"}{2}$ is traceless, we obtain using
the projection (\ref{5z}):

\be
 {\rm Tr}\left(\frac{``\lambda_8"}{2} {\cal
P}\sqrt{\frac{N}{2(N-1)}}\right)
= - \frac{1}{2(N-1)} + \frac{N}{2(N-1)}{\left|\psi_1\right|}^2.
                                              \lb{9z}
\ee The value of the square ${\left|\psi_1\right|}^2$ over the simplex is
proportional to one of the heights in this simplex. It is obvious
from the geometry of a simplex that the distribution of
${\left|\psi_1\right|}^2$ is \be
     d{\mbox P} = (N-1){\left(1 - {\left|\psi_1\right|}^2\right)}^{(N-2)}
d\left({\left|\psi_1\right|}^2\right),
                                          \lb{10z}
\ee where, of course, $0 \le {\left|{\psi}_1\right|}^2 \le 1$ only is
allowed. In Eq.~(\ref{10z}) the quantity P is a probability.

By definition: \be
     {\rm average}\left\{f({\left|{\psi}_1\right|}^2)\right\}
= (N-1)\int_0^1 f\left({\left|{\psi}_1\right|}^2\right)\left(1 -
{|\psi_1|}^2\right)^{(N-2)}d\left({|\psi_1|}^2\right).
                                            \lb{11z}
\ee Then
\bee
\frac{{\tilde g}^4_{eff}}{{\tilde g}^4_{{\rm genuine}}} &=&
\frac{N^2}{(N-1)} \int_0^1
{\left(\frac{1}{N} - {\left|\psi_1\right|}^2\right)}^2{\left(1 -
{\left|\psi_1\right|}^2\right)}^{N-2} d\left({\left|\psi_1\right|}^2\right)\\
&=& \frac{N^2}{N-1}\int_0^1 {\left(1-y-\frac{1}{N}\right)}^2y^{N-2}dy\\
&=& \frac{1}{N^2-1}                \lb{12z}
\eee
and we have confirmed our crude estimate (\ref{3z}).

\subsubsection*{Relative normalization of couplings}

Now we are interested in how ${\tilde g}^2_{\rm genuine}$ is related
to $\alpha_N=g^2_N/{4\pi}$.

We would get the simple Dirac relation: \be
          g_{(1)}\cdot {\tilde g}_{\rm genuine} = 2\pi,
                                              \lb{13z}
\ee if $g_{(1)}\equiv g_{U(1)-{\rm subgroup}}$ is the coupling for
the $U(1)$--subgroup of $SU(N)$ normalized in such a way that the
charge quantum $g_{(1)}$ corresponds to a covariant derivative
$\partial_{\mu} - g_{(1)}A_{\mu}^{U(1)}$.

Then we shall follow the convention --- usually used to define
$\alpha_N={g^2_N}/{4\pi}$ --- that the covariant derivative for the
{\un{N}}--plet representation is: \be
        D_{\mu} = \partial_{\mu} - g_N \frac{\lambda^a}{2}A_{\mu}^a
                                                     \lb{14z}
\ee with \be
 Tr(\frac{\lambda^a}{2}\frac{\lambda^b}{2}) = \frac{1}{2}\delta^{ab},
                                                   \lb{15z}
\ee and the gauge field kinetic term is \be
       L = - \frac{1}{4}F_{\mu\nu}^aF^{a\mu\nu},   \lb{16z}
\ee where \be
    F_{\mu\nu}^a = \partial_{\mu}A_{\nu}^a - \partial_{\nu}A_{\mu}^a
                    - g_Nf^{abc}A_{\mu}^bA_{\nu}^c.
                                         \lb{17z}
\ee Especially if we want to choose a basis for our generalized
Gell--Mann matrices so that one basic vector is our
$\frac{``\lambda_8"}{2}$, then for $A_{\mu}^{``8"}$ we have the
covariant derivation $\partial_{\mu} -
g_N\frac{``\lambda_8"}{2}A_{\mu}^{``8"}$. If this covariant
derivative is written in terms of the $U(1)$--subgroup,
corresponding to monopoles with the Dirac relation (\ref{13z}),
then the covariant derivative has a form $\partial_{\mu} -
g_{(1)}A_{\mu}^8\cdot{\un{\un{M}}}$. Here ${\un{\un{M}}}$ has
the property that $\exp\left(i2\pi{\un{\un{M}}}\right)$ corresponds to the
elements of the group $SU(N)/Z_N$ going all around and back to the
unit element. Of course, ${\un{\un{M}}}
=\frac{g_N}{g_{(1)}}\cdot\frac{``\lambda_8"}{2}$ and the ratio
$g_N/g_{(1)}$ must be such one that
$\exp\left(i2\pi\frac{g_N}{g_{(1)}}\frac{``\lambda_8"}{2}\right)$ shall
represent --- after first return --- the unit element of the group
$SU(N)/Z_N$. Now this unit element really means the coset
consisting of the center elements $\exp\left(i\frac{2\pi k}{N}\right)\in
SU(N),\, (k\in Z)$, and the requirement of the normalization of
$g_{(1)}$ ensuring the Dirac relation (\ref{13z}) is: \be
    \exp\left(i2\pi\frac{g_N}{g_{(1)}}\frac{``\lambda_8"}{2}\right)
                      = \exp\left(i\frac{2\pi}{N}\right){\bf 1\!\!\!1}.
                                           \lb{18z}
\ee This requirement is satisfied if the eigenvalues of
$\frac{g_{N}}{g_{(1)}}\frac{``\lambda_8"}{2}$ are modulo 1 equal
to $-1/N$, i.e. formally we might write: \be
   \frac{g_{N}}{g_{(1)}}\frac{``\lambda_8"}{2}
     = - \frac{1}{N} \quad (\rm{mod}\, 1).                  \lb{19z}
\ee According to (\ref{1z}), we have: \be \frac{g_N}{g_{(1)}}\cdot
\frac {1}{\sqrt{2N(N-1)}} \left(\ba{*{4}{c}}
N-1 & 0 & \cdots & 0\\
0 & -1 & \cdots & 0 \\
\vdots & \vdots &\ddots & \vdots\\
0 & 0 &  \cdots & -1 \\
\ea
\right) = - \frac{1}{N} \quad (\rm{mod}\, 1),
\lb{20z} \ee what implies: \be
   \frac{g_N}{g_{(1)}} = \sqrt{\frac{2(N-1)}{N}},         \lb{21z}
\ee or \be
   \frac{g_N^2}{g_{(1)}^2} = \frac{2(N-1)}{N}.         \lb{22z}
\ee

\subsection{The relation between $U(1)$ and $SU(N)$ critical couplings}
\label{subsec7.3}

Collecting the relations (\ref{13z}, \ref{22z}) and (\ref{3z}),
we get:
\bee
   \alpha_N^{-1} &=& \frac{4\pi}{g_N^2}
= \frac{N}{2(N-1)}\cdot \frac
    {4\pi}{g^2_{(1)}}
= \frac{N}{2(N-1)}\cdot \frac{{\tilde g}^2_{\rm genuine}}{\pi}
\nonumber\\
& =& \frac{N}{2(N-1)}\sqrt{N^2-1}\cdot \frac{{\tilde
g}^2_{eff}}{\pi} = \frac{N}{2(N-1)}\sqrt{N^2-1}\cdot
\frac{4\pi}{g_{U(1)}^2}
\nonumber\\
&=& \frac{N}{2}\sqrt{\frac{N+1}{N-1}}\cdot
\alpha_{U(1)}^{-1},
                                                            \lb{23z}
\eee where \be
          g_{U(1)}{\tilde g}_{eff} = 2\pi                 \lb{24z}
\ee and $\alpha_{U(1)}=g_{U(1)}^2/{4\pi}$.

The meaning of this result is that provided that we have ${\tilde
g}_{eff}$ the same for $SU(N)/Z_N$ and $U(1)$ gauge theories the
couplings are related according to Eq.~(\ref{23z}).

We have a use for this relation when we want to calculate the
phase transition couplings considering the scalar monopole field
responsible for the phase transition in the gauge groups
$SU(N)/Z_N$. Having in mind the ``Abelian" dominance in the $SU(N)$
monopole vacuum, we must think that ${\tilde g}_{eff}^{crit}$
coincides with $g_{crit}$ of the $U(1)$ gauge theory. Of course,
here we have an approximation taking into account only monopoles
interaction and ignoring the relatively small self--interactions
of the Yang--Mills fields. In this approximation we obtain the
same phase transition (triple point, or critical) ${\tilde
g}_{eff}$--coupling which is equal to $g_{crit}$ of $U(1)$ whatever
the gauge group $SU(N)$ might be. Thus we conclude that for the
various groups $U(1)$ and $SU(N)/{Z_N}$, according to
Eq.~(\ref{23z}), we have the following relation between the phase
transition couplings: \be
      \alpha_{N,\, crit}^{-1}
           = \frac{N}{2}\sqrt{\frac{N+1}{N-1}}
                          \alpha_{U(1),\, crit}^{-1}.
                                            \lb{25z}
\ee Using the relation (\ref{25z}), we obtain: \be
    \alpha_{U(1),\, crit}^{-1} : \alpha_{2,\, crit}^{-1} : \alpha_{3,\,
crit}^{-1}
           = 1 : \sqrt{3} : \frac{3}{\sqrt{2}} = 1 : 1.73 : 2.12.
                                                            \lb{26z}
\ee These relations are used below for the explanation of
predictions of the MPP.

\section{G--theory, or Anti--grand unification theory (AGUT)}
\label{sec8}

Having an interest in the fundamental laws of physics, we can
consider the two possibilities:

\begin{enumerate}
\item[1.] At very small (Planck length) distances {\it our space--time is
continuous} and there exists the fundamental theory (maybe with a
very high symmetry) which we do not know at present time.

\item[2.] At very small distances {\it our space--time is discrete}, and
this discreteness influences on the Planck scale physics.
\end{enumerate}

The theory of Scale Relativity (SR) \ct{10na,10nb} predicts
that there exists a minimal scale of the space--time resolution
equal to the Planck length $\lambda_P$, which can be considered as
a fundamental scale of our Nature. This gives us a reason to make
an assumption that {\it our $(3+1)$--dimensional space is discrete
on the fundamental level.}

This may be an initial (basic) point of view of the theory, which
takes a discreteness as existing, not as the lattice computation
trick in QCD, say. In the simplest case we can imagine our $(3+1)$
space--time as a regular hypercubic lattice with a parameter
$a=\lambda_P$. Then the lattice artifact monopoles can play an
essential role near the Planck scale. But of course, it is
necessary to comment that we do not know (at least, on the level
of our today knowledge), what lattice--like structure (random
lattice, or foam, or string lattice, etc.) is realized in the
description of physical processes at very small distances even if
there should be a lattice.

Investigating the phase transition in the dual Higgs monopole
model, we have pursued two objects. From one side, we had an aim
to explain the lattice results. But we had also another aim.

According to the Multiple Point Model (MPM), at the Planck scale
there exists a multiple critical point (MCP), which is a boundary
point of the phase transitions in $U(1)$, $SU(2)$ and $SU(3)$ sectors of
the fundamental regularized gauge theory $G$. It is natural to
assume that the objects responsible for these transitions are the
physically existing Higgs scalar monopoles, which have to be
introduced into the theory as fundamental fields.  Our
calculations indicate that the corresponding critical couplings
coincide with the lattice ones, confirming the idea of Ref.~\ct{8}.

The results reviewed in the present paper are very encouraging for
the Anti--Grand Unification Theory (AGUT), which always is used in
conjunction with the MPM.

Most efforts to explain the Standard Model (SM) describing well
all experimental results known today are devoted to Grand
Unification Theories (GUTs). The supersymmetric extension of the
SM consists of taking the SM and adding the corresponding
supersymmetric partners \ct{32a}.  The Minimal Supersymmetric
Standard Model (MSSM) shows \ct{33a} the possibility of the
existence of the grand unification point at \bc $\mu_{\rm GUT}\sim
10^{16}$ GeV. \ec Unfortunately, at present time the experiment
does not indicate any manifestation of Supersymmetry. In this
connection, the Anti--Grand Unification Theory (AGUT) was
developed in Refs.~\ct{11}, \ct{15,15a,16,16a,17} as an alternative to
SUSY GUTs. According to this theory, supersymmetry does not come
into the existence up to the Planck energy scale:
\be
M_{Pl}\approx 1.22\cdot 10^{19}\, {\mbox{GeV}}. \lb{2i}
\ee
The Standard Model (SM) is based on the group SMG described by
Eq.~(\ref{1i}). AGUT suggests that at the Planck scale: $ \mu_G\sim
\mu_{Pl}=M_{Pl}$ there exists the more fundamental group $G$
containing a number of copies of the Standard Model Group SMG.

\section{Family Replicated Gauge Group Model (FRGGM) as an extension of the SM}
\label{sec9}

The extension of the Standard Model with the Family Replicated
Gauge Group :
\be
G = (SMG)^{N_{fam}} = \left[SU(3)_c\right]^{N_{fam}}\times
\left[SU(2)_L\right]^{N_{fam}} \times \left[U(1)_Y\right]^{N_{fam}} \lb{2t}
\ee
was first suggested in the paper \ct{11} and developed in the book
\ct{12} (see also the review \ct{13}). Here $N_{fam}$ designates
the number of quark and lepton families.

If $N_{fam}=3$ (as our theory predicts and experiment confirms),
then the fundamental gauge group $G$ is:
\be
G = (SMG)^3 = SMG_{1st\; fam.}\times SMG_{2nd\; fam.}\times
SMG_{3rd\; fam.}, \label{3t}
\ee
or
\be
G = (SMG)^3 = {\left[SU(3)_c\right]}^3\times {\left[SU(2)_L\right]}^3\times
{\left[U(1)_Y\right]}^3. \label{4t}
\ee
The generalized fundamental group:
\be
  G_f = (SMG)^3\times U(1)_f
\lb{5t}
\ee
was suggested by fitting the SM charged fermion masses and mixing
angles in papers \ct{15,16}.

A new generalization of our FRGG--model was suggested in paper
\ct{16a}, where:
\bee
G_{ext} &=& \left(SMG\times U(1)_{(B-L)}\right)^3 \nonumber \\
&\equiv& \left[SU(3)_c\right]^3\times \left[SU(2)_L\right]^3\times
\left[U(1)_Y\right]^3 \times
\left[U(1)_{(B-L)}\right]^3 \lb{6t}
\eee
is the fundamental gauge group, which takes right--handed neutrinos
and the see--saw mechanism into account. This extended model can
describe all modern neutrino experiments, giving a reasonable fit
to all the quark-lepton masses and mixing angles.

The gauge group $G=G_{ext}$ contains: $3\times8=24$ gluons,
$3\times3=9$ $W$--bosons, and $3\times1+3\times1=6$ Abelian gauge
bosons.

At first sight, this $\left(SMG\times U(1)_{(B-L)}\right)^3$ group with
its 39 generators seems to be just one among many possible SM
gauge group extensions. However, it is not such an arbitrary
choice. There are at least reasonable requirements (postulates) on
the gauge group $G$ (or $G_f$, or $G_{ext}$) which have uniquely to
specify this group.  It should obey the following postulates (the
first two are also valid for $SU(5)$ GUT):

\begin{enumerate}

\item[1.] $G$ or $G_f$ should only contain transformations, transforming
the known 45 Weyl fermions (= 3 generations of 15 Weyl particles
each) --- counted as left handed, say --- into each other unitarily,
so that $G$ (or $G_f$) must be a subgroup of $U(45)$: $G\subseteq
U(45)$.

\item[2.] No anomalies, neither gauge nor mixed. AGUT assumes that only
straightforward anomaly cancellation takes place and forbids the
Green--Schwarz type anomaly cancellation \ct{39}.

\item[3.] AGUT should NOT UNIFY the irreducible representations under the
SM gauge group, called here SMG (see Eq.~(\ref{1i})).

\item[4.] $G$ is the maximal group satisfying the above--mentioned
postulates.

\end{enumerate}

There are five Higgs fields named $\phi_{WS},\, S,\, W,\, T,\, \xi$ in
AGUT extended by Froggatt and Nielsen \ct{15} with the group of
symmetry $G_f$ given by Eq.~(\ref{5t}). These fields break AGUT to
the SM what means that their vacuum expectation values (VEV) are
active. The field $\phi_{WS}$ corresponds to the Weinberg--Salam
theory, $\langle S\rangle =1$, so that we have only three free parameters ---
three VEVs $\langle W\rangle,\, \langle T\rangle$ and $\langle\xi\rangle$ to fit
the experiment in
the
framework of this model. The authors of Ref.~\ct{15} used them
with aim to find the best fit to conventional experimental data
for all fermion masses and mixing angles in the SM (see Table \ref{table1}).

The result is encouraging. The fit is given by the ${\chi}^2$
function (called here ${\tilde \chi}^2$). The lowest value of
${\tilde \chi}^2 (\approx 1.87)$ gives the following VEVs: \be
\langle S\rangle=1; \quad\langle W\rangle=0.179; \quad\langle T\rangle=0.071;
\quad
\langle\xi\rangle =0.099.
                                                     \lb{80y}
\ee The extended AGUT by Nielsen and Takanishi \ct{16a}, having
the group of symmetry $G_{ext}$ (see Eq.~(\ref{6t})), was suggested
with aim to explain the neutrino oscillations. Introducing the
right--handed neutrino in the model, the authors replaced the
assumption 1 and considered $U(48)$ group instead of $U(45)$, so that
$G_{ext}$ is a subgroup of $U(48)$: $G_{ext}\subseteq U(48)$. This
group ends up having 7 Higgs fields falling into 4 classes
according to the order of magnitude of the expectation values:

\begin{enumerate}

\item[1.] The smallest VEV Higgs field plays role of the SM
Weinberg--Salam Higgs field $\phi_{WS}$ having the weak scale
value $\langle\phi_{WS}\rangle= 246\, \mbox{GeV}/{\sqrt 2}$.

\item[2.] The next smallest VEV Higgs field breaks all families
$U(1)_{(B-L)}~$ group, which is broken at the see--saw scale. This
VEV is $\langle\phi_{(B-L)}\rangle \sim 10^{12}$ GeV. Such a field is absent
in the ``old" extended AGUT.

\item[3.] The next 4 Higgs fields are $W,\, T,\, \xi$ and $\chi$, which have
VEVs of the order of a factor 10 to 50 under the Planck unit. That
means that if intermediate propagators have scales given by the
Planck scale, as it is assumed in AGUT in general, then they will
give rise to suppression factors of the order 1/10 each time they
are needed to cause a transition. The field $\chi$ is absent in
the ``old" $G_f$--AGUT. It was introduced in Refs.~\ct{16,16a}
for the purpose of the study of neutrinos.

\item[4.] The last one, with VEV of the same order as the Planck scale,
is the Higgs field S. It had VEV $\langle S\rangle =1$ in the ``old"
extended
AGUT \ct{15} by Froggatt and Nielsen (with $G_f$ group of symmetry), but
this VEV is not equal to unity in the ``new" extended AGUT \ct{16a}.
Therefore there is a possibility to observe phenomenological
consequences of the field S in the Nielsen--Takanishi model \ct{16a}.
\end{enumerate}

In contrast to the ``old" extended AGUT by Froggatt--Nielsen
(called here as $G_f$--theory), the new results of
$G_{ext}$--theory by Nielsen--Takanishi \ct{16a} are more
encouraging.

We conclude that the $G$--theory, in general, is successful in
describing of the SM experiment.

The gauge group $G_{ext}$ undergoes spontaneous breakdown
(at some orders of magnitude below the Planck scale) to the
Standard Model Group SMG which is the diagonal subgroup of the
non--Abelian sector of the group $G_{ext}$. As was shown in
Ref.~\ct{16a}, 6 different Higgs fields: $\omega$, $\rho$, $W$,
$T$, $\phi_{WS}$, $\phi_{(B-L)}$ break our FRGG--model to the SM. The
field $\phi_{WS}$ corresponds to the Weinberg--Salam Higgs field of
Electroweak theory. Its vacuum expectation value (VEV) is fixed by
the Fermi constant: $\left\langle\phi_{WS}\right\rangle=246$ GeV,
so that we have only 5 free parameters --- five VEVs:
$\left\langle\omega\right\rangle$,
$\left\langle\rho\right\rangle$, $\left\langle W \right\rangle$,
$\left\langle T \right\rangle$, $\left\langle \phi_{(B-L)}
\right\rangle$ to fit the experiment in the framework of the SM.
These five adjustable parameters were used with the aim of finding
the best fit to experimental data for all fermion masses and
mixing angles in the SM, and also to explain the neutrino
oscillation experiments.

Typical fit to the masses and mixing angles for the SM leptons and
quarks in the framework of the $G_{ext}$--AGUT is given in Table
\ref{table2}.

Experimental results on solar neutrino and atmospheric neutrino
oscillations from Sudbury Neutrino Observatory (SNO Collaboration)
and the Super--Kamiokande Collaboration have been used to extract
the following parameters:
$$
\Delta m^2_{\odot} = m_2^2 - m_1^2, \quad\Delta m^2_{\rm atm} =
m_3^2 - m_2^2,
$$
\be \tan^2 \theta_{\odot} = \tan^2 \theta_{12}, \quad\tan^2
\theta_{\rm atm} = \tan^2 \theta_{23}, \ee where
$m_1,\, m_2,\, m_3$ are the hierarchical left-handed neutrino
effective masses for the three families. Also the CHOOZ
reactor results were used. It is assumed that the fundamental Yukawa
couplings in this model are of order unity and so the authors make order of
magnitude predictions. The typical fit is shown in Table \ref{table2}. As we
can see, the 5 parameter order of magnitude fit is very
encouraging.

There are also 3 see--saw heavy neutrinos in this model (one
right--handed neutrino in each family) with masses:
$M_1,\, M_2,\, M_3.$ The model predicts the following neutrino
masses:
\be m_1\approx 1.4\times 10^{-3} \; {\mbox{eV}}, \quad
m_2\approx 9.6\times 10^{-3} \; {\mbox{eV}}, \quad m_3\approx
4.2\times 10^{-2} \; {\mbox{eV}}
\ee
--- for left-handed neutrinos, and
\be
M_1\approx 1.0\times 10^6 \; {\mbox{GeV}}, \quad M_2\approx
6.1\times 10^9 \; {\mbox{GeV}},  \quad M_3\approx 7.8\times 10^9 \;
{\mbox{GeV}}
\ee
--- for right-handed (heavy) neutrinos.

Finally, we conclude that theory with the FRGG--symmetry is very
successful in describing experiment.

The best fit gave the
following values for VEVs:
\be
  \left\langle W\right\rangle\approx 0.157, \quad
  \left\langle T \right\rangle\approx 0.077, \quad
   \left\langle\omega\right\rangle\approx 0.244, \quad
\left\langle\rho\right\rangle\approx 0.265
\ee
in the ``fundamental units'', $M_{Pl}=1$, and
\be
  \left\langle\phi_{B-L}\right\rangle
\approx 5.25\times 10^{15} \, {\mbox{GeV}}
\ee which gives the see--saw scale: the scale of breakdown of the
$U(1)_{(B-L)}$ groups ($\sim 5\times 10^{15}$ GeV).

\section{Evolution of running fine structure constants}
\label{sec10}

Let us consider now the evolution of the SM running fine structure
constants. The usual definition of the SM coupling constants: \be
  \alpha_1 = \frac{5}{3}\frac{\alpha}{\cos^2\theta_{\ov{MS}}}, \quad
  \alpha_2 = \frac{\alpha}{\sin^2\theta_{\ov{MS}}}, \quad
  \alpha_3 \equiv \alpha_s = \frac {g^2_s}{4\pi},     \lb{81y}
\ee where $\alpha$ and $\alpha_s$ are the electromagnetic and
$SU(3)$ fine structure constants, respectively, is given in the
Modified minimal subtraction scheme ($\ov{MS}$). Here
$\theta_{\ov{MS}}$ is the Weinberg weak angle in $\ov{MS}$ scheme.
Using RGE with experimentally established parameters, it is
possible to extrapolate the experimental values of three inverse
running constants $\alpha_i^{-1}(\mu)$ (here $\mu$ is an energy
scale and $i=1,\, 2,\, 3$ correspond to $U(1)$, $SU(2)$ and $SU(3)$ groups of
the SM) from the Electroweak scale to the Planck scale. The
precision of the LEP data allows to make this extrapolation with
small errors (see \ct{33a}). Assuming that these RGEs for
$\alpha_i^{-1}(\mu)$ contain only the contributions of the SM
particles up to $\mu\approx \mu_{Pl}$ and doing the extrapolation
with one Higgs doublet under the assumption of a ``desert", the
following results for the inverses $\alpha_{Y,\, 2,\, 3}^{-1}$ (here
$\alpha_Y\equiv (3/5) \alpha_1$) were obtained in Ref.~\ct{8}
(compare with \ct{33a}): \be
   \alpha_Y^{-1}\left(\mu_{Pl}\right)\approx 55.5;  \quad
   \alpha_2^{-1}\left(\mu_{Pl}\right)\approx 49.5;  \quad
   \alpha_3^{-1}\left(\mu_{Pl}\right)\approx 54.0.
                                                        \lb{82y}
\ee The extrapolation of $\alpha_{Y,\, 2,\, 3}^{-1}(\mu)$ up to the
point $\mu=\mu_{Pl}$ is shown in Fig.~\ref{f15}.

\subsection{``Gravitational fine structure constant" evolution}
\label{subsec10.1}

In this connection, it is very attractive to include gravity. The
quantity:
\be
  \alpha_g = \left(\frac{\mu}{\mu_{Pl}}\right)^2
\label{2x}
\ee
plays the role of the running ``gravitational fine structure
constant" (see Ref.~\ct{41}) and the evolution of its inverse is
presented in Fig.~\ref{f16} together with the evolutions of
$\alpha_i^{-1}(\mu)$ where $i=1,\, 2,\, 3$.

Then we see the intersection of $\alpha_g^{-1}(\mu)$ with
$\alpha_1^{-1}(\mu)$ at the point:
$$
               \left(x_0, \alpha_0^{-1}\right),
$$
where $\left(x_0 = \log_{10}(\mu_{int.})\right)$:
\be
      x_0 \approx 18.3,   \quad
       \alpha_0^{-1} \approx 34.4.                   \lb{3x}
\ee

\section{Monopoles in the SM and FRGGM}
\label{sec11}

\subsection{Renormalization group equations for electric and
magnetic fine structure constants}
\label{subsec11.1}

J.~Schwinger was first \ct{41r} who investigated the problem of
renormalization of the magnetic charge in Quantum
Electro-Magneto Dynamics (QEMD), i.e. in the Abelian quantum field
theory of electrically and magnetically charged particles (with
charges $e$ and $g$, respectively).

Considering the ``bare" charges $e_0$ and $g_0$ and renormalised
(effective) charges $e$ and $g$, Schwinger (and later the authors
of Refs.~\ct{42r} and \ct{43r}) obtained:
\be
               \frac{e}{g} = \frac{e_0}{g_0},              \lb{4ar}
\ee
what means the absence of the Dirac relation \ct{43ar} for the
renormalised electric and magnetic charges.

But there exists another solution of this problem (see
Refs.~\ct{44r,45r,46r,47r} and review \ct{48r}), which gives:
\be
                       eg = e_0g_0 = 2\pi n, \quad n\in Z,
                                                        \lb{4br}
\ee
i.e. the existence of the Dirac relation (charge quantization
condition) for both, bare and renormalised electric and magnetic
charges. Here we have $n=1$ for the minimal (elementary) charges.

These two cases lead to the two possibilities for the
renormalization group equations (RGEs) describing the evolution of
electric and magnetic fine structure constants (\ref{47a}),
which obey the following RGEs containing the electric and magnetic
$\beta$--functions:
\be
\frac {d\left(\log (\alpha(\mu))\right)}{dt} = \pm \frac {d\left(\log
\left(\tilde
\alpha(\mu)\right)\right)}{dt} = \beta^{(e)}(\alpha) \pm
\beta^{(m)}\left(\tilde
\alpha\right).
                                                       \lb{6ar}
\ee
In Eq.~(\ref{6ar}) we have:
\be
             t = \log\left(\frac{\mu^2}{\mu_R^2}\right),            \lb{6r}
\ee
where $\mu$ is the energy scale and $\mu_R$ is the renormalization
point.

The second possibility (with minuses) in Eq.~(\ref{6ar})
corresponds to the validity of the Dirac relation (\ref{4br}) for
the renormalised charges.  We believe only in this case considered
by authors in Ref.~\ct{47r} where it was used the Zwanziger
formalism of QEMD \ct{49r,50r,51r}. In the present paper,
excluding the Schwinger's renormalization condition (\ref{4ar}),
we assume only the Dirac relation for running $\alpha$ and $\tilde
\alpha$: $\alpha \tilde \alpha = 1/4.$

It is necessary to comment that RGEs (\ref{6ar}) are valid only
for $\mu > \mu_{\rm threshold} = m_{mon}$, where $m_{mon}$ is the
monopole mass.

In contrast to the method given in Ref.~\ct{47r}, there exists a
simple way \ct{21d} to obtain Eq.~(\ref{6ar}) for single electric and
magnetic charges of the same type (scalar or fermionic). The
general expressions for RGEs are:
\bee
\frac {d\left(\log (\alpha(\mu))\right)}{dt} &=&
   \beta_1(\alpha) + \beta_2\left(\tilde \alpha\right) + C,
                                                           \lb{16Ar}
\\
\frac {d\left(\log\left( \tilde \alpha(\mu)\right)\right)}{dt} &=&
\tilde \beta_1(\alpha) +
\tilde \beta_2\left(\tilde \alpha\right) + \tilde C,
                                                           \lb{16Br}
\eee
The Dirac relation (\ref{54y}) gives:
\be
\frac {d\left(\log( \alpha(\mu))\right)}{dt} = - \frac {d\left(\log\left(
\tilde
\alpha(\mu)\right)\right)}{dt}.
                                                           \lb{16Cr}
\ee
Using Eq.~ (\ref{16Cr}) and the duality symmetry of QEMD, i.e. the
symmetry under the interchange:
\be
          \alpha \longleftrightarrow \tilde \alpha,
                                    \lb{16Dr}
\ee
it is not difficult to obtain:
\be
              C = \tilde C = 0, \quad
   \beta_1(\alpha) = -  \beta_2(\alpha) =
\tilde \beta_1(\alpha) = - \tilde \beta_2(\alpha) = \beta(\alpha),
                                                           \lb{16Er}
\ee
and we have the following RGE:
\be
\frac {d\left(\log( \alpha(\mu))\right)}{dt} = - \frac {d\left(\log\left(
\tilde
\alpha(\mu)\right)\right)}{dt} =
   \beta(\alpha) - \beta\left(\tilde \alpha\right).
                                                           \lb{16Fr}
\ee
If monopole charges, together with electric ones, are sufficiently
small, then $\beta$--functions can be considered by the
perturbation theory:
\be
   \beta(\alpha) = \beta_2 \left(\frac{\alpha} {4\pi}\right) + \beta_4
{\left(\frac{\alpha}{4\pi}\right)}^2 + ...
                                                         \lb{18xr}
\ee
and
\be
  \beta(\tilde\alpha) = \beta_2 \left(\frac{\tilde\alpha} {4\pi}\right) +
\beta_4 {\left(\frac{\tilde\alpha}{4\pi}\right)}^2 + ...
                                                        \lb{19xr}
\ee
with (see paper \ct{47r} and references there):
\be
     \beta_2 = \frac 13   \quad{\mbox{and}} \quad \beta_4 =1  \quad
- \quad {\mbox {for scalar particles}},
                           \lb{20xr}
\ee
and
\be
     \beta_2 = \frac 43   \quad{\mbox{and}} \quad \beta_4 \approx 4
 \quad- \quad {\mbox {for fermions}}.
                           \lb{21xr}
\ee
These cases were investigated in Ref.~\ct{47r}. For scalar electric
and magnetic charges we have \ct{47r}:
\be
   \frac {d\left(\log( \alpha(\mu))\right)}{dt} = - \frac {d\left(\log\left(
   \tilde \alpha(\mu)\right)\right)}{dt} = \beta_2 \frac{\alpha - \tilde
\alpha}{4\pi}\left(1 + 3
   \frac{\alpha + \tilde \alpha}{4\pi} + ...\right)
                                                      \lb{17xr}
\ee
with $\beta_2 = 1/3$, and approximately the same result is valid
for fermionic particles with $\beta_2 = 4/3$.
Eq.~(\ref{17xr}) shows that there exists a region when both fine
structure constants are perturbative. Approximately this region is
given by the following inequalities:
\be
    0.2 \stackrel{<}{\sim }\left(\alpha,\, \tilde \alpha\right)
                    \stackrel{<}{\sim }1.
                                               \lb{22xr}
\ee
Using the Dirac relation (\ref{54y}), we see from Eq.~(\ref{17xr})
that in the region (\ref{22xr}) the two--loop contribution is not
larger than 30\% of the one--loop contribution, and the
perturbation theory can be realized in this case (see
Refs.~\ct{21,21aa,21a,21b,21c,21cc,21d}).

It is necessary to comment that the region (\ref{22xr}) almost
coincides with the region of phase transition couplings obtained
in the lattice $U(1)$--gauge theory (see Subsection \ref{subsec4.4}).

The Zwanziger--like formalism for non--Abelian theories was
considered in Ref.~\ct{21dd}.

Here we want to discuss the comment given in Ref.~\ct{53r}
where the authors argue that the Dirac relation is a consequence of
the non--perturbative effects and cannot be calculated
perturbatively. We insist that the Dirac relation exists always
when we have vortices in the phase. In the region of charges
(\ref{22xr}), obtained in QED, the vortices are created for
perturbative values of non--dual and dual charges.
We always have the Dirac relation, because we always have
the confinement phase: in dual, or non--dual sector, where
the different types of vortices exist.

\subsection{Diminishing of the monopole charge in the FRGGM}
\label{subsec11.2}

There is an interesting way out of this problem if one wants to
have the existence of monopoles, namely to extend the SM gauge
group so cleverly that certain selected linear combinations of
charges get bigger electric couplings than the corresponding SM
couplings. That could make the monopoles which, for these certain
linear combinations of charges, couple more weakly and thus have a
better chance of being allowed ``to exist" \ct{40}.

An example of such an extension of the SM that can impose the
possibility of allowing the existence of free monopoles is just
Family Replicated Gauge Group Model (FRGGM).

FRGGs of type $\left[SU(N)\right]^{N_{fam}}$ lead to the lowering of the
magnetic charge of the monopole belonging to one family:
\be
\tilde \alpha_{\rm one\, family} = \frac{\tilde \alpha}{N_{fam}}.
\lb{*t}
\ee
{}For $N_{fam} = 3$  (for $\left[SU(2)\right]^3$ and $\left[SU(3)\right]^3$) we
have:
$$
\tilde \alpha_{\rm one\, family}^{(2,3)}=\frac{{\tilde\alpha}^{(2,\, 3)}}{3}.$$
For the family replicated group $\left[U(1)\right]^{N_{fam}}$ we obtain:
\be
\tilde \alpha_{\rm one\, family} = \frac{\tilde \alpha}{N^{*}},
\lb{**t}
\ee
where $$N^{*}=\frac{1}{2}N_{fam}\left(N_{fam}+1\right).$$ For $N_{fam}=3$ and
$\left[U(1)\right]^3$, we have: $$\tilde \alpha_{\rm one\, family}^{(1)}=\frac
{{\tilde \alpha}^{(1)}}{6}$$ --- six times smaller! This result was
obtained previously in Ref.~\ct{8}.

\subsection{The FRGGM prediction for the values of
electric and magnetic charges}
\label{subsec11.3}

By reasons considered at the end of this review, we prefer not to
use the terminology ``Anti--grand unification theory, i.e. AGUT",
but call the FRGG--theory with the group of symmetry $G$, or
$G_f$, or $G_{ext}$, given by Eqs.~(\ref{3t}--\ref{6t}) as
``$G$--theory", because as it is shown below, we have a
possibility of the Grand Unification near the Planck scale using
just this theory.

According to the FRGGM, at some point $\mu=\mu_G < \mu_{Pl}$ (but near
$\mu_{Pl}$) the fundamental group $G$ (or $G_f$, or
$G_{ext}$) undergoes spontaneous breakdown to its diagonal
subgroup: \be
      G \longrightarrow G_{diag.\, subgr.} = \left\{g,g,g || g\in SMG\right\},
                                                          \lb{83y}
\ee which is identified with the usual (low--energy) group SMG.
The point $\mu_G\sim 10^{18}$ GeV also is shown in Fig.~\ref{f15},
together with a region of $G$--theory, where AGUT works.

The AGUT prediction of the values of $\alpha_i(\mu)$ at
$\mu=\mu_{Pl}$ is based on the MPM assumption about the existence
of the phase transition boundary point MCP at the Planck scale,
and gives these values in terms of the corresponding critical
couplings $\alpha_{i,crit}$ \ct{8,14,40} (see
Eqs.~(\ref{*t}, \ref{**t})): \be
            \alpha_i(\mu_{Pl}) = \frac {\alpha_{i,\, crit}}{N_{gen}}
                       = \frac{\alpha_{i,\, crit}}{3}
                 \quad{\mbox{for}} \quad i=2,3,       \lb{84y}
\ee and \be \alpha_1(\mu_{Pl}) =
\frac{\alpha_{1,\, crit}}{\frac{1}{2}N_{gen}\left(N_{gen} + 1\right)}
                   = \frac{\alpha_{1,\, crit}}{6}  \quad{\mbox{for}} \quad
U(1).
                                      \lb{85y}
\ee There exists a simple explanation of the relations (\ref{84y})
and (\ref{85y}). As it was mentioned above, the group $G$ breaks
down at $\mu=\mu_G$. It should be said that at the very high
energies $\mu_G \le \mu \le \mu_{Pl}$ (see Fig.~\ref{f15}) each
generation has its own gluons, own $W$'s, etc. The breaking makes
only linear combination of a certain color combination of gluons
which exists in the SM below $\mu=\mu_G$ and down to the low
energies. We can say that the phenomenological gluon is a linear
combination (with amplitude $1/\sqrt 3$ for $N_{gen}=3$) for each
of the AGUT gluons of the same color combination. This means that
coupling constant for the phenomenological gluon has a strength
that is $\sqrt 3$ times smaller, if as we effectively assume that
three AGUT $SU(3)$ couplings are equal to each other. Then we have
the following formula connecting the fine structure constants of
$G$--theory (e.g. AGUT) and low energy surviving diagonal subgroup
$G_{diag.\, subg.}\subseteq {(SMG)}^3$ given by
Eq.~(\ref{83y}): \be \alpha_{diag.,\, i}^{-1} =
\alpha_{1st\, gen.,\, i}^ {-1} + \alpha_{2nd\,
gen.,\, i}^{-1} + \alpha_{3rd\, gen.,\, i}^{-1}.
                                                      \lb{86y}
\ee Here $i = U(1)$, $SU(2)$, $SU(3)$, and $i=3$ means that we talk about
the gluon couplings. For non--Abelian theories we immediately
obtain Eq.~(\ref{84y}) from Eq.~(\ref{86y}) at the critical point MCP.

In contrast to non--Abelian theories, in which the gauge invariance
forbids the mixed (in generations) terms in the Lagrangian of
$G$--theory, the $U(1)$--sector of AGUT contains such mixed terms: \be
   \frac{1}{g^2}\sum_{p,\, q} F_{\mu\nu,\, pq}^2 =
   \frac{1}{g^2_{11}}F_{\mu\nu,\, 11}^2 +
   \frac{1}{g^2_{12}}F_{\mu\nu,\, 12}^2 +
   ...
   + \frac{1}{g^2_{23}}F_{\mu\nu,\, 23}^2 +
   \frac{1}{g^2_{33}}F_{\mu\nu,\,  33}^2,              \lb{87y}
\ee where $p,\, q = 1,\, 2,\, 3$ are the indices of three generations of
the AGUT group $(SMG)^3$. The last equation explains the
difference between the expressions (\ref{84y}) and (\ref{85y}).

It was assumed in Ref.~\ct{8} that the MCP values $\alpha_{i,\, crit}$
in Eqs.~(\ref{84y}) and (\ref{85y}) coincide with the triple point
values of the effective fine structure constants given by the
generalized lattice $SU(3)$--, $SU(2)$-- and $U(1)$--gauge theories
described by Eqs.~(\ref{37}) and (\ref{38}). Also it was used a
natural assumption that the effective $\alpha_{crit}$ does not
change its value (at least too much) along the whole bordeline ``3"
of Fig.~\ref{f7} for the phase transition ``Coulomb--confinement" in the
$U(1)$ lattice gauge theory with the generalized (two parameters)
lattice Wilson action (\ref{38}).

Now let us consider $\alpha_Y^{-1}(\approx \alpha^{-1})$ at the
point $\mu=\mu_G\sim 10^{18}$ GeV shown in Fig.~\ref{f15}. If the point
$\mu=\mu_G$ is very close to the Planck scale $\mu=\mu_{Pl}$, then
according to Eqs.~(\ref{82y}) and (\ref{85y}), we have: \be
         \alpha_{1st\,  gen.}^{-1}\approx
    \alpha_{2nd\,  gen.}^{-1}\approx \alpha_{3rd\,  gen.}^{-1}\approx
    \frac{\alpha_Y^{-1}\left(\mu_G\right)}{6}\approx 9,        \lb{88y}
\ee what is almost equal to the value (\ref{50a}):
\be
            \alpha_{crit.,\, theor}^{-1}\approx 8.5 \lb{4e}
\ee
obtained by the ``Parisi improvement method" (see Fig.~14). This
means that in the $U(1)$ sector of AGUT we have $\alpha_Y $ (or
$\alpha_1$) near the critical point. Therefore, we can expect the
existence of MCP at the Planck scale.

\subsection{ Evolution of the running fine structure constant in the
$U(1)$ theory with monopoles}
\label{subsec11.4}

Considering the evolution of the running $U(1)$ fine structure
constant
$$\alpha_Y(\mu)=\frac{\alpha(\mu)}{cos^2\theta_{\ov{MS}}},$$
we have in the SM the following one--loop approximation:
\be
  \alpha_Y^{-1}(\mu) =
  \alpha_Y^{-1}\left(\mu_R\right) + \frac{b_Y}{4\pi}t,
                                                \lb{4ea}
\ee
where $t$ is the evolution variable (\ref{3}), and $b_Y$ is given
for $N_{gen}$ generations and $N_S$ Higgs bosons by the following
expression:
\be
        b_Y = - \frac{20}{9}N_{gen} - \frac{1}{6} N_S.
                                                     \lb{4ee}
\ee
The evolution of $\alpha_Y^{-1}(\mu)$ is represented in Fig.~\ref{f15}
for $N_{gen}=3$ and $N_S=1$ by the straight line going up to $\mu
= \mu_{Pl}$.

Let us consider now the exotic (not existing in reality) case when
we have, for example, the cut--off energy $\mu_{\rm cut-off}\sim
10^{42}$ GeV \ct{21d}. In this case the evolution of
${\alpha_Y}^{-1}(\mu)$
is given by Fig.~\ref{f17a}, where the straight line 1 (one--loop
approximation) goes to the Landau pole at $\alpha_Y^{-1} = 0$. But
it is obvious that in the vicinity of the Landau pole, when
$\alpha_Y^{-1}(\mu)\to 0$, the charge of $U(1)$ group becomes larger
and larger with increasing of $\mu$. This means that the one--loop
approximation for $\beta$--function is not valid for large $\mu$,
and the straight line 1 may change its behaviour. In general, the
two--loop approximation for $\beta$--function of QED (see
Refs.~\ct{22y,27y,28} and \ct{42}) shows that this
straight line turns and goes down.

Exotically we can consider our space-time as a lattice with
parameter $a$ smaller than the Planck scale value $\lambda_P$. For
example, we can imagine a lattice with $a\sim
\left(10^{42}\, {\mbox{GeV}}\right)^{-1}$, or the existence of the fundamental
magnetically charged Higgs scalar field in the vicinity of large
$\mu_{crit}\sim 10^{38}$ GeV, when we have the phase transition
point with $\alpha_{Y,\, crit}^{-1}\approx 5$ (see below (\ref{5xe})
and Fig.~\ref{f17a}).

The artifact monopoles, responsible for the confinement of
electric charges at the very small distances, can be approximated
by the magnetically charged Higgs scalar field, which leads to the
confinement--deconfinement phase transition, as it was shown in
Refs.~\ct{21,21aa,21a,21b,21c,21cc,21d}. If we have this phase
transition, then there exists a rapid fall in the evolution of
$\alpha^{-1}(\mu)$  $(\alpha_Y^{-1}$ or $\alpha_{Y,\, G}^{-1}$,
etc.) near the phase transition (critical) point. This ``fall"
always accompanies the phase transition from the Coulomb--like
phase to the confinement one.

Indeed, we can present the effective Lagrangian of our field
system as a function of the variable
\be
F^2\equiv F_{\mu\nu}^2,     \lb{5xde}
\ee
where
\be
 F_{\mu\nu}= \partial_{\mu}A_{\nu}-\partial_{\nu}A_{\mu}   \lb{6xe}
\ee
is the field strength tensor:
\be
L_{eff}  =
       - \frac {\alpha_{eff}^{-1}\left(F^2\right)}{16\pi} F^2.
                                   \lb{6xabe}
\ee
When $F^2={\vec B}^2$ ($\vec B$ is the magnetic field) and ${\vec
B}^2$ is independent of space--time coordinates (see
Refs.~\ct{22q,23q,23qq}), we can write the effective potential:
\be
V_{eff}  =
         \frac {\alpha_{eff}^{-1}\left({\vec B}^2\right)}{16\pi} {\vec B}^2 =
           A\alpha_{eff}^{-1}(t)e^{2t}.
                                               \lb{6xae}
\ee
Here, choosing ${\vec B}^2={\mu}^4$ and $\mu_R=\mu_{\rm cut-off}$, we
have:
$$
A=\frac{\mu_{\rm cut-off}^4}{16\pi},
$$
and
\be
            t = \frac{1}{2}\log\left(\frac{F^2}{\mu_{\rm cut-off}^4}\right).
                                                      \lb{6xde}
\ee
It is well--known \ct{22q,23q} that $\alpha_{eff}^{-1}(t)$
has the same evolution over $t=\log \left(\mu^2/\mu_{\rm cut-off}^2\right)$
whether we consider $\mu^2=p^2$ (where p is the 4-momentum), or
$\mu^4=F^2={\vec B}^2$ (at least, up to the second order
perturbation).

In the confinement region $\left(t > t_{crit}\right)$ the effective
potential
(\ref{6xae}) has a minimum, given by the requirement \ct{23q}:
\be
         \left.\left[\frac {d\alpha_{eff}^{-1}}{dt} +
                           2\alpha_{eff}^{-1}(t)\right]\right|_{t=t_{min}} = 0.
                                     \lb{7xe}
\ee
Of course, we need this minimum for $t$--values above $t_{crit}$
in order to have confinement which namely means that we have a
nonzero $F^2 ={\vec B}_0^2=const$ in the vacuum \ct{24p,14p,14pa,14pb}.

This minimum of the effective potential can exist only if
$\alpha_{eff}^{-1}(\mu)$ has a rapid fall near the phase
transition point which is illustrated in Figs.~\ref{f17a}, \ref{f17b}
by curve 2.

The existence of minimum of the effective potential (\ref{6xae})
explains why the straight line 1 changes its behaviour and rapidly
falls: $\alpha_{eff}^{-1}(t)$ is multiplied by $\exp(2t)$ in
Eq.~(\ref{6xae}).

After this ``fall" $\alpha_Y^{-1}(\mu)$ has a crook and goes to
the constant value, as it is shown in Fig.~\ref{f17a} by solid curve 2,
demonstrating the phase transition from Coulomb--like phase to the
confinement one.

The next step is to give the explanation why
$\alpha_{eff}^{-1}(t)$ is arrested when $t\to t_{\rm cut-off}$.

The process of formation of strings in the confinement phase
(considered in Ref.~\ct{21}) leads to the ``freezing" of $\alpha$:
in the confinement phase the effective electric fine structure
constant is almost unchanged and approaches its maximal value
$\alpha=\alpha_{max}$ when ($\mu \to \infty$). It was shown
in Subsection \ref{subsec4.4} that the authors of
Ref.~\ct{14p} predicted the maximal value:
$ \alpha_{max} \approx \pi/12\approx 0.26,$
due to the Casimir effect for strings.

Fig.~\ref{f9a} demonstrates the tendency to freezing of $\alpha$ in the
compact QED for $\beta < \beta_T$ (i.e. for ``bare" constant $e_0 >
1$, what means $\alpha_0^{-1} < 4\pi \approx 12.56$).

Choosing the lattice result (\ref{47}), which almost coincides
with the HMM result (\ref{55y}), we have:
\be
     \alpha_{Y}^{-1}\left(\mu_{crit}\right) \equiv \alpha_{U(1),\,
crit}^{-1}\approx 5.
                                                            \lb{5xe}
\ee
Analogously using (\ref{50c}) we have the maximal value for
$\alpha_Y^{-1}$:
\be
       \alpha_{Y,\, max}^{-1}
        \approx \frac{1}{0.26} \approx 3.8.              \lb{5xae}
\ee
An interesting situation arises in the theory with FRGG--symmetry,
when it begins to work at $\mu=\mu_G \left(< \mu_{Pl}\right)$. As it was
shown in Subsection \ref{subsec11.2}, in the vicinity of the phase transition
point the $U(1)$--sector of FRGG has $\alpha_{Y,\, G}\equiv
\alpha_{Y,\, one\, fam.}$, which is 6 times larger than $\alpha_Y$.
Now the phase transition ``deconfinement--confinement" occurs at
$\mu_{crit}=\mu_{Pl}$ (but not at $\mu_{crit}\sim 10^{38}$ GeV as
it was in the SM prolonged up to the scale $\mu_{\rm cut-off}\sim
10^{42}$ GeV). This case is not ``exotic" more, and confirms the
MPP idea \ct{8}.

The evolution of the one family inverse fine structure constant
$\alpha_{Y,\, G}(\mu )^{-1}(x)$  is given in Fig.~\ref{f17b}, which
demonstrates the existence of critical point at the Planck scale.

Here it is necessary to comment that we have given a qualitative
behaviour of the fall in Figs.~\ref{f17a}, \ref{f17b} believing in
the confinement
existing in the $U(1)_Y$ theory. In general, the existence of the
confinement, as well as the shape of the fall
(wide or narrow), depends on the type of theory considered. If the
cut--off energy is not very high $\left(\mu_{\rm cut-off} <
\mu_{crit}\right)$, or
the lattice spacing $a$ is not too small, then there is no
confinement region in such a theory (for example, in the $U(1)$
sector of the SM we have $\mu_{\rm cut-off}=\mu_{Pl}$ which is smaller
than $\mu_{crit}$ given by Fig.~\ref{f17a} and the confinement phase
is not available).

According to MPP {\footnote{We call it MPP-II to distinguish this
definition from the version in which MPP-I is defined only as a
phase transition depending on bare couplings while no scale
involved}}, Nature has to have the phase transition point at the
Planck scale not only for the Abelian $U(1)$ theory, but also for
non--Abelian theories. This means that the effective potential of
the $SU(3)$ gauge theory has the second minimum at the Planck scale
(the first one corresponds to the low--energy hadron physics).
String states of this second confinement phase are not observed in
Nature, because the FRGG--theory approaches the confinement phase
at the Planck scale, but does not reach it.

\subsection{Olive's monopoles}
\label{subsec11.5}

The fact that we have one special monopole for each of the three
groups $SU(3)$, $SU(2)$ and $U(1)$, which we have considered calculating
the phase transition (critical) couplings (for the confinement due
to monopole condensation, either in the SM, or in FRGGM) is indeed
not consistent with quarks and leptons as phenomenologically found
objects. The point is that the $SU(3)$ monopole, for instance,
radiates a flux corresponding to a path in the gauge group $SU(3)$
from the unit element to one of the non--trivial center elements.
Such a monopole gives rise to a phase factor $\exp(2\pi i/3)$ when
a quark encircles its Dirac string. Therefore, it does not allow
quarks.

What is allowed consistently with the SM representations is the
object which was proposed by D.~Olive \ct{40ao} and called ``the
Olive--monopole". These monopoles have the magnetic charge under
{\it all three subgroups: $SU(3)$, $SU(2)$ and $U(1)$} of SMG. Their
total magnetic charge corresponds to the center element $
\left(Ie^{i2\pi /3},\, -I,\, 2\pi \right)\in SU(3)\times SU(2)\times {\bf R}$
contained in the covering group $SU(3)\times SU(2)\times {\bf R}$
of SMG, which according to the interpretation of Ref.~\ct{40bo},
has a meaning of the {\it Lie group}, rather than just Lie
algebra, fitting the SM representations. That is, Olive--monopole
in the SM has at the same time three different magnetic charges
with the following sizes:

\begin{enumerate}
\item[1.] An $SU(3)$ magnetic charge identical to the one that would allow
only the representations of the {\un{group}} $SU(3)/Z_{3}$,
i.e. only the representations with triality t=0.

\item[2.] An $SU(2)$ magnetic charge identical to the one that would allow
only representations of the {\un{group}} $SU(2)/Z_{2}=
SO(3)$, i.e. only the representations with integer weak isospin.

\item[3.] And finally, a $U(1)$ weak hypercharge monopolic charge of a size
that if alone would allow only integer values of the weak
hypercharge half, i.e. of $y/2=$ integer.
\end{enumerate}

These three magnetic charge contributions would, if alone, not
allow  the existence neither fermions, nor the Higgs bosons in the
SM. However, considering the phase of a quark or lepton field
along a little circle encircling the Dirac string for the Olive's
SM--monopole, one gets typically a phase rotation from each of the
three contributions to the magnetic charge. The consistency
condition to have the Dirac string without visible effect is that
these phase contributions {\em together} make up a multiple of
$2\pi $. It can be checked that the quark and lepton
representations, as well as the Weinberg--Salam Higgs boson
representation, lead to the full phase rotations which are indeed
a multiple of $2\pi $.

Thus, as it was already mentioned above, if we imagine monopoles
with each of these contributions alone they would not allow
neither the phenomenologically observed quarks and leptons, nor
the Higgs bosons.

In going to the FRGGM we can, without problem, postulate one
Olive--monopole for each proto--family since the proto--family
representations are just analogous to the ones in the SM.

Considering the Olive--monopoles condensation (causing a
confinement-deconfinement phase transition) for different
families, we assume that as long as we consider, for example, only
the $SU(3)$--coupling to cause the phase transition, the
Olive--monopole functions as if it is the $SU(3)$--monopole
consistent only with the representations of $SU(3)/Z_3$. But this
is just what gives the phase transition couplings derived with
help of Eqs.~(\ref{5xe}) and (\ref{4e}). Similarly, it is easy to
see that the use of the Olive--monopole for all gauge groups
$SU(3)$, $SU(2)$, $U(1)$ leads to the phase transition couplings
obtained by combining Eqs.~(\ref{4e}) and (\ref{5xe}).

\section{Anti--GUT prediction of coupling constants near the
Planck scale}
\label{sec12}

As it was mentioned above, the lattice investigators were not able
to obtain the lattice triple point values of $\alpha_{i,\, crit}$
$(i=1,\, 2,\, 3$ correspond to $U(1)$, $SU(2)$ and $SU(3)$ groups)
by Monte Carlo
simulation methods. These values were calculated theoretically by
Bennett and Nielsen in Ref.~\ct{8}. Using the lattice triple point
values of $\left(\beta_A;\, \beta_f\right)$ and $\left(\beta^{lat}
;\, \gamma^{lat}\right)$ (see Figs.~\ref{f5a}, \ref{f5b} and Fig.~\ref{f7}),
they have obtained
$\alpha_{i,\, crit}$ by the ``Parisi improvement method":
\be
    \alpha_{Y,\, crit}^{-1}\approx 9.2\pm 1,
     \quad \alpha_{2,\, crit}^{-1}\approx 16.5\pm 1,  \quad
    \alpha_{3,\, crit}^{-1}\approx 18.9\pm 1.                 \lb{89y}
\ee
Assuming the existence of MCP at $\mu=\mu_{Pl}$ and
substituting the last results in Eqs.~(\ref{84y}) and (\ref{85y}),
we have the following prediction of AGUT \ct{8}:
\be
   \alpha_Y^{-1}\left(\mu_{Pl}\right)\approx 55\pm 6;  \quad
   \alpha_2^{-1}\left(\mu_{Pl}\right)\approx 49.5\pm 3;  \quad
   \alpha_3^{-1}\left(\mu_{Pl}\right)\approx 57.0\pm 3.
                                                          \lb{90y}
\ee
These results coincide with the results (\ref{82y}) obtained
by the extrapolation of experimental data to the Planck scale in
the framework of the pure SM (without any new particles) \ct{8,33a}.

Using the relation (\ref{25z}), we obtained the result
(\ref{26z}), which in our case gives the following relations
\ct{21d}: \be
    \alpha_{Y,\, crit}^{-1} : \alpha_{2,\, crit}^{-1} : \alpha_{3,\, crit}^{-1}
           = 1 : \sqrt{3} : \frac{3}{\sqrt{2}} = 1 : 1.73 : 2.12.
                                                     \lb{91y}
\ee Let us compare now these relations with the MPM prediction.

For $\alpha_{Y,\, crit}^{-1}\approx 9.2$  given by the first equation
of (\ref{89y}), we have:
\be
 \alpha_{Y,\, crit}^{-1} : \alpha_{2,\, crit}^{-1} : \alpha_{3,\, crit}^{-1}
    = 9.2 : 15.9 : 19.5.                                     \lb{92y}
\ee In the framework of errors, the last result coincides with the
AGUT--MPM prediction (\ref{89y}). Of course, it is necessary to
take into account an approximate description of the confinement
dynamics in the $SU(N)$ gauge theories, developed by our
investigations.

\section{The possibility of the Grand Unification near the Planck scale}
\label{sec13}

In the Anti-grand unified theory (AGUT)
\ct{11,12,13,14,15,15a,16,16a,17,18,19} the FRGG
breakdown was considered at $\mu_G\sim 10^{18}$ GeV. It is a
significant point for MPM. In this case the evolutions of the fine
structure constants $\alpha_i(\mu )$ exclude the existence of the
unification point up to the Planck scale (also in the region $\mu
> \mu_G$).

The aim of this Section is to show, as in Ref.~\ct{21d}, that we
have quite different consequences of the extension of the SM to
FRGGM if $G$--group undergoes the breakdown to its diagonal
subgroup (i.e. SM) not at $\mu_G\sim 10^{18}$ {GeV}, but at
$\mu_G\sim 10^{14}$ or $10^{15}$ {GeV}, i.e. before the
intersection of $\alpha_{2}^{-1}(\mu)$ with $\alpha_{3}^{-1}(\mu)$
at $\mu\approx 10^{16}$ GeV.

In fact, here we are going to illustrate the idea that with
monopoles we can modify the running of the fine structure
constants so much that unification can be arranged without needing
SUSY. To avoid confinement of monopoles at the cut--off scale we
need FRGG or some replacement for it (see Subsection \ref{subsec11.2}).

If we want to realize the behaviour shown in Fig.~\ref{f17b} for the
function $\alpha_{Y,\, FRGG}^{-1}(\mu )= \alpha_{Y,\, G}^{-1}(\mu )$
near the Planck scale, then we need the evolution of all
$\alpha_i^{-1}(\mu)$ to turn away from asymptotic freedom. This
could be achieved by having more fermions, i.e. generations,
appearing above the $\mu_G$--scale.

Now we shall suggest that such appearance of more fermions above
$\mu_G$ is not so unlikely.

\subsection{Guessing more particles in the FRGGM}
\label{subsec13.1}

Once we have learned about the not so extremely simple SM, it is
not looking likely that the fundamental theory --- the true model
of everything --- should be so simple as to have only one single
type of particle --- the ``urparticle" --- unless this ``urparticle"
should be a particle that can be in many states internally such as
say the superstring. Therefore, we should not necessarily assume
that the number of species of particles is minimal anymore
--- as could have been reasonable in a period of science where one
had only electron, proton and perhaps neutron, ignoring the photon
exchanges which bind the atom together, so that only three
particles really existed there. Since now there are too many
particles.

Looking at the problem of guessing the physical laws beyond the
SM, we should rather attempt to guess a set of species of
particles to exist and the order of magnitudes of the numbers of
such species which one should find at the various scales of
energy. Indeed, the historical learning about the species of
particles in Nature has rather been that physicists have learned
about many types of particles not much called for at first: It is
only rather few of particle types, which physicists know today
that have so great significance in the building of matter or other
obviously important applications that one could not almost equally
well imagine a world without these particles. They have just been
found flavour after flavour experimentally studies often as a
surprise, and if for instance the charmed quark was needed for
making left handed doublets, that could be considered as a very
little detail in the weak interactions which maybe was not needed
itself.

These remarks are meant to suggest that if we should make our
expectations to be more ``realistic" in the sense that we should
get less surprised next time when the Nature provides us with new
and seemingly not needed particles, we should rather than guessing
on the minimal system of particles seek to make some more
statistical considerations as to how many particle types we should
really expect to find in different ranges of energy or mass.

To even crudely attack this problem of guessing it is important to
have in mind the reasons for particles having the mass order of
magnitudes. In this connection, a quite crucial feature of the SM
``zoo" is that except for the Higgs particle itself, are mass
protected particles, which have no mass in the limit when the
Higgs field has no vacuum expectation value (VEV). In principle,
you would therefore expect that the SM particles have masses of
order of the Higgs VEV. Actually they are mostly lighter than that
by up to five orders of magnitude.

In the light of this mass protection phenomenon it would be really
very strange to assume that there should be no other particles
than in the SM if one went up the energy scale and looked for
heavier particles, because then what one could ask: Why should
there be only mass protected particles ? After all, there are lots
of possibilities for making vector coupled Dirac particles, say.
It would be a strange accident if Nature should only have mass
protected particles and not a single true Dirac particle being
vector coupled to even the weak gauge particles. It is much more
natural to think that there are at higher masses lots of different
particles, mass protected as well as not mass protected. But
because we until now only could ``see" the lightest ones among
them, we only ``saw" the mass protected ones.

In this light the estimate of how many particles are to be found
with higher masses should now be a question of estimating how many
particles turn out mass protected and getting masses which we can
afford to ``see" today.

We are already in the present article having the picture that as
one goes up in the energy scale $\mu $ there will be bigger and
bigger gauge group, which will thus be able to mass protect more
and more particle types. Each time one passes a breaking scale at
which a part of the gauge group breaks down --- or thought the way
from infrared towards the ultraviolet: each time we get into
having a new set of gauge particles --- there will be a bunch of
fermions which are mass protected to get --- modulo small Yukawa
couplings --- masses just of the order of magnitude of the Higgs
scale corresponding to that scale of diminishing of the gauge
group.

If we for example think of the scale of breaking of our FRGG down
to its diagonal subgroup, then we must expect that there should be
some fermions just mass protected to that scale. However, these
particles should be vector coupled w.r.t. the SM gauge fields, and
only mass protected by the gauge fields of the FRGG, which are not
diagonal. We would really like to say that it would be rather
strange if indeed there were no such particles just mass protected
to that scale.

\subsection{Quantitative estimate of number of particles in the FRGGM}
\label{subsec13.2}

We might even make an attempt to perform a quantitative estimate
of how many particle species we should expect to appear when we
pass the scale $\mu = \mu_G$ going above $\mu_G $. Of course, such
an estimate can be expected to be very crude and statistical, but
we hope that anyway it would be better than the unjustified guess
that there should be nothing, although this guess could be in some
sense the simplest one.

Since it is going to be dependent on the detailed way of arguing,
we should like to make a couple of such estimates:

\begin{enumerate}
\item[1.] The first estimate is the guessing that, in analogy with the
type of particle combinations which we have had in FRGGM already,
we find the particles grouped into families which are just copies
of the SM families. If only one of the gauge group families is
considered, then two others are represented trivially on that
family. We shall, however, allow that these families can easily be
mirror families, in the sense that they have the weak doublets
being the right handed particles and actually every gauge quantum
number parity are reflected. But, by some principle, only small
representations are realized in Nature and we could assume away
the higher representations. Now let us call the number of mirror
plus ordinary families which are present above the scale $\mu_G$
of the breakdown of FRGG to the diagonal subgroup as
$N_{fam,\, tot}$, and assume that we have a statistical distribution
as if these families or mirror families had been made one by one
independently of each other with a probability of 50\% for it
being a mirror family and 50\% for it being an ordinary left
handed one. The order of the number of families survived under the
scale $\mu_G$ should then be equal to the order of the difference
between two samples of the Poisson distributed numbers with
average $N_{fam,\, tot}/2$. In fact, we might consider respectively
the number of mirror families and the number of genuine (left)
families as such Poisson distributed numbers. The excess of the
one type over the other one is then the number of low energy
scales surviving families, and the physicists living today can
afford to see them. It is well known that crudely this difference
is of the order of $\sqrt{N_{fam,\, tot}}$.  But we know that this
number of surviving families has already been measured to be 3,
and so we expect that $\sqrt{N_{fam,\, tot}}=3$, what gives
$N_{fam,\, tot}\approx 9$. This would mean that there are 6 more
families to be found above the diagonal subgroup breaking scale
$\mu_G$.

\item[2.] As an alternative way of estimating, we can say very crudely
that the fermions above the scale $\mu_G$ could be mass protected
by 3 times as many possible gauge quantum numbers, as far as there
are three families of gauge boson systems in our FRGGM. If we use
the ``small representations assumption", then going from $\mu >
\mu_G$ to $\mu < \mu_G$, two of three fermions loose their mass
protection, and these two fermions obtain masses of order of the
scale $\mu_G$. But this means that 1/3 of all fermions survive to
get masses below the scale $\mu_G$ and become ``observable in
practice". Again we got that there should be two times as many
particles with masses at the diagonal subgroup breaking scale
$\mu_G$ than at the EW scale. We got the same result by two
different ways. Notice though that in both case we have used a
phenomenologically supported assumption that Nature prefers very
small representations. In fact, it seems to be true that the SM
representations are typically the smallest ones allowed by the
charge quantization rule: \be
     \frac{Y}{2} + \frac{d}{2} + \frac{t}{3} = 0  \quad ({\mbox{mod}}\, 1),
                                 \lb{G1}
\ee where $d$ and $t$ are duality and triality, respectively.
\end{enumerate}

\subsection{The FRGGM prediction of RGEs. The evolution of
fine structure constants near the Planck scale}
\label{subsec13.3}

Let us consider now, in contrast to the AGUT having the breakdown
of G--group at $\mu_G\sim 10^{18}$ GeV, a new possibility of the
FRGG breakdown at $\mu_G\sim 10^{14}$ or $10^{15}$ GeV (that is,
before the intersection of $\alpha_{2}^{-1}(\mu )$ with
$\alpha_{3}^{-1}(\mu )$, taking place at $\mu \sim 10^{16}$ GeV in
SM). This possibility was considered in Ref.~\ct{21d}. Then in the region
$\mu_G < \mu < \mu_{Pl}$ we have three
SMG $\times U(1)_{(B-L)}$ groups for three FRGG families as in
Refs.~\ct{11,12,13,14,15,15a,16,16a,17,18,19}.

In this region we have, according to the statistical estimates
made in Subsection \ref{subsec13.2}, a lot of fermions, mass protected or not
mass protected, belonging to usual families or to mirror ones. We
designate the total number of these fermions $N_F$, maybe
different with $N_{fam.\, tot}$.

Also monopoles can be important in the vicinity of the Planck
scale: they can give essential contributions to RGEs for
$\alpha_i(\mu )$ and change the previously considered evolution of
the fine structure constants.

Analogously to Eq.~(\ref{16Fr}), obtained in Ref.~\ct{47r}, we can
write the following RGEs for $\alpha_i(\mu)$ containing
$\beta$--functions for monopoles:
\be
\frac {d\left(\log\left( \alpha_i(\mu)\right)\right)}{dt} =
   \beta(\alpha_i) - \beta^{(m)}\left(\tilde \alpha_i\right), \quad i=1,2,3.
                                                           \lb{G2}
\ee
We can use the one--loop approximation for $\beta(\alpha_i)$
because $\alpha_i$ are small, and the two--loop approximation for
dual $\beta$--function $\beta^{(m)}\left(\tilde \alpha_i\right)$ by reason that
$\tilde \alpha_i$ are not very small.  Finally, taking into
account that in the non--Abelian sectors of FRGG we have the
Abelian artifact monopoles (see Subsection \ref{subsec7.1}), we obtain the
following RGEs:
\be
   \frac {d\left(\alpha_i^{-1}(\mu)\right)}{dt} = \frac{b_i}{4\pi } +
   \frac{N_M^{(i)}}{\alpha_i}\beta^{(m)}\left(\tilde \alpha_{U(1)}\right),
\lb{G3}
\ee
where $b_i$ are given by the following values:
\bee
   b_i &=& \left(b_1, b_2, b_3\right)
\nonumber\\
&=&\left( - \frac{4N_F}{3} -\frac{1}{10}N_S,\,
      \frac{22}{3}N_V - \frac{4N_F}{3} -\frac{1}{6}N_S,\,
      11 N_V - \frac{4N_F}{3} \right).                   \lb{G4}
\eee
The integers $N_F,\, N_S,\, N_V$ are respectively the total numbers
of fermions, Higgs bosons and vector gauge fields in FRGGM
considered in our theory, while the integers $N_M^{(i)}$ describe
the amount of contributions of scalar monopoles.

The Abelian monopole $\beta$--function in the two--loop approximation
is:
\be
      \beta^{(m)}\left(\tilde \alpha_{U(1)}\right) = \frac{\tilde
\alpha_{U(1)}}{12\pi }
      \left(1 + 3 \frac{\tilde \alpha_{U(1)}}{4\pi }\right).      \lb{G5}
\ee
Using the Dirac relation (\ref{54y}) we have: \be
      \beta^{(m)} =
      \frac{\alpha_{U(1)}^{-1}}{48\pi }
      \left(1 + 3 \frac{\alpha_{U(1)}^{-1}}{16\pi }\right),      \lb{G6}
\ee and the group dependence relation (\ref{25z}) gives: \be
      \beta^{(m)} =
      \frac{{C_i\alpha_i}^{-1}}{48\pi }
      \left(1 + 3 \frac{{C_i\alpha_i}^{-1}}{16\pi }\right),
                                                     \lb{G7}
\ee where
\be
   C_i = \left(C_1, C_2, C_3\right) =
        \left(\frac{5}{3}, \frac{1}{\sqrt{3}}, \frac{\sqrt{2}}{3}\right).
                                                          \lb{G8}
\ee
Finally we have the following RGEs:
\be
   \frac {d\left(\alpha_i^{-1}(\mu)\right)}{dt} = \frac{b_i}{4\pi } +
   N_M^{(i)} \frac{{C_i\alpha_i}^{-2}}{48\pi }
      \left(1 + 3 \frac{{C_i\alpha_i}^{-1}}{16\pi }\right),         \lb{G9}
\ee
where $b_i$ and $C_i$ are given by Eqs.~(\ref{G4}) and (\ref{G8}),
respectively.

In our FRGG model:
\be
      N_V = 3,                       \lb{G10}
\ee
because we have 3 times more gauge fields in comparison with the
SM $\left(N_{fam}=3\right)$.

As an illustration of the {\em possibility} of unification in an
FRGG--scheme with monopoles we propose some--strictly speaking
adjusted--parameter choices within the likely ranges already
suggested.

In fact, we take the total number of fermions $N_F=2N_{fam.\, tot}$
(usual and mirror families), $N_{fam.\, tot}=N_{fam}N_{gen}=3\times
3=9$ (three SMG groups with three generations in each group), we
have obtained (see Fig.~\ref{f18}) the evolutions of $\alpha_i^{-1}(\mu
)$ near the Planck scale by numerical calculations for $N_F=18$,
$N_S=6$, $N_M^{(1)}=6$, $N_M^{(2,\, 3)}=3$ and the following
$\alpha_i^{-1}(\mu_{Pl})$:
\be
   \alpha_1^{-1}\left(\mu_{Pl}\right)\approx 13, \quad
\alpha_2^{-1}\left(\mu_{Pl}\right)\approx 19, \quad
\alpha_1^{-1}\left(\mu_{Pl}\right)\approx 24,                  \lb{G10a}
\ee
which were considered instead of Eq.~(\ref{89y}).

We think that the values $N_M^{(i)}$,  which we have used here,
are in agreement with Eqs.~(\ref{84y}) and (\ref{85y}), and $N_S=6$
shows the the existence of the six scalar Higgs bosons breaking
FRGG to SMG (compare with the similar descriptions in
Refs.~\ct{15,15a,16,16a}).

Fig.~\ref{f18} shows the existence of the unification point.

We see that a lot of new fermions in the region $\mu > \mu_G$ and
monopoles near the Planck scale change the one--loop approximation
behaviour of $\alpha_i^{-1}(\mu)$ in SM. In the vicinity of the
Planck scale these evolutions begin to decrease, approaching the
phase transition (multiple critical) point at $\mu = \mu_{Pl}$
what means the suppression of the asymptotic freedom in the
non--Abelian theories.

Here it is necessary to emphasize that these results do not depend
on the fact, whether we have in Nature lattice artifact monopoles,
or the fundamental Higgs scalar particles with a magnetic charge
(scalar monopoles).

Fig.~\ref{f19} demonstrates the unification of all gauge interactions
including gravity (the intersection of $\alpha_g^{-1}$ with
$\alpha_i^{-1}$) at
\be
\alpha_{\rm GUT}^{-1}\approx 27 \quad {\mbox{and}}  \quad x_{\rm GUT}\approx
18.4.                                     \lb{G11a}
\ee

Here we can expect the existence of $\left[SU(5)\right]^3$ or
$\left[SO(10)\right]^3$
unification. Of course, the results obtained
in Ref.~\ct{21d} are preliminary and in future it is desirable to
perform the spacious investigation of the unification possibility.

Calculating the GUT--values for one family fine structure
constants considered in this paper we have for $i=5$:
\be
       \alpha_{GUT,\, one\, fam}^{-1} = \frac{\alpha_{\rm GUT}}{3}\approx 9,
                                                             \lb{G11b}
\ee
what corresponds to the Abelian monopole (for the $SU(5)/Z_5$
lattice artifact) coupling with the total average monopolic fine
structure constant $\tilde \alpha_{eff}$ and the ``genuine"
monopole fine structure constant $\tilde \alpha_{\rm genuine}$, as
defined in Ref.~\ct{21d}, determined from Eq.~(\ref{25z}) and the Dirac
relation by
\be
        \alpha_N^{-1}=\frac{N}{2}\sqrt{\frac{N+1}{N-1}}\cdot 4\tilde
\alpha_{eff} = \frac{2N}{N-1}\tilde \alpha_{\rm genuine}
                                                           \lb{G11c}
\ee
leading to
\be
    \tilde \alpha_{eff}= 9\cdot
\frac{2}{5\cdot 4}\sqrt{\frac{4}{6}} = 0.7,
                                        \lb{G11d}
\ee
and
\be
   \tilde \alpha_{\rm genuine}\approx \frac{2\cdot 4}{4\cdot 5}\cdot 9
\approx 3.6.               \lb{G11e}
\ee
This value ${\tilde \alpha}_{eff}$ suggests that we may apply crude
perturbation theory both for monopoles and charges if accepting
that in the region (\ref{22xr}).

Critical coupling corresponds (see Section \ref{sec5}) to $\tilde
\alpha_{eff,\,  crit} = 1.20$ giving for $SU(5)/Z_5$
\be
   \alpha_{5,\, crit}^{-1}\approx \frac{5}{2}\sqrt{\frac{6}{4}}\cdot 4\cdot
1.20 \approx 14.5,               \lb{G11f}
\ee
meaning that the unified couplings for $\left[SU(5)\right]^3$ are already of
confinement strength.

Within the uncertainties we might, however, consider from
(\ref{G11b}) the coupling strength $\alpha_{\rm GUT}\approx 1/9$ as
being equal to the critical value $\alpha_{5,\, crit}\approx 1/14$
from Eq.~(\ref{G11f}).

If indeed the $\alpha_{\rm GUT}$ were so strong as to suggest
confinement at the unification point, it will cause the problem
that the fundamental fermions in $SU(5)$ representations would be
confined and never show up at lower energy scales.

Assuming the appearance of SUSY, we can expect to see sparticles
at the GUT--scale with masses:
\be
                 M\approx 10^{18.4}\, {\mbox{GeV}}.         \lb{G11}
\ee
Then the scale $\mu_{\rm GUT}=M$ given by Eq.~(\ref{G11}) can be
considered as a SUSY breaking scale.

The unification theory with $\left[SU(5)\right]^3$--symmetry was suggested
first by S.~Rajpoot \ct{44} (see also \ct{45}).

Note that since our possibility of unification has the very high
scale (\ref{G11}), it allows for much longer proton lifetime than
corresponding models with more usual unification scales, around
$10^{16}$ GeV. This is true not only for proton decay caused by
the gauge boson exchange, but also by the triplet Higgs exchange,
since then the mass of the latter also may be put up in the scale.

Considering the predictions of such a theory for the low--energy
physics and cosmology, maybe in future we shall be able to answer
the question: ``Does the unification of $\left[SU(5)\right]^3$ or
$\left[SO(10)\right]^3$
type (SUSY or not SUSY) really exist near the Planck scale ?"

Recently F.S.~Ling and P.~Ramond \ct{46} considered the group of
symmetry $\left[SO(10)\right]^3$ and showed that it explains the observed
hierarchies of fermion masses and mixings.

\section{Discussion of some various scenarios of working MPP}
\label{sec14}

In the present article, except for Section \ref{sec12}, we have taken the
picture that the inverse fine structure constants $\alpha_i^{-1}$
run very fast being smaller as $\mu$ gets bigger in the interval
$\left[\mu_G,\, \mu_{Pl}\right]$. The old literature \ct{8} (see Section
\ref{sec12})
considers that the running of the fine structure constants between
$\mu_G$ and $\mu_{Pl}$ is minute. It could be taken, for instance,
by the philosophy of Ref.~\ct{8}, as lowest order of the
perturbation theory gives it because the scale ratio logarithm
$\log\left(\mu_G/\mu_{Pl}\right)$ is supposed to be so small that the details
of $\beta$--functions are hardly of any importance. This point of
view is in disagreement with the expectation of quick strong jump
put forward in Section \ref{sec12}. The argument for there having to be
such a jump in $\alpha_i^{-1}$ just before $t=0$ (when
$\mu=\mu_{Pl}$) is based on {\it the assumption that there is a
phase transition at the Planck scale as a function of $t$}. In
principle, however, the occurrence of the jump depends on hard
computations and on (precisely) {\it how big the running
$\alpha_i^{-1}(t)$ are when approaching the Planck scale}. Also
that depends on what (matter) particles of Nature exist and
influence the $\beta$--function (monopoles, extra fermions above
$\mu_G$, etc.).

We would like here to list some options for obtaining
MCP--agreement in one version or the other one combined with
pictures of associated matter:

\begin{enumerate}
\item[1.] The option of Ref.~\ct{8} is to use
\begin{enumerate}
\item[a.] the MCP--I definition;

\item[b.] the approximation of very little running between $\mu_G$ and
$\mu_{Pl}$.
\end{enumerate}

In this interpretation MCP--I, we do not really think of phases as
a function of the scale, but only as a function of the bare
parameters, e.g. the bare fine structure constants. Rather we have
to estimate what are the bare (or the Planck scale) couplings at
the phase transition point conceived of as a transition point in
the coupling constant space, but not as a function of scale. Here
we may simply claim that the Parisi improvement approximation
(\ref{49}) is close to the calculation of the ``bare" coupling. The
Parisi improvement namely calculates an effective coupling as it
would be measured by making small tests of the effective action of
the theory on a very local basis just around one plaquette.
Indeed, we expect that if in the lattice model one seeks to
measure the running coupling at a scale only tinily under the
lattice scale, one should really get the result to be the Parisi
improved value corrected by a tiny running only.

Since even in this option 1), corresponding to Ref.~\ct{8}, we want
a phase transition which we here like to interpret as due to
monopole condensation --- in an other phase though --- we need to
have monopoles. Thus, we get the problem of avoiding these
monopoles in the $\beta$--function except for an extremely small
amount of scales. In the phase in which we live it is suggested
that the monopoles are made unimportant in the $\beta$--function
below $\mu_G$ by being {\it confined} by the Higgs field VEVs
which break the FRGG down to the SMG (possibly extended with
$U(1)_{(B-L)}$). Since the monopoles, we need, are monopoles for
the separate family-gauge-groups and since most of the latter are
Higgsed at the $\mu_G$--scale, we expect the monopoles to be
confined into hadron--like combinations/bound states by the Higgs
fields at the scale $\mu_G$. Thus if $\mu_G$ is close to
$\mu_{Pl}$ --- in logarithm --- there will be very little
contribution of the $\beta$--function running due to monopoles.

In such a picture even extra fermions, as suggested in Subsections
\ref{subsec13.1} and \ref{subsec13.2}, and the SM particles will not
give much running between $\mu_G$ and $\mu_{Pl}$.

Now there seems to be a discrepancy with calculations in the
MCP-II approach (Refs.~\ct{21,21aa,21a,21b,21c,21cc,21d}
and \ct{10s}) which gave
the phase transition point what in (\ref{50a}) is called
$\alpha_{crit.\, lat}^{-1}\approx 5$. But now we must remember that
this value was calculated as a long distance value, i.e. not a
``bare" value of the fine structure constant in the lattice
calculations of Ref.~\ct{10s}.  Also our Coleman--Weinberg type
calculation of it \ct{21,21aa,21a,21b,21c,21cc,21d}
was rather a calculation of the
renormalised (or dressed) coupling than of a bare coupling.

The disagreement between $\alpha_{crit.\, lat}^{-1}\approx 5$ (= the
critical coupling) and the bare coupling in the picture 1 sketched
here is suggested to be due to the renormgroup running in other
phases caused by monopoles there.

In fact, we have in this picture 1 other phases --- existing
somewhere else or at some other time --- in which there is no
breaking down to the diagonal subgroup, as in our phase. In such
phases the monopoles for the family groups can be active in the
renormalization group over a longer range of scales provided they
are sufficiently light. Assuming that the Schwinger's
renormalization scheme is wrong and that the running due to
monopoles make the coupling weaker at the higher scales than at
the lower energies, it could be {\it in the other phase} a value
corresponding to $\alpha_{crit.\, lat}^{-1}\approx 5$ in
Eq.~(\ref{50b}) for some lower $\mu$, say at the mass of monopoles,
while it is still the Parisi improved value at the bare or
fundamental (Planck) scale.

Preliminary calculations indicate that requiring a large positive
value of the running $\lambda(\mu)$ at the cut--off scale, i.e. a
large positive bare $\lambda_0$ in a coupling constant combination
at the phase transition makes the value $\alpha_{crit.\, lat}\approx
0.2$ (or $\alpha_{crit}\approx 0.208$ given by
Refs.~\ct{21aa,21a,21b}) run to a bare $\alpha$ rather close to
the Parisi improvement phase transition coupling value (\ref{49}),
giving $\alpha_{crit}^{-1}\approx 8.5.$

In order to explain this picture we need to talk about at least
three phases, namely, two phases without the Higgs fields
performing the breakdown of the group $G$ to the diagonal subgroup
and our own phase. In one of these phases the monopoles condense
and provide the ``electric" confinement, while the other one has
essentially massless gauge particles in the family gauge group
discussed. In reality, we need a lot of phases in addition to our
own because we need the different combinations of the monopoles
being condensed or not for the different family groups.

\item[2.] The second picture uses MCP--II, which interprets the phase
transition required by any MCP--version as being a phase
transition as a function of the scale parameter, $\mu$ or $t$, and
the requirement of MCP--II is that it occurs just at the
fundamental scale, identified with the Planck scale.

Since the gauge couplings, if their running is provided
(perturbatively) only by the SM particles, would need
$\alpha_{crit}^{-1}\approx 9$ rather than
$\alpha_{crit}^{-1}\approx 5$, further $\beta$--function effects
are needed.

Once the couplings get --- as function of $t$ --- sufficiently
strong, then, of course, perturbation theory gets unjustified,
however, and higher orders or non--perturbative effects are
important. Indeed, higher order seems to help strengthening of the
electric couplings approaching the Planck scale. But most
crucially it is argued that the very fact of finding a phase
transition at $\mu_{Pl}$, as postulated by MCP-II in itself (see
Section 13), suggests that there must be a rather quick
running just below $\mu_{Pl}$. This picture 2 has such a property
because of the extra fermions --- or whatever --- which bring the
strength of the fine structure constants up for $\mu > \mu_G$, so
that non=perturbative and higher order effects can take over and
manage to realize MCP-II.
\end{enumerate}

\section{Conclusions}
\label{sec15}

In the present review we have developed an idea of the Multiple
Point Principle (MPP), according to which several vacuum states
with the same energy density exist in Nature. Here the MPP is
implemented to the Standard Model (SM), Family replicated gauge
group model (FRGGM) and phase transitions in gauge theories (with
and without monopoles).

We have shown that the existence of monopoles in Nature leads to
the consideration of the FRGGM as an extension of the SM, in the
sense that the use of monopoles corresponding to the family
replicated gauge fields can bring the monopole charge down from
the unbelievably large value which it gets in the simple SM.

In this review:
\begin{enumerate}
\item[1.] The MPP was put forward as a fine-tuning mechanism predicting
the ratio between the fundamental and electroweak (EW) scales in the
SM. It was shown that this ratio is exponentially huge:
$$\frac{\mu_{fund}}{\mu_{EW}}\sim e^{40}.$$

\item[2.] Using renormalization group equations (RGEs) for the SM, we
obtained
the effective potential in the two--loop approximation and
investigated the existence of its postulated second minimum at the
fundamental scale.

\item[3.] Lattice gauge theories and phase transitions on the lattice
are reviewed.

\item[4.] The Dual Abelian Higgs Model of scalar
monopoles (Higgs Monopole Model --- HMM) is considered as a simplest
effective dynamics reproducing a confinement mechanism in the pure gauge
lattice theories. It is developed a theory where lattice artifact monopoles
are approximated as fundamental point-like particles and described by the
Higgs scalar fields.

\item[5.] Following the Coleman--Weinberg idea, the RG improvement
of the effective potential is used  in the HMM with $\beta$--functions
calculated in the two--loop approximation.

\item[6.] The phase transition between the Coulomb--like and confinement
phases has been investigated in the $U(1)$ gauge theory. Critical coupling
constants were calculated: it was shown that $\alpha_{crit}\approx 0.17$
and ${\tilde \alpha}_{crit}\approx 1.48$ --- in the one--loop
approximation for $\beta$--functions, and $\alpha_{crit}\approx 0.208$
and ${\tilde \alpha}_{crit}\approx 1.20$ --- in the two--loop
approximation, in agreement with the lattice result:
$$
 \alpha_{crit}^{lat} =
0.20\pm 0.015 \quad {\mbox{and}}  \quad {\tilde \alpha}_{crit}^{lat}
= 1.25\pm 0.10  \quad{\mbox{at}} \quad
\beta_T\equiv\beta_{crit}\approx{1.011}.
$$

\item[7.]  The most significant conclusion for the Multiple Point Model
(MPM) is possibly the validity of an approximate universality of the
critical couplings. It was shown that one can crudely calculate the phase
transition couplings without using any specific lattice. The
details of the lattice, also the details of the regularization,
do not matter for values of the phase transition couplings so much.
Critical couplings depend only on groups with any regularization.
Such an approximate universality
is absolutely needed if we want to compare the
lattice phase transition couplings with the experimental couplings
observed in Nature.

\item[8.] The 't Hooft idea about the Abelian dominance in
the monopole vacuum of non--Abelian theories is discussed:
monopoles of the Yang--Mills theories are the solutions of the
$U(1)$--subgroups, arbitrary embedded into the $SU(N)$ group, and
belong to the Cartain algebra: $U(1)^{N-1}\in SU(N)$.

\item[9.] Choosing the Abelian gauge and taking into account that the
direction in the Lie algebra of monopole fields are gauge
dependent, it is found an average over these directions and
obtained \un{the group dependence relation} between the phase
transition fine structure constants for the groups $U(1)$ and
$SU(N)/Z_N$:
$$
\alpha_{N,\, crit}^{-1} =
           \frac{N}{2}\sqrt{\frac{N+1}{N-1}} \alpha_{U(1),\, crit}^{-1}.
$$

\item[10.]
The Family replicated gauge group model (FRGGM) is reviewed. It is
shown that monopoles have $N_{fam}$, or $N^*$, times smaller
magnetic charge for the gauge group $SU(N)$, or $U(1)$, in the FRGGM than in the
SM. Here $N_{fam}$ is a number of families and $N^*=\frac 12
N_{fam}\left(N_{fam} + 1\right)$.

\item[11.]
Investigating the phase transition in the dual Higgs monopole
model, we have pursued two objects: the first aim was to explain
the lattice results, but the second one was to confirm the MPM
prediction, according to which at the Planck scale there exists a
Multiple Critical Point (MCP).

\item[12.]
The Anti--grand unification theory (AGUT) is reviewed.
It is considered that the breakdown of the FRGG at
$\mu_G\sim 10^{18}$ GeV leads to the Anti--GUT
with the absence of any unification up to the scale $\mu \sim
10^{18}$ GeV.

\item[13.]
Using the group dependence of critical couplings,
we have obtained the following relations:
$$
    \alpha_{Y,\, crit}^{-1} : \alpha_{2,\, crit}^{-1} : \alpha_{3,\, crit}^{-1}
           = 1 : \sqrt{3} : \frac{3}{\sqrt{2}} = 1 : 1.73 : 2.12.
$$
For $\alpha_{Y,\, crit}^{-1}\approx 9.2$ the last
equations give the following result:
$$
 \alpha_{Y,\, crit}^{-1} : \alpha_{2,\, crit}^{-1} : \alpha_{3,\, crit}^{-1}
    = 9.2 : 15.9 : 19.5,
$$
what confirms the Bennett--Froggatt--Nielsen AGUT--MPM
prediction for the SM fine structure constants at the Planck scale:
$$
  \alpha_Y^{-1}\left(\mu_{Pl}\right)\approx 55\pm 6;  \quad
   \alpha_2^{-1}\left(\mu_{Pl}\right)\approx 49.5\pm 3;  \quad
   \alpha_3^{-1}\left(\mu_{Pl}\right)\approx 57.0\pm 3.
$$

\item[14.] We have considered the gravitational interaction between two
particles of equal masses M, given by the Newtonian potential, and
presented the evolution of the quantity:
$$
     \alpha_g = \left(\frac{\mu}{\mu_{Pl}}\right)^2
$$
as a ``gravitational fine structure constant".

\item[15.] We have shown that the intersection of $\alpha_g^{-1}(\mu)$
with $\alpha_1^{-1}(\mu)$ occurs
at the point $\left(x_0, \alpha_0^{-1}\right)$ with the
following values:
$$
       \alpha_0^{-1} \approx 34.4, \quad
                x_0 \approx 18.3,
$$
where $ x = \log_{10}(\mu) $ { GeV}.

\item[16.] It is discussed two scenarios implementing Multiple Critical
Point (MCP) existence.

\item[17.] We have considered the case when our $(3+1)$--dimensional
space--time is discrete and has a lattice--like structure. As a
consequence of such an assumption, we have seen that the lattice
artifact monopoles play an essential role in the FRGGM near the Planck
scale: then these FRGG--monopoles give perturbative contributions
to the $\beta$--functions of RGEs written for both, electric and
magnetic fine structure constants, and change the evolution of
$\alpha_i^{-1}(\mu )$ in the vicinity of the Planck scale.

\item[18.] Finally, we have investigated the case when the breakdown of
FRGG undergoes at $\mu_G\sim 10^{14},\, 10^{15}$ GeV and is accompanied by
a lot of extra fermions in the region $\mu_G < \mu < \mu_{Pl}$.
These extra fermions suppressing the asymptotic freedom of
non--Abelian theories lead, together with monopoles, to the
possible existence of unification of all interactions including
gravity at $\mu_{\rm GUT}=10^{18.4}$ GeV and $\alpha_{\rm GUT}^{-1}=27$.

\item[19.] It is discussed the possibility of the existence of
the family replicated unifications $\left[SU(5)\right]^3$ SUSY
or $\left[SO(10)\right]^3$ SUSY.
\end{enumerate}

In this review we have considered a special case of the new type of
unification suggested in Ref.~\ct{21d}. The realistic family replicated
unification theory needs a serious program of investigations giving the
predictions
for the low--energy physics and cosmology what may be developed
in future.

\vspace{1cm}

{\bf ACKNOWLEDGEMENTS:}\\

We would like to express special thanks to Prof. H.B.~Nielsen for
the fruitful collaboration and interesting discussions.

One of the authors (L.V.L.) thanks the Institute of Mathematical
Sciences of Chennai (India) and personally Prof. N.D.~Hari Dass
for hospitality and financial support. It is a pleasure to thank
all participants of the seminars of this Institute for useful
discussions and interest.

We are deeply thankful to D.L.~Bennett, C.D.~Froggatt, N.D.~Hari
Dass, R.B.~Nevzorov and Y.~Takanishi for help, discussions and
comments.

This work was supported by the Russian Foundation for Basic
Research (RFBR), project $N^o$ 05-02-17642.

\clearpage\newpage
\bfi
\epsfxsize=14cm
{\epsfbox{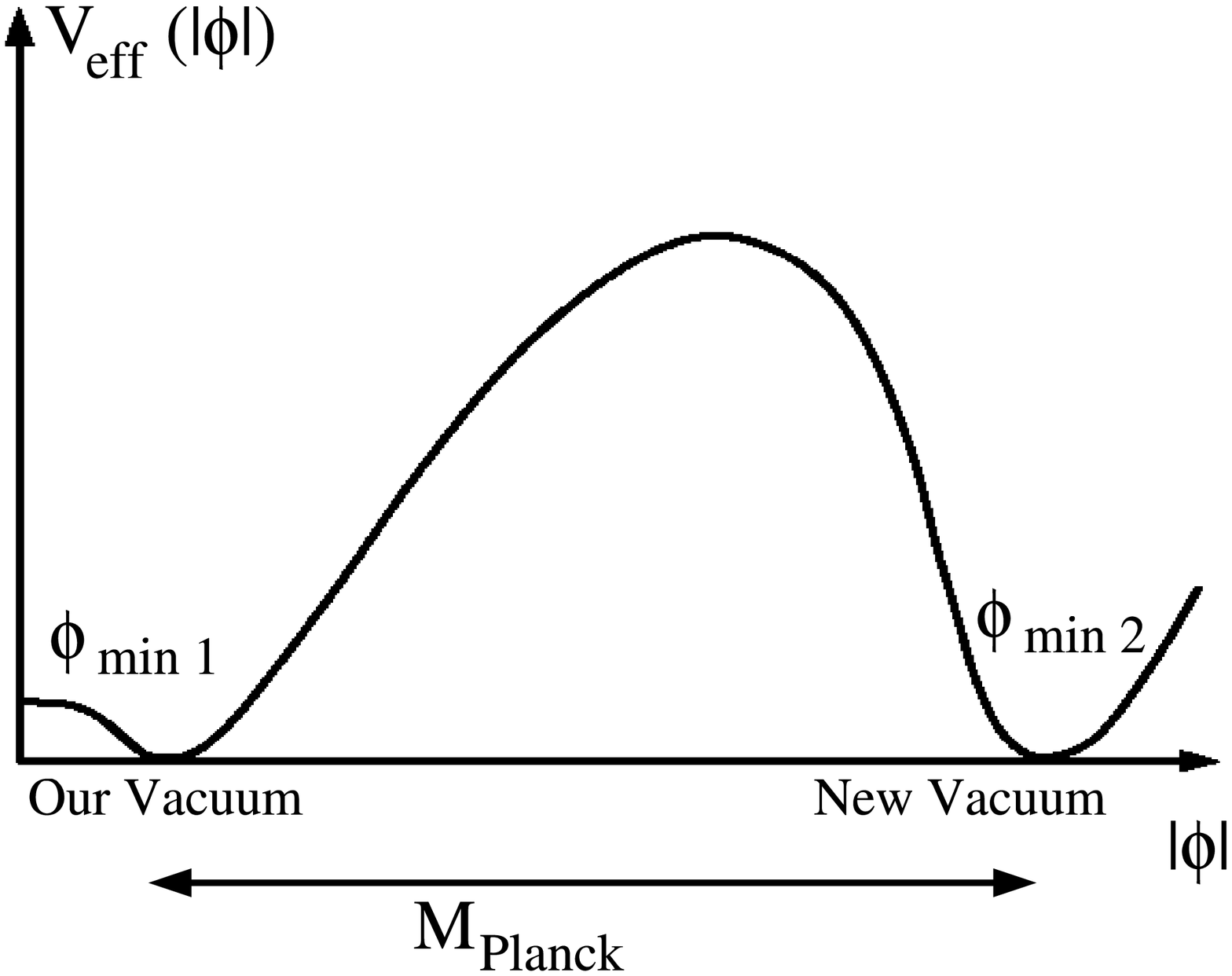}}
\caption{2nd vacuum
degenerate with usual SM vacuum. SM valid up to Planck scale except
$\phi_{min2}\approx M_{\rm Planck}$}
\lb{f1}
\efi

\clearpage\newpage
\bfi
\epsfxsize=14cm
{\epsfbox{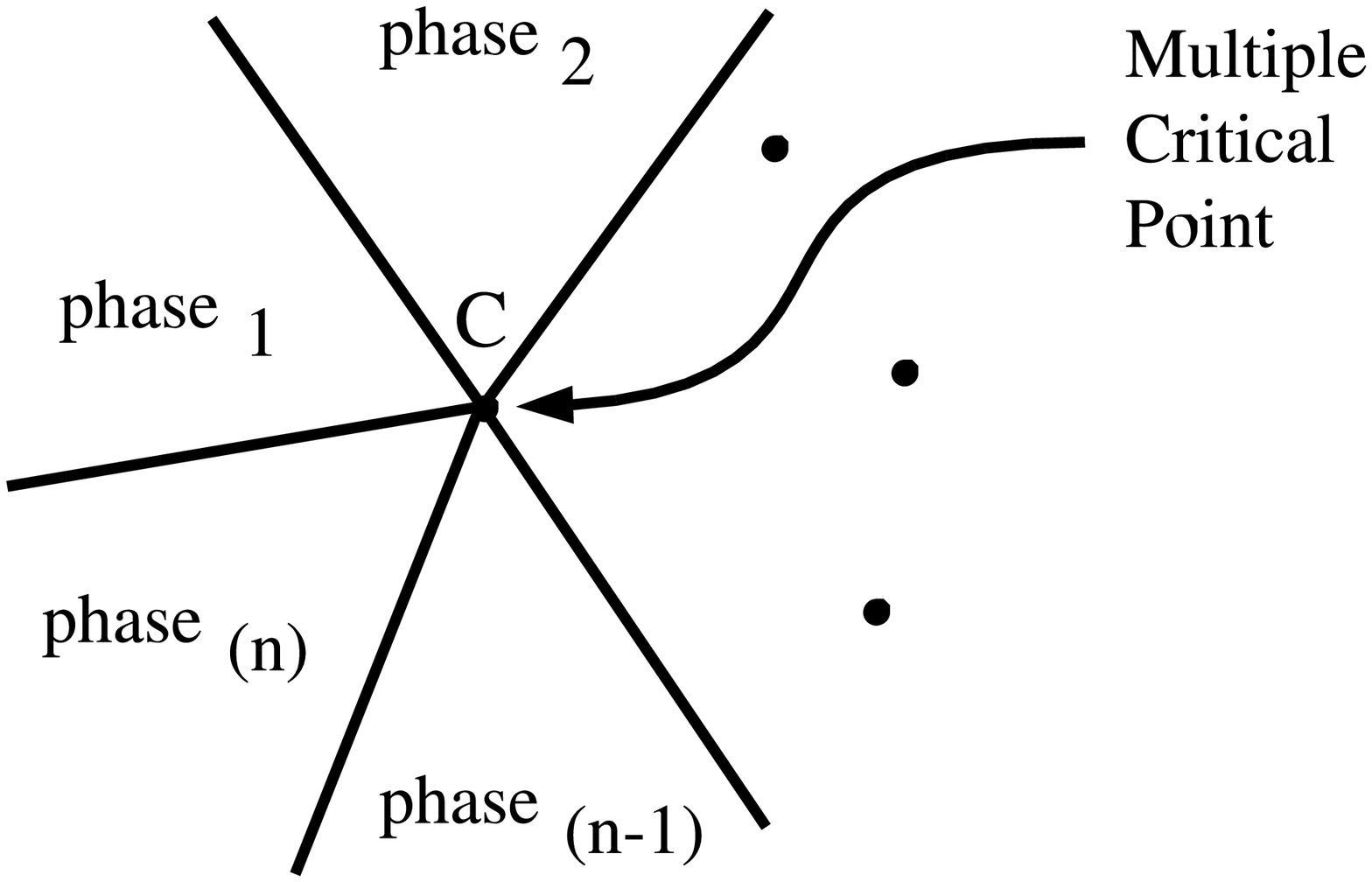}}
\caption{The schematic representation of the phase diagram of a gauge theory
having n phases. The point C is the Multiple Critical Point.}
\lb{f2}
\efi

\clearpage\newpage
\bfi
\epsfxsize=14cm
{\epsfbox{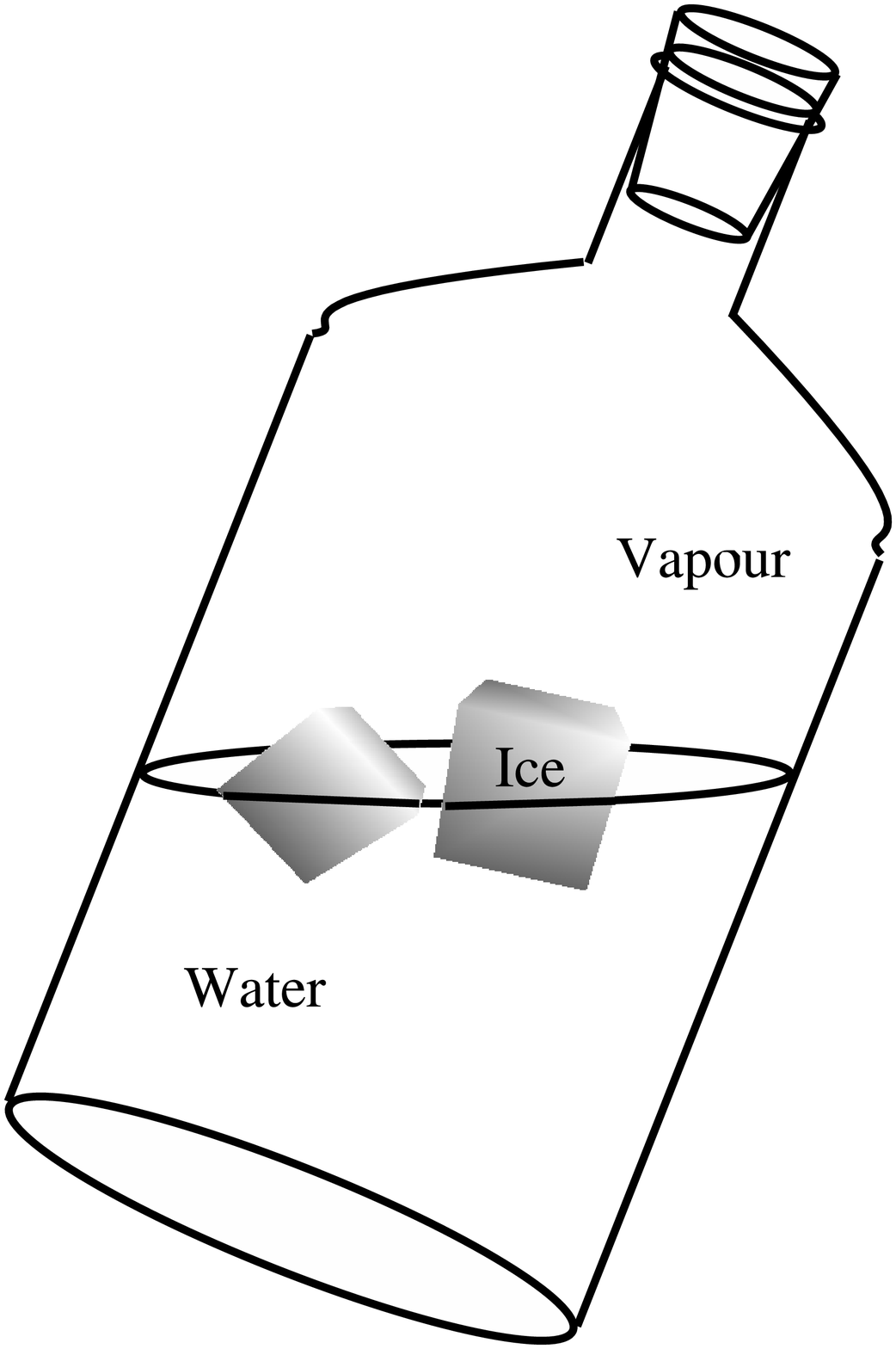}}
\caption{Triple point of water analogy: Ice, Water and Vapour.
             The volume, energy and a number of moles are fixed
             in the system.}
\lb{f3}
\efi

\clearpage\newpage
\bfi
\epsfxsize=14cm
{\epsfbox{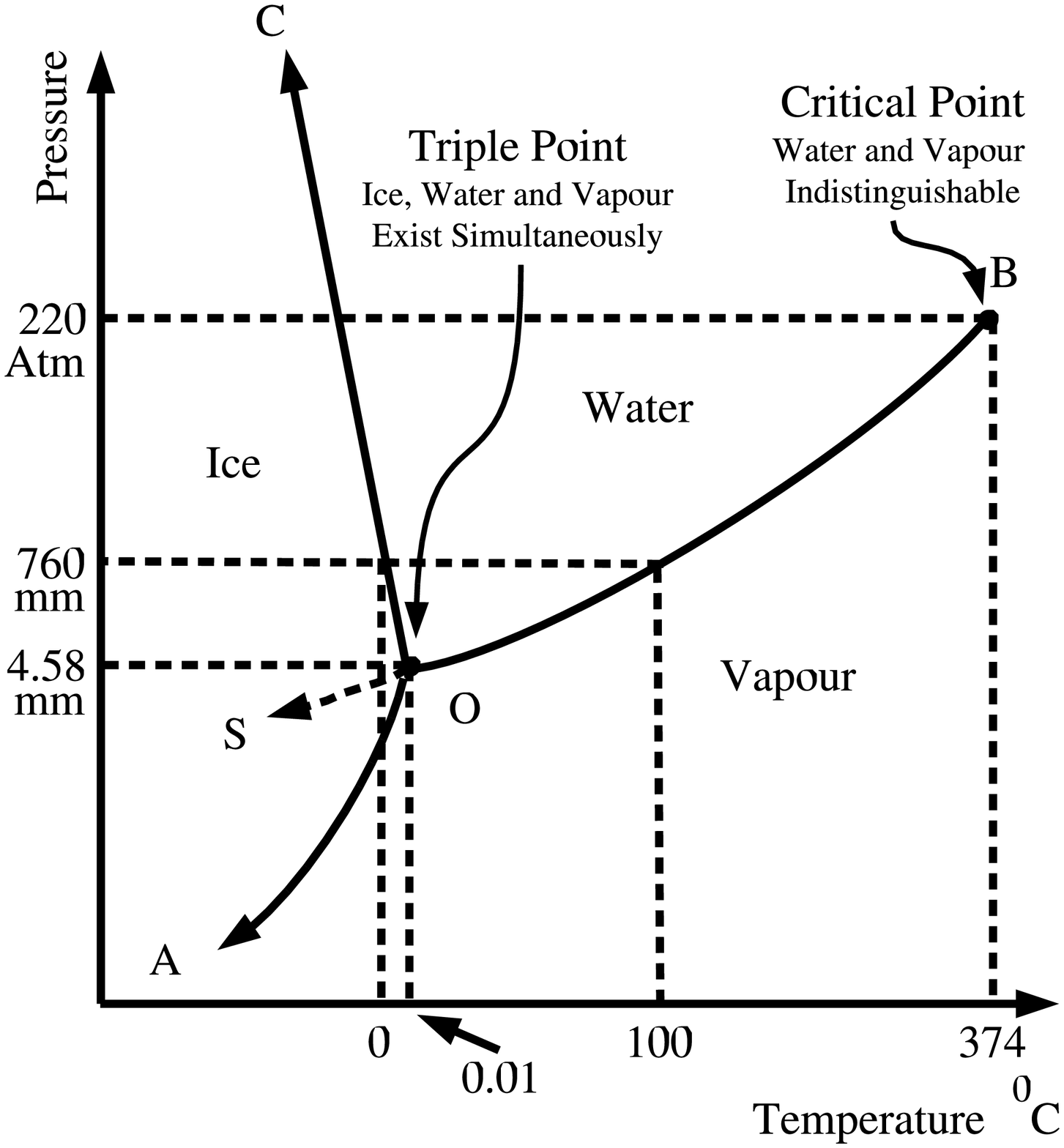}}
\caption{Triple point of water analogy. Fine--tuned intensive
            variables: critical temperature $T_c = 0.01\, ^{\rm o}$C,
            critical pressure $P_c = 4.58$ mm Hg.}
\lb{f4}
\efi

\clearpage\newpage
\bfi
\epsfxsize=14cm
{\epsfbox{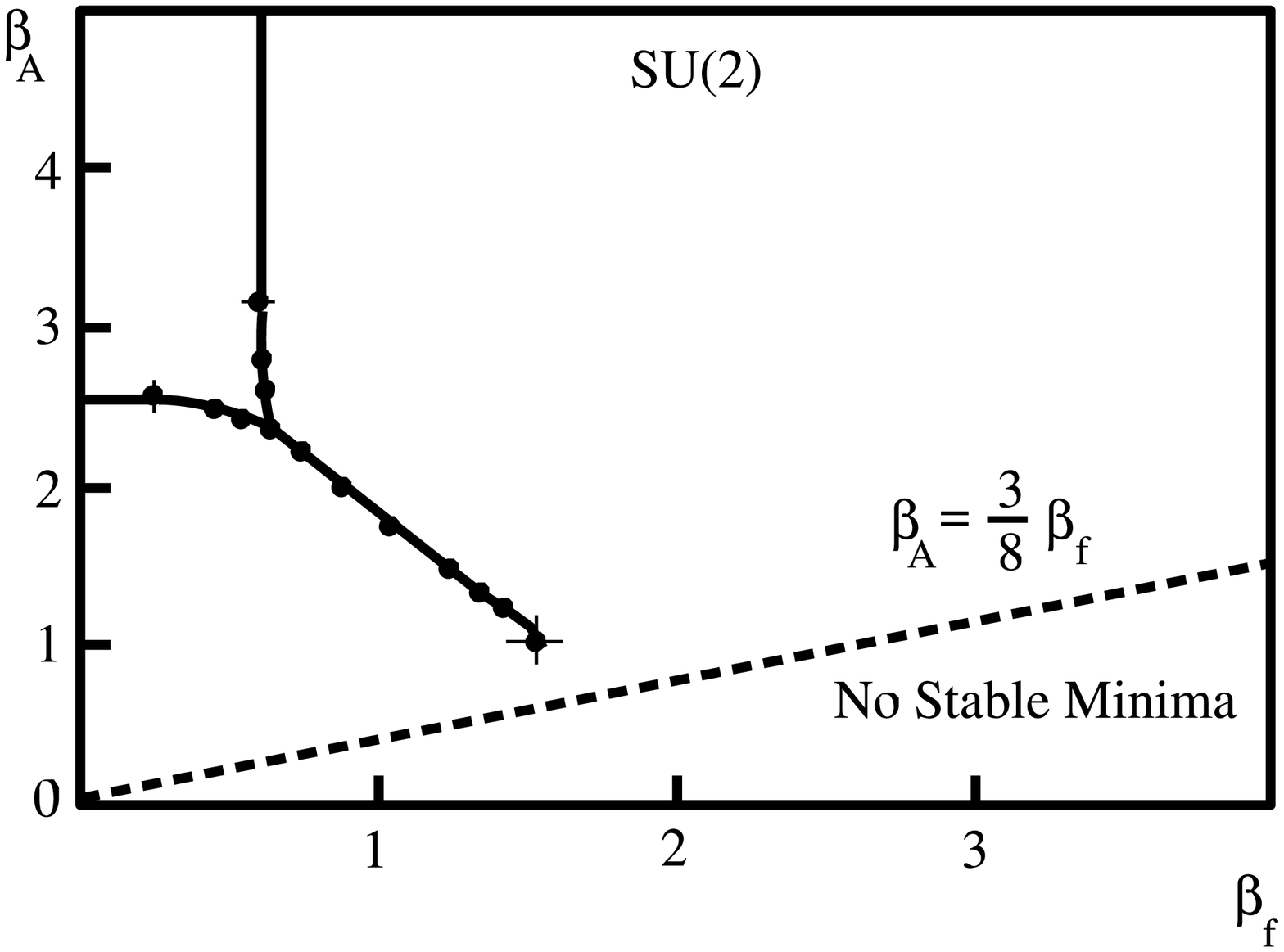}}
\caption{Phase diagram for the $SU(2)$ lattice gauge theory with the
generalized Wilson lattice action. The result of Monte--Carlo simulations.
Here $(\beta_{\rm f};\, \beta_{\rm A})_{crit}=(0.54;\, 2.4)$}
\lb{f5a}
\efi

\clearpage\newpage
\bfi
\epsfxsize=14cm
{\epsfbox{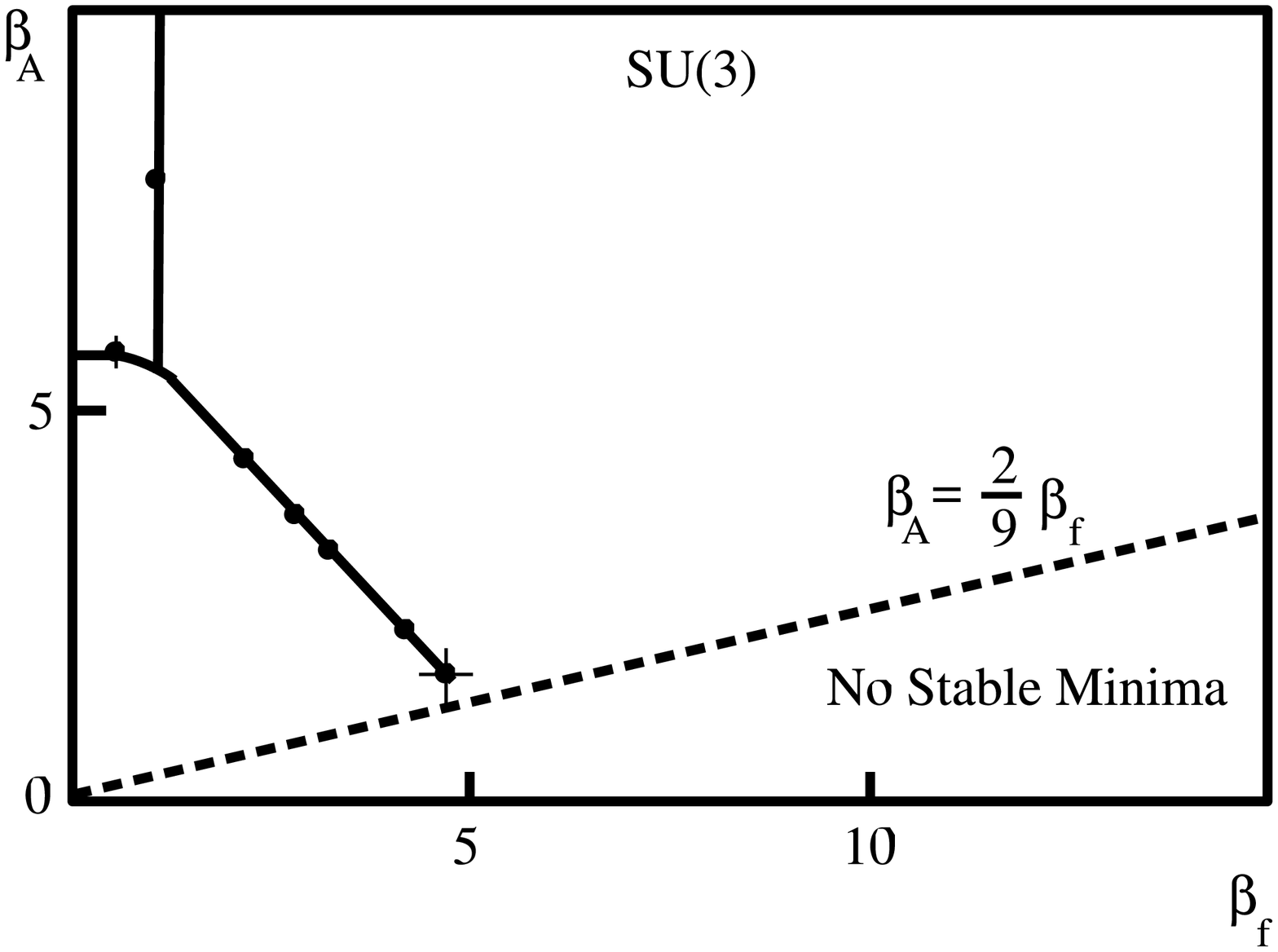}}
\caption{Phase diagram for the $SU(3)$ lattice gauge theory with the
generalized Wilson lattice action. The result of Monte--Carlo simulations.
Here $(\beta_{\rm f};\, \beta_{\rm A})_{crit}=(0.80;\, 5.4)$}
\lb{f5b}
\efi

\clearpage\newpage
\bfi
\epsfxsize=14cm
{\epsfbox{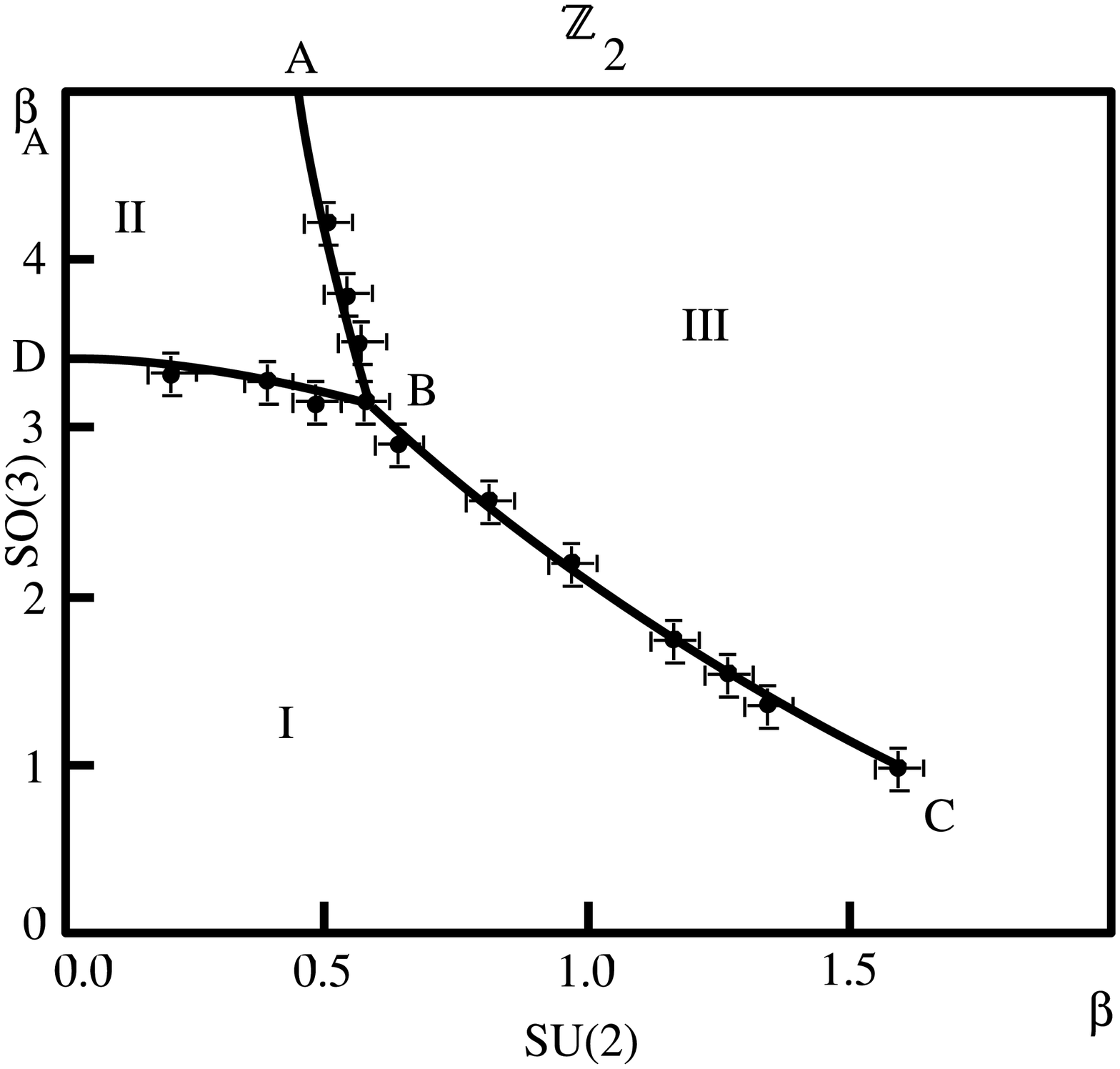}}
\caption{The phase diagram for the lattice $SU(2)$--$SO(3)$ gauge theory.
The range I contains $Z_2$--vortices with density $E$ and
$Z_2$--monopoles with density $M$. Here $E\approx M\approx 0.5$.
The range II corresponds to $E\approx 0.5, M\approx 0,$
and in the range III we have $E\approx M\approx 0$.}
\lb{f6}
\efi

\clearpage\newpage
\bfi
\epsfxsize=14cm
{\epsfbox{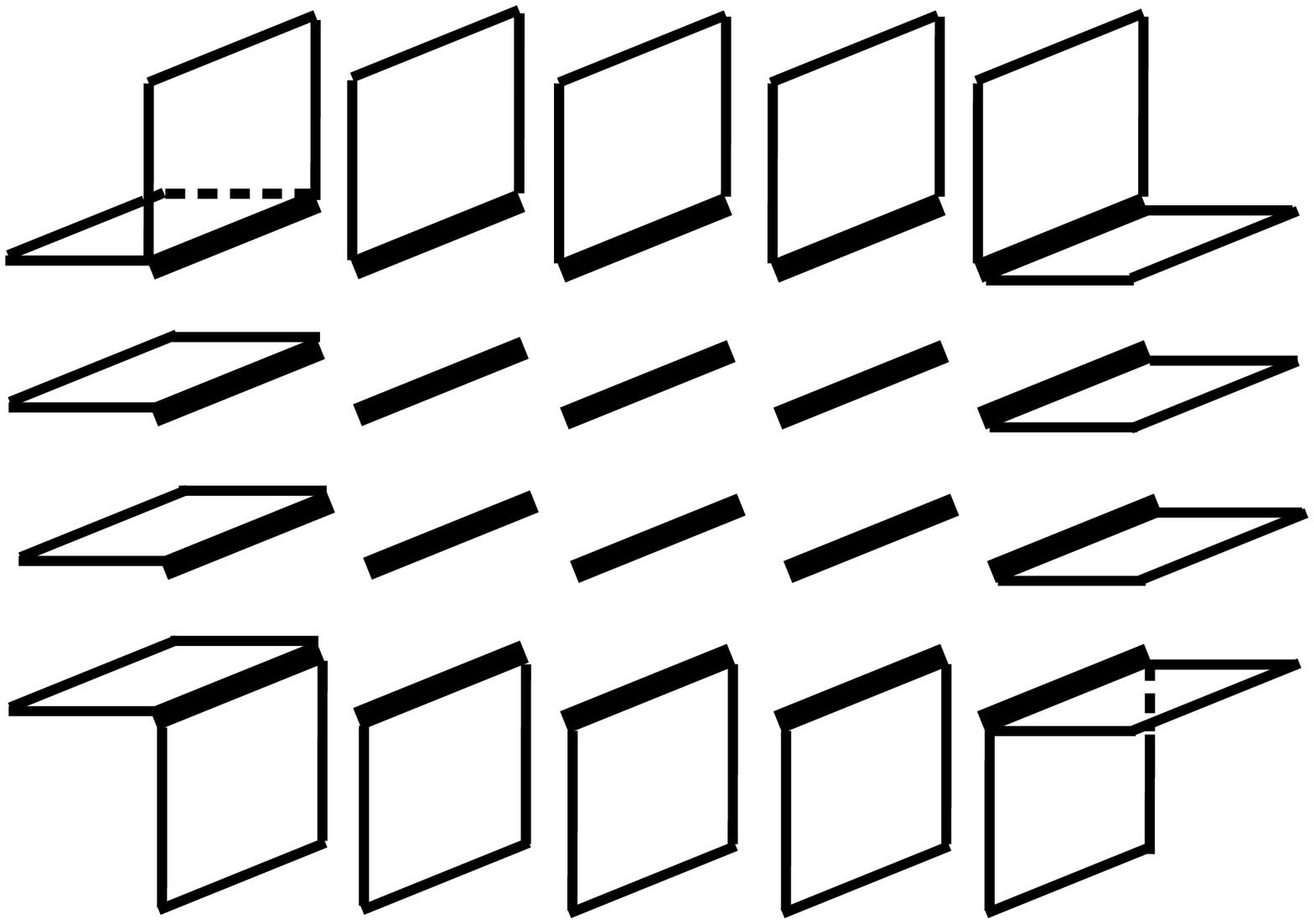}}
\caption{The closed $Z_2$--vortex of the 3--dimensional lattice.}
\lb{f6a}
\efi

\clearpage\newpage
\bfi
\epsfxsize=14cm
{\epsfbox{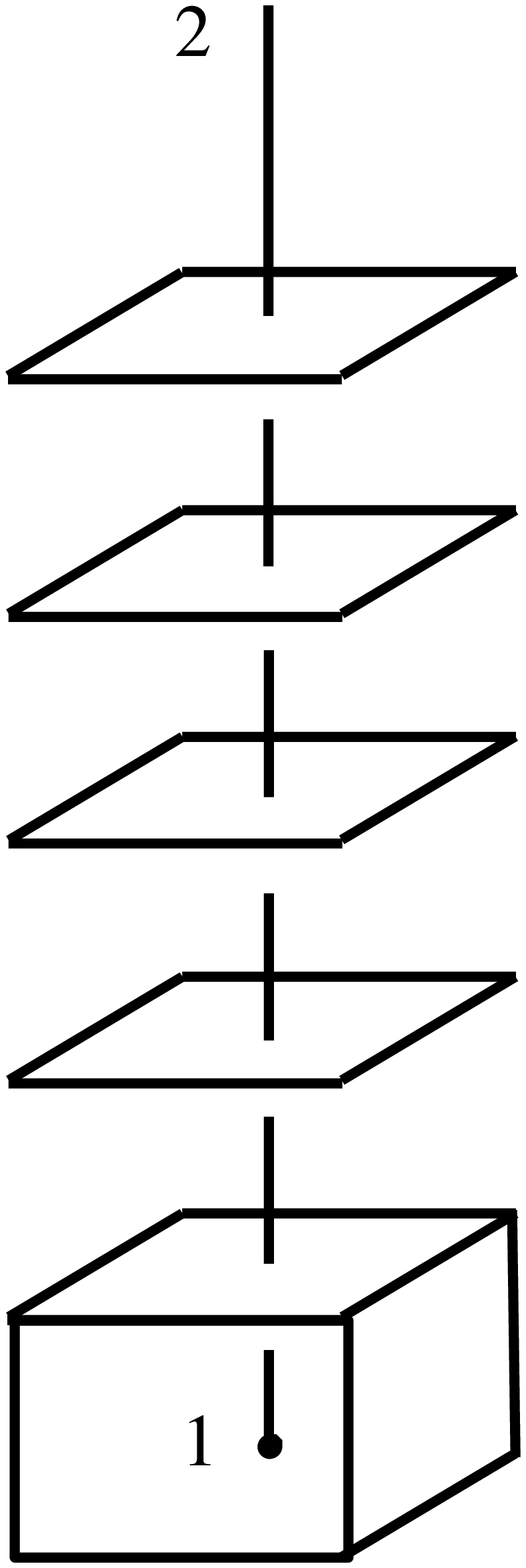}}
\caption{$Z_2$--monopole of the 3--dimensional lattice.}
\lb{f6b}
\efi

\clearpage\newpage
\bfi
\epsfxsize=14cm
{\epsfbox{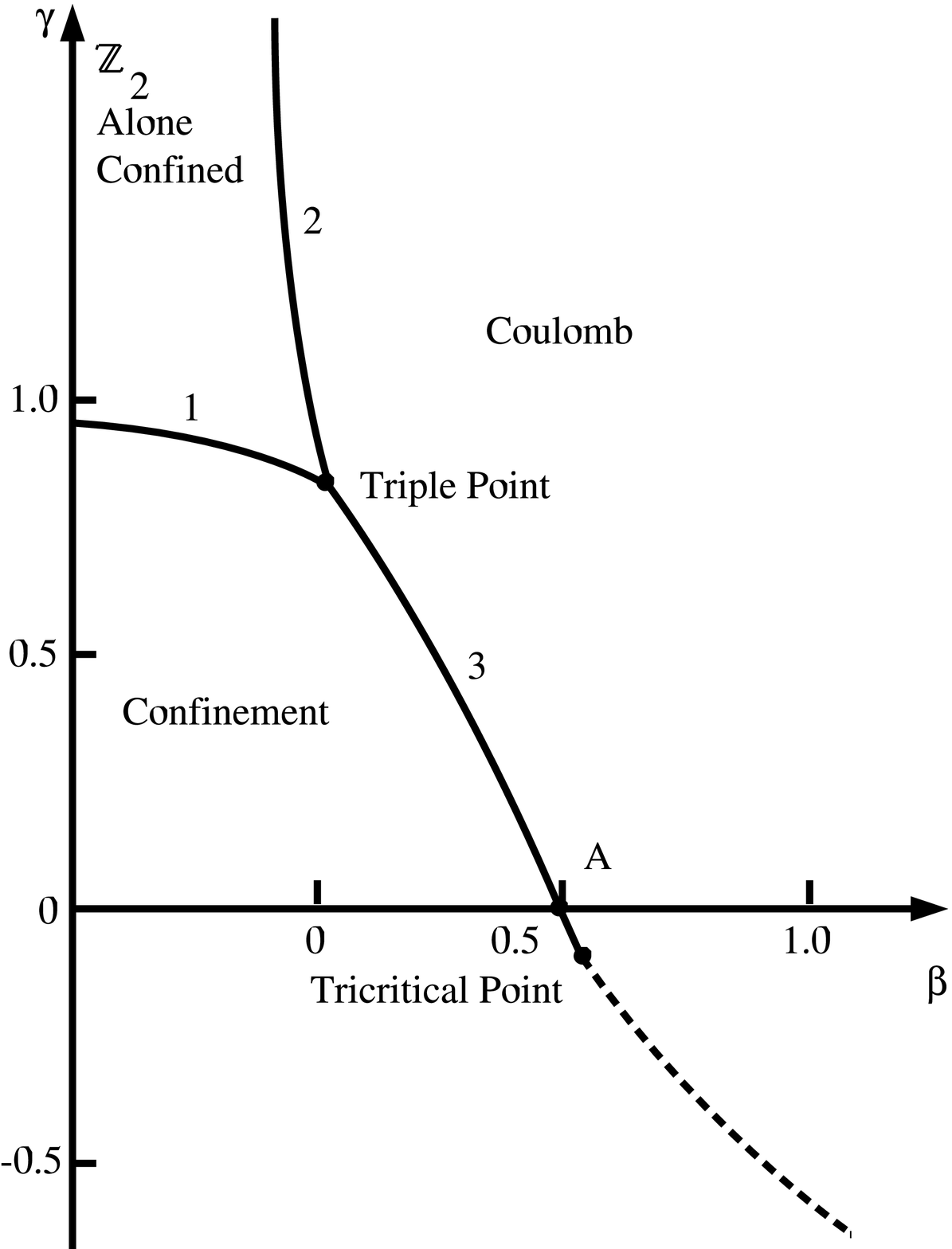}}
\caption{The phase diagram for $U(1)$ when the two--parameter lattice action is
used. This type of action makes it possible to provoke the confinement $Z_2$
or ($Z_3$) alone. The diagram shows the existence of a triple (critical) point.
From this triple point emanate three phase borders: the phase border ``1''
separates the totally confining phase from the phase where only the discrete
subgroup $Z_2$ is confined; the phase border ``2'' separates the latter phase
from the totally Coulomb--like phase; and the phase border ``3'' separates the
totally confining and totally Coulomb--like phases.}
\lb{f7}
\efi

\clearpage\newpage
\bfi
\epsfxsize=14cm
{\epsfbox{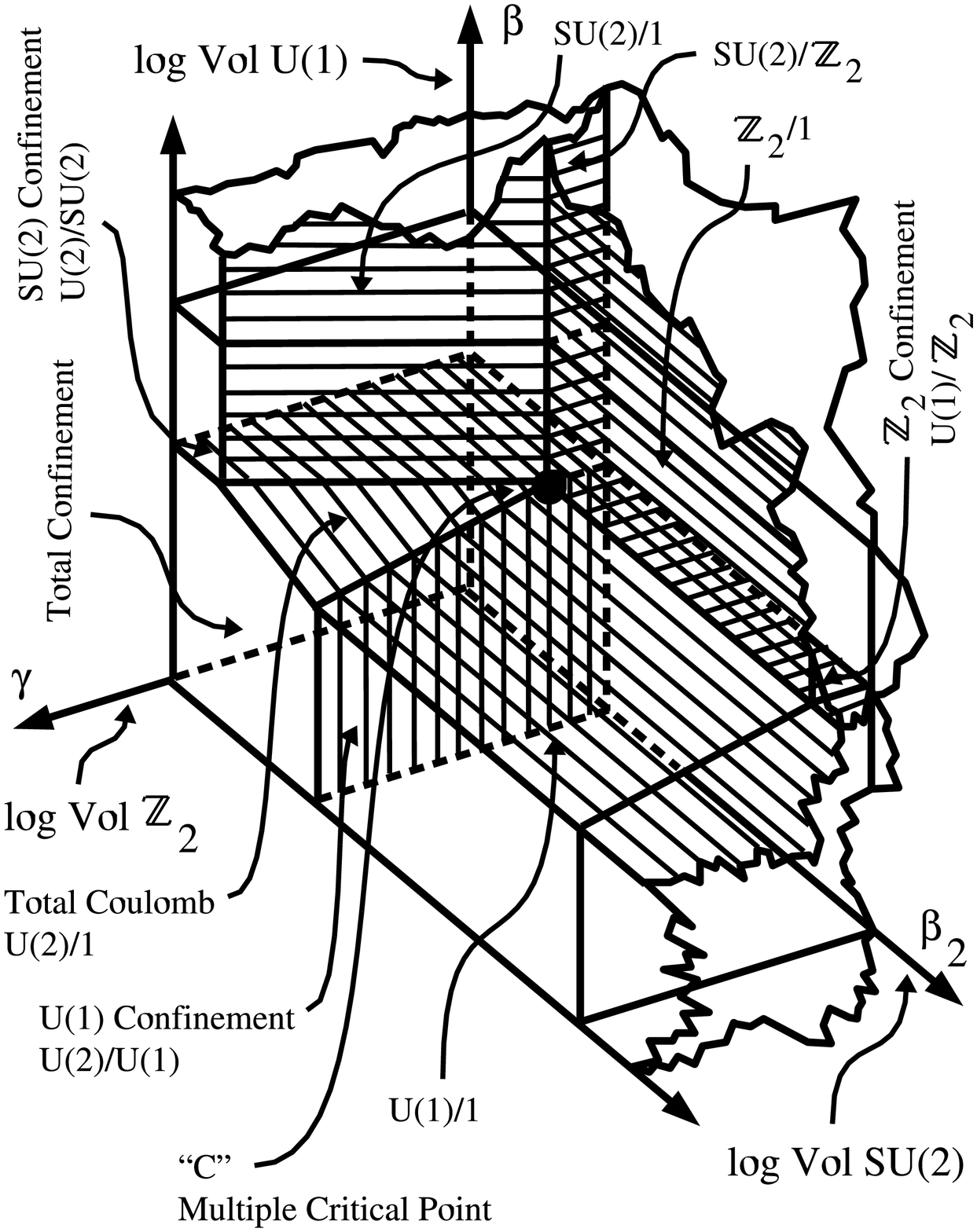}}
\caption{Phase diagram for the lattice $U(1)\times SU(2)$ gauge theory.
Five phases meet at the multiple critical point.}
\lb{f8}
\efi

\clearpage\newpage
\bfi
\epsfxsize=14cm
{\epsfbox{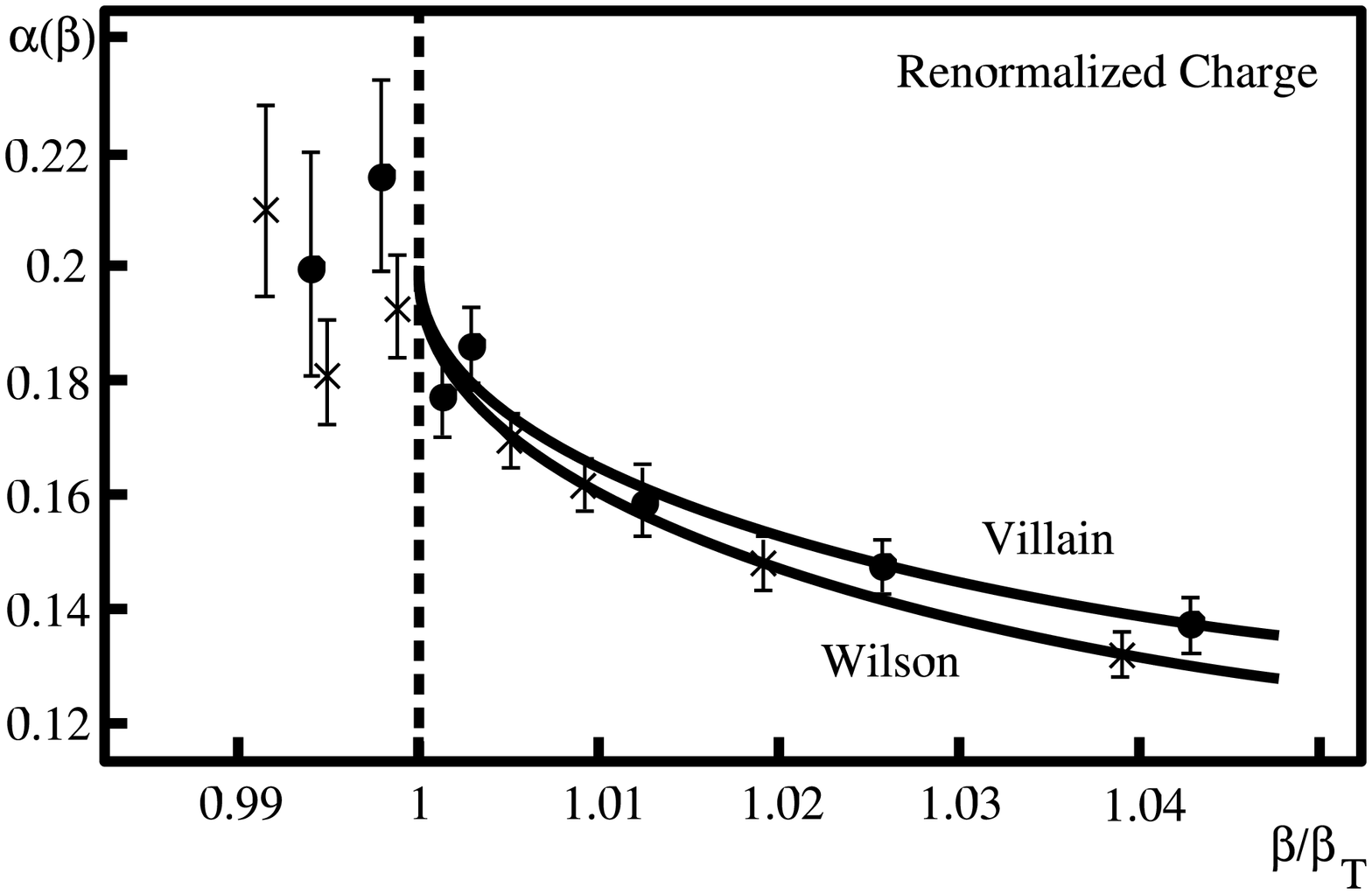}}
\caption{The renormalized electric fine structure constant plotted versus
$\beta/\beta_T$ for the Villain action (full circles) and the Wilson action
(crosses). The points are obtained by the Monte--Carlo simulations method for
the compact QED.}
\lb{f9a}
\efi

\clearpage\newpage
\bfi
\epsfxsize=14cm
{\epsfbox{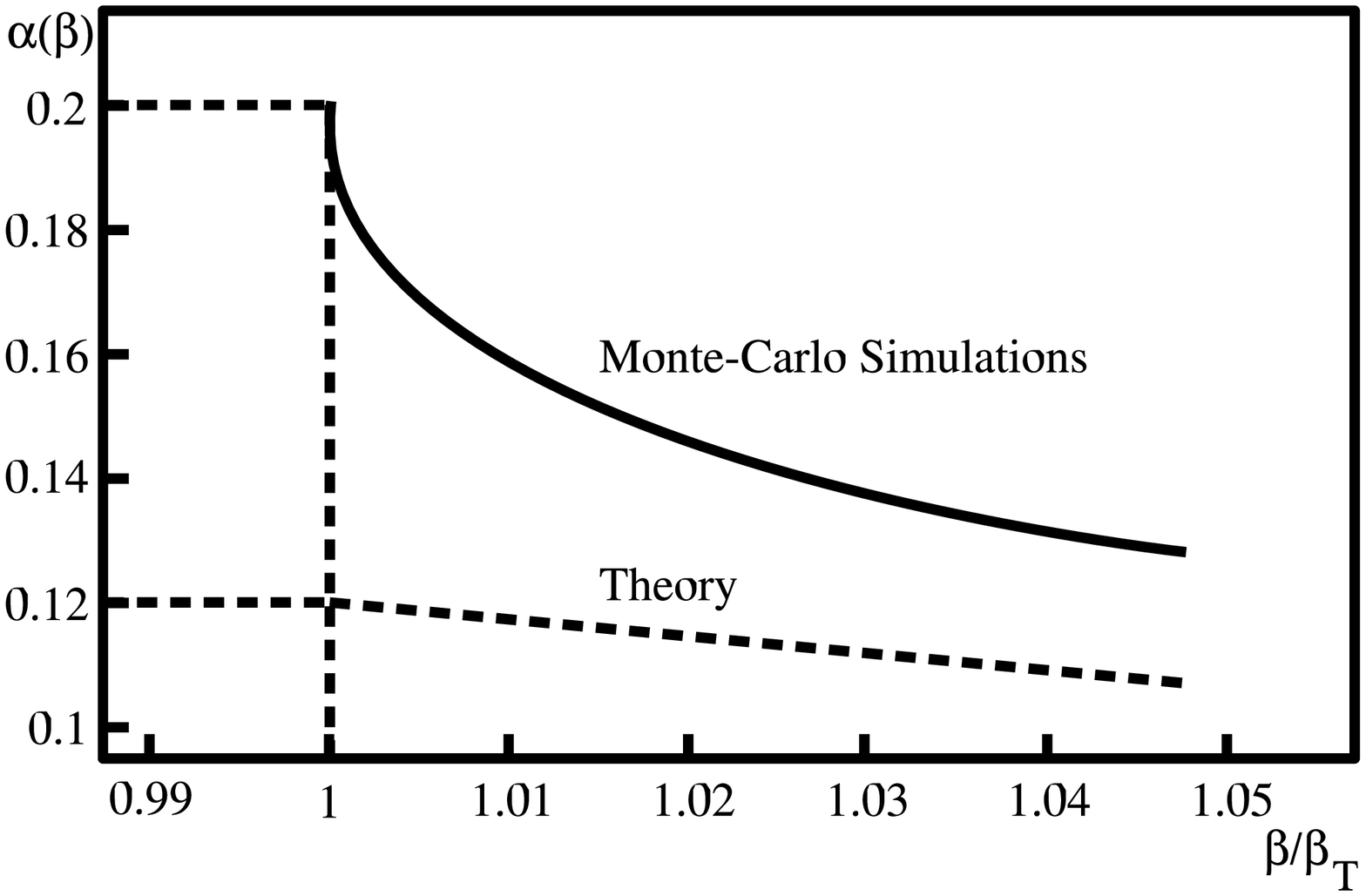}}
\caption{The behavior of the effective electric fine structure constant in the
vicinity of the phase transition point obtained with the lattice Wilson action.
The dashed curve corresponds to the theoretical calculations by the ``Parisi
improvement method''.}
\lb{f9b}
\efi

\clearpage\newpage
\bfi
\epsfxsize=14cm
{\epsfbox{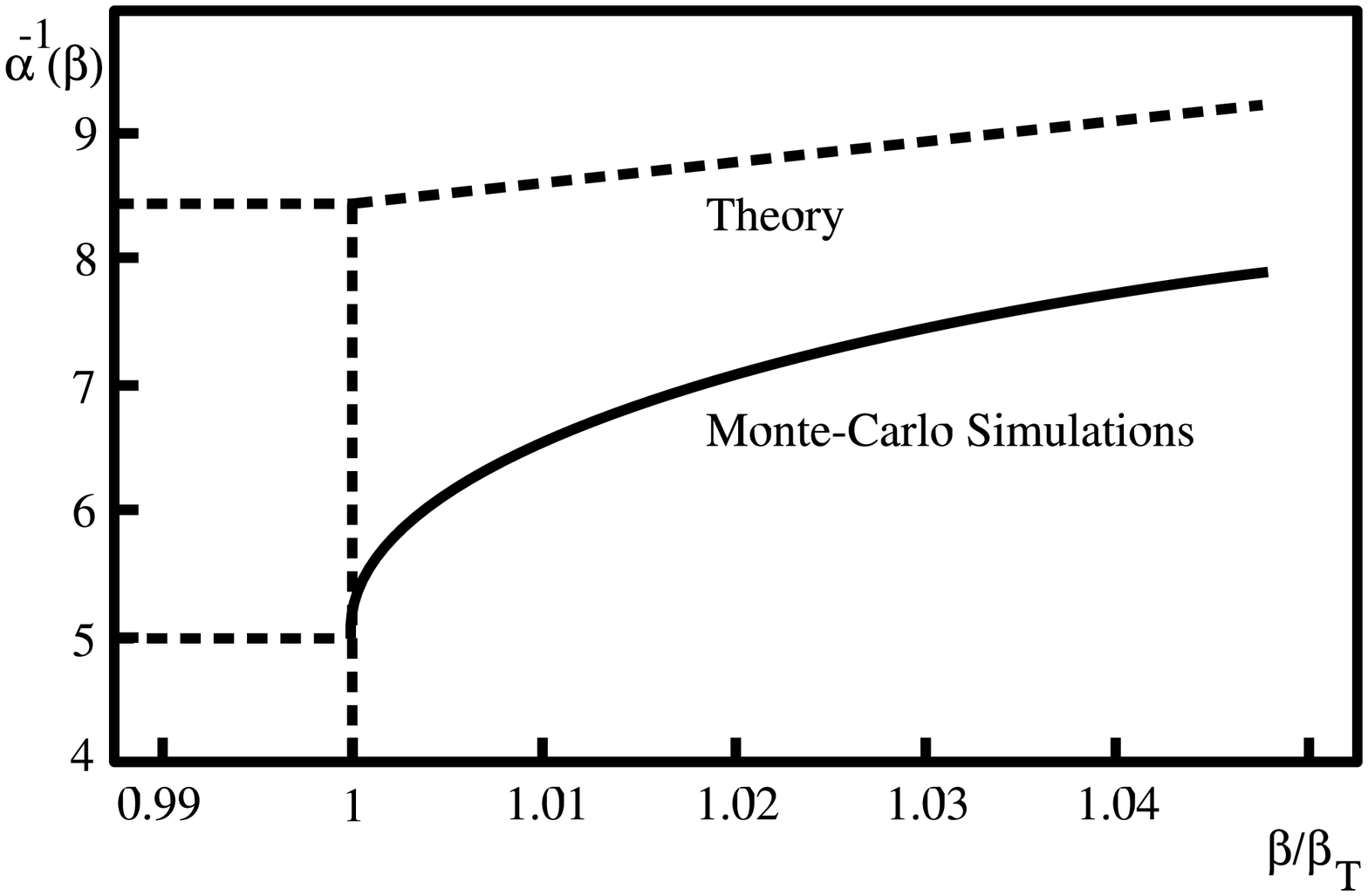}}
\caption{The behavior of the inverse effective electric fine structure
constant in the vicinity of the phase transition point ploted versus
$\beta/\beta_T$ for the simple Wilson lattice action. The dashed curve
corresponds to the theoretical calculations by the ``Parisi improvement
method''.}
\lb{f9c}
\efi

\clearpage\newpage
\bfi
\epsfxsize=14cm
{\epsfbox{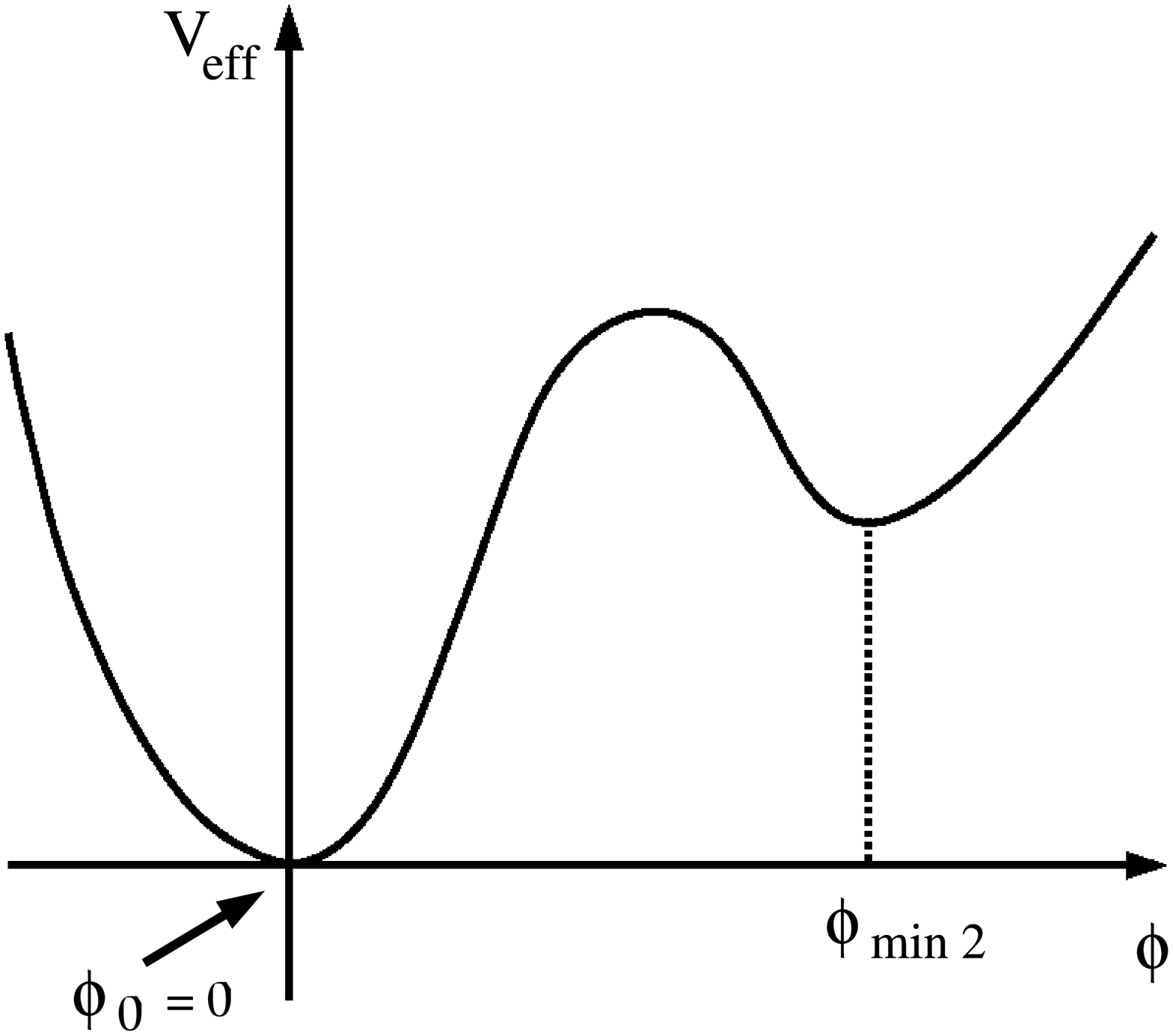}}
\caption{The effective potential $V_{eff}(\phi)$ having the first
local minimum at $\phi_0 = 0$ and $V_{eff}(0) = 0$. If the second
minimum occurs at $V_{eff}\left(\phi_{min2}\right) > 0$, this case
corresponds to the ``symmetric", or Coulomb--like phase.}
\lb{f10}
\efi

\clearpage\newpage
\bfi
\epsfxsize=14cm
{\epsfbox{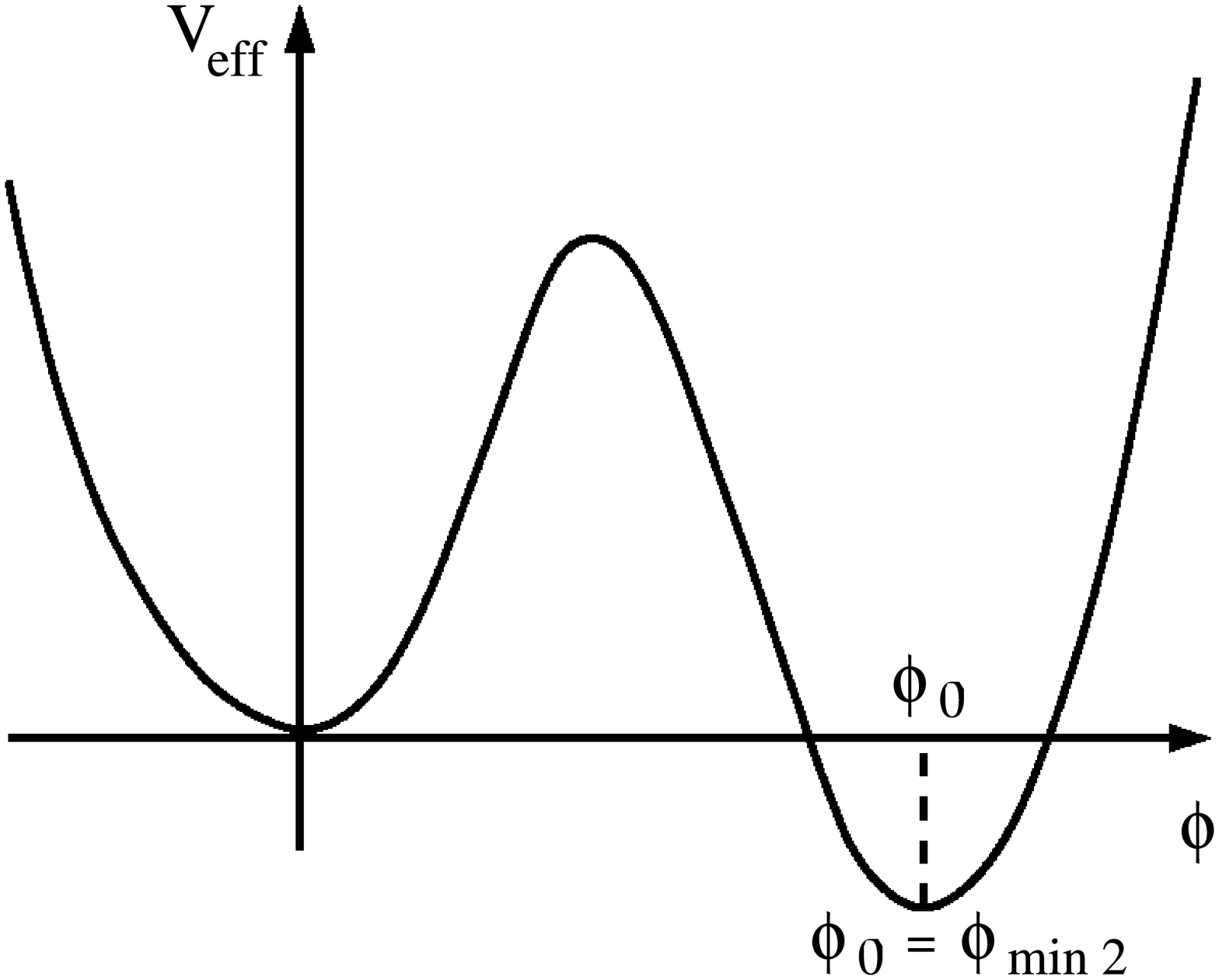}}
\caption{The effective potential $V_{eff}(\phi)$ for the confinement phase.
In this case the second local minimum occurs at $\phi_0 =
\phi_{min2}\neq 0$ and $V_{eff}^{min}\left(\phi_{min2}\right) <
0$.}
\lb{f11}
\efi

\clearpage\newpage
\bfi
\epsfxsize=14cm
{\epsfbox{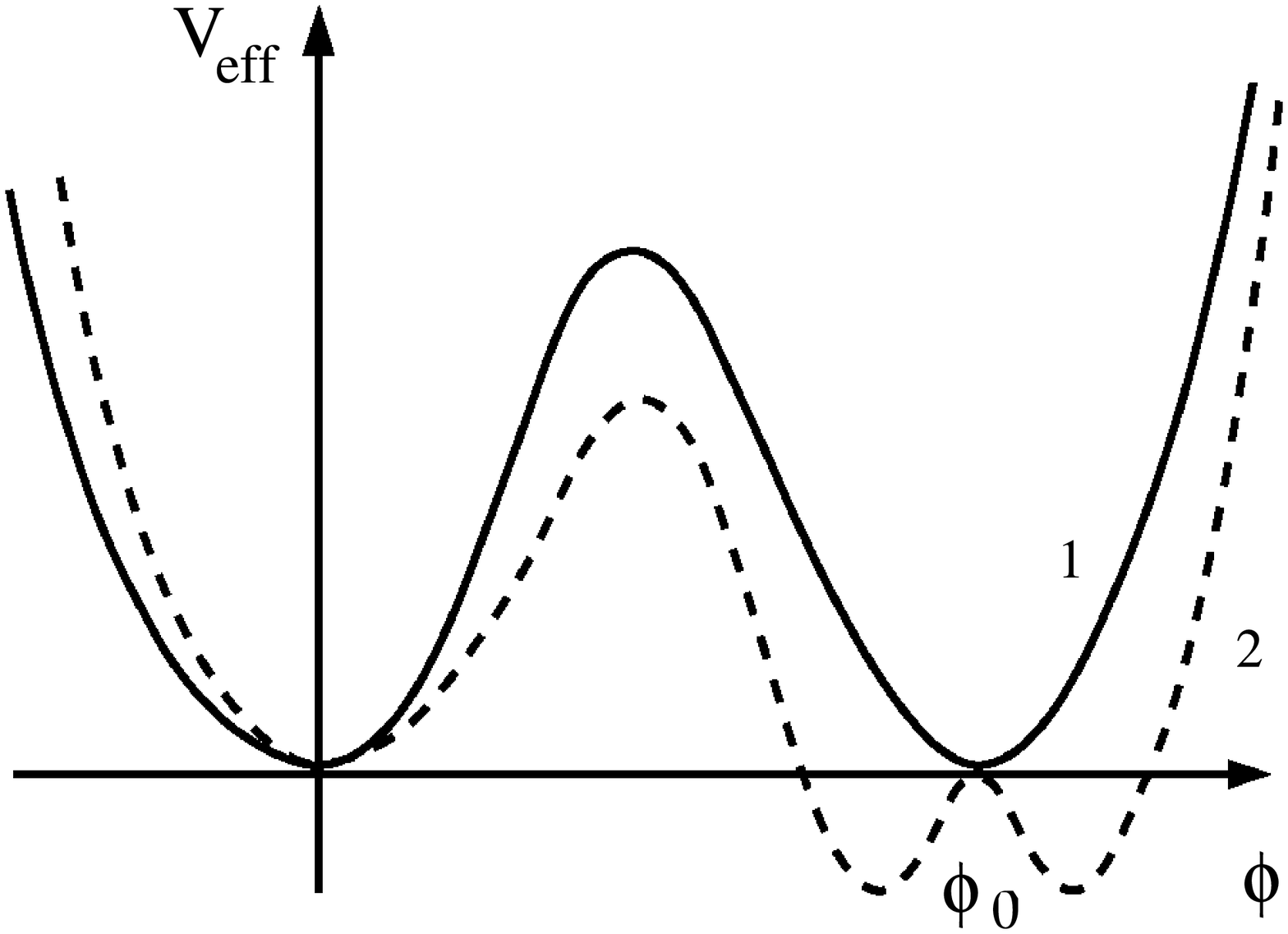}}
\caption{The effective potential $V_{eff}$: the curve ``1''
corresponds to the ``Coulomb--confinement'' phase transition; curve ``2''
describes the existence of two minima corresponding to the confinement phases.}
\lb{f12}
\efi

\clearpage\newpage
\bfi
\epsfxsize=14cm
{\epsfbox{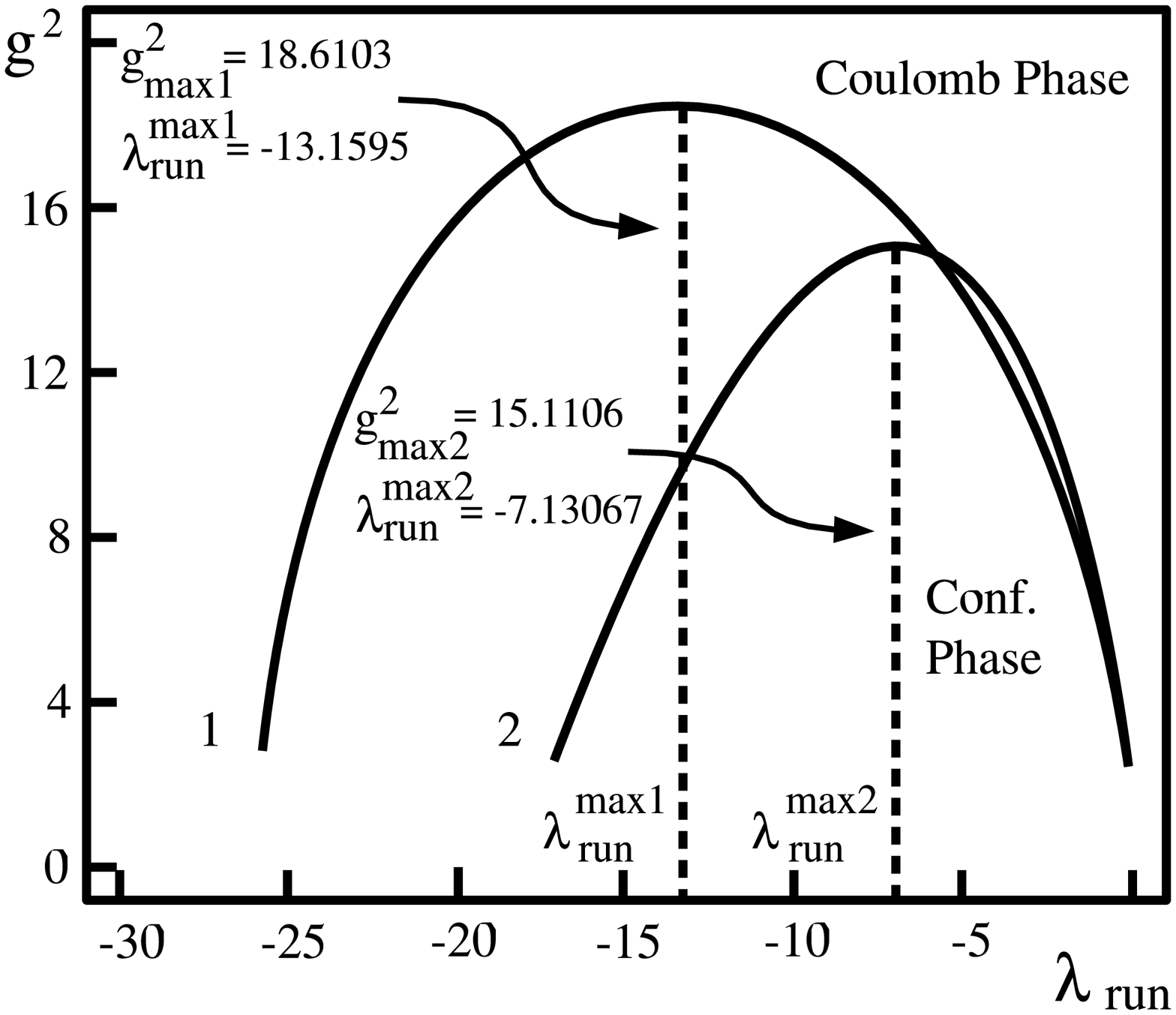}}
\caption{The one--loop (curve ``1'') and two--loop (curve ``2'') approximation
phase diagram in the dual Abelian Higgs model of scalar monopoles.}
\lb{f13}
\efi

\clearpage\newpage
\bfi
\epsfxsize=14cm
{\epsfbox{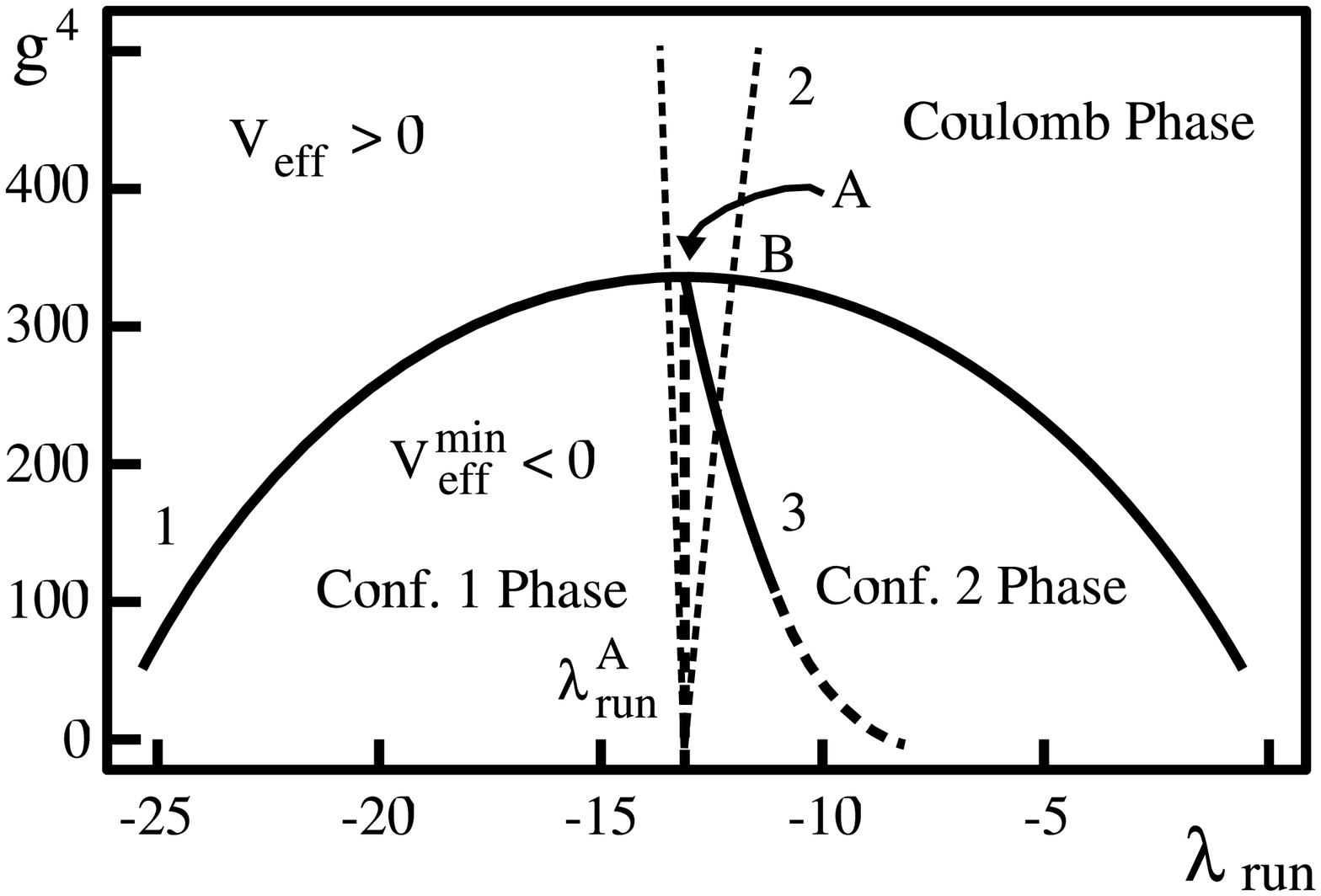}}
\caption{The phase diagram $\left(\lambda_{run};\, g^4\equiv g^4_{run}\right)$,
corresponding to the Higgs monopole model in the one--loop approximation,
shows the existence of a triple point A
$\left(\lambda_{(A)}\approx -13.4;\, g^2_{(A)}\approx 18.6\right)$.
This triple point is a boundary point of three
phase transitions: the ``Coulomb--like'' phase and two confinement phases
(``Conf. 1'' and ``Conf. 2'') meet together at the triple point A. The dashed
curve ``2'' shows the requirement: $V_{eff}(\phi_0^2)=V''_{eff}(\phi_0^2)=0$.
Monopole condensation leads to the
confinement of the electric charges: ANO electric vortices (with electric
charges at their ends, or closed) are created in the confinement phases
``Conf. 1'' and ``Conf. 2''.}
\lb{f14}
\efi

\clearpage\newpage
\bfi
\epsfxsize=14cm
{\epsfbox{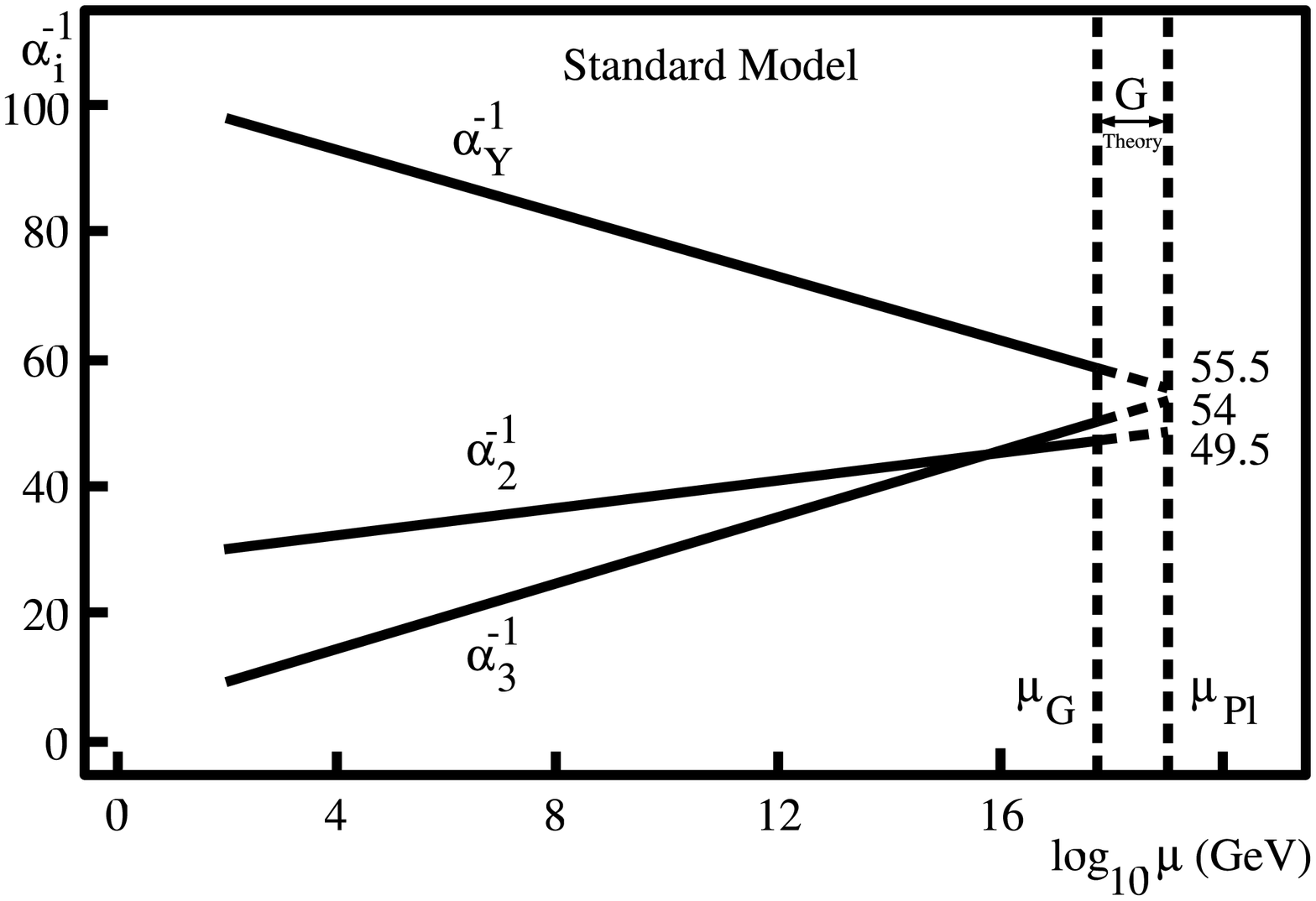}}
\caption{The evolution of three inverse running constants $\alpha_i^{-1}(\mu)$,
where $i=Y,\, 2,\, 3$ correspond to $U(1)_Y$, $SU(2)$ and $SU(3)$
groups of the SM.
The extrapolation of their experimental values from the Electroweak scale to
the Planck scale was obtained by using the renormalization group eqations with
one Higgs doublet under the assumption of a ``desert''. The precision of the
LEP data allows to make this extrapolation with small errors. AGUT works in
the region $\mu_{\rm G}\le\mu\le\mu_{\rm Pl}$.}
\lb{f15}
\efi

\clearpage\newpage
\bfi
\epsfxsize=14cm
{\epsfbox{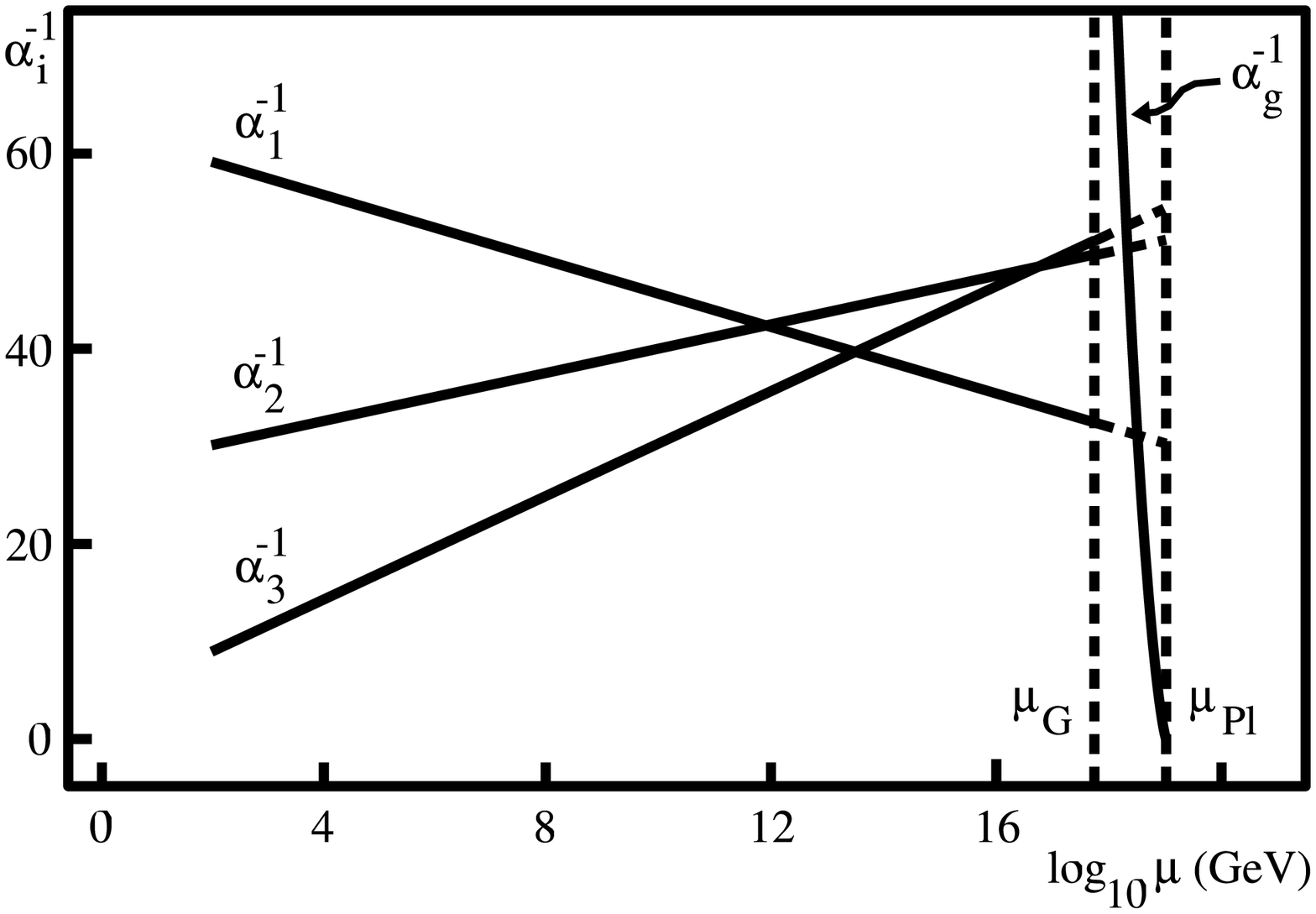}}
\caption{The intersection of the inverse ``gravitational finestructure
constant'' $\alpha_g^{-1}(\mu)$ with $\alpha_1^{-1}(\mu)$ occurs at the
point $\left(x_0,\alpha_0^{-1}\right):\, \alpha_0^{-1}\approx 34.4$ and
$x_0\approx 18.3$, where $x=\log_{10}(\mu)$ (GeV).}
\lb{f16}
\efi

\clearpage\newpage
\bfi
\epsfxsize=14cm
{\epsfbox{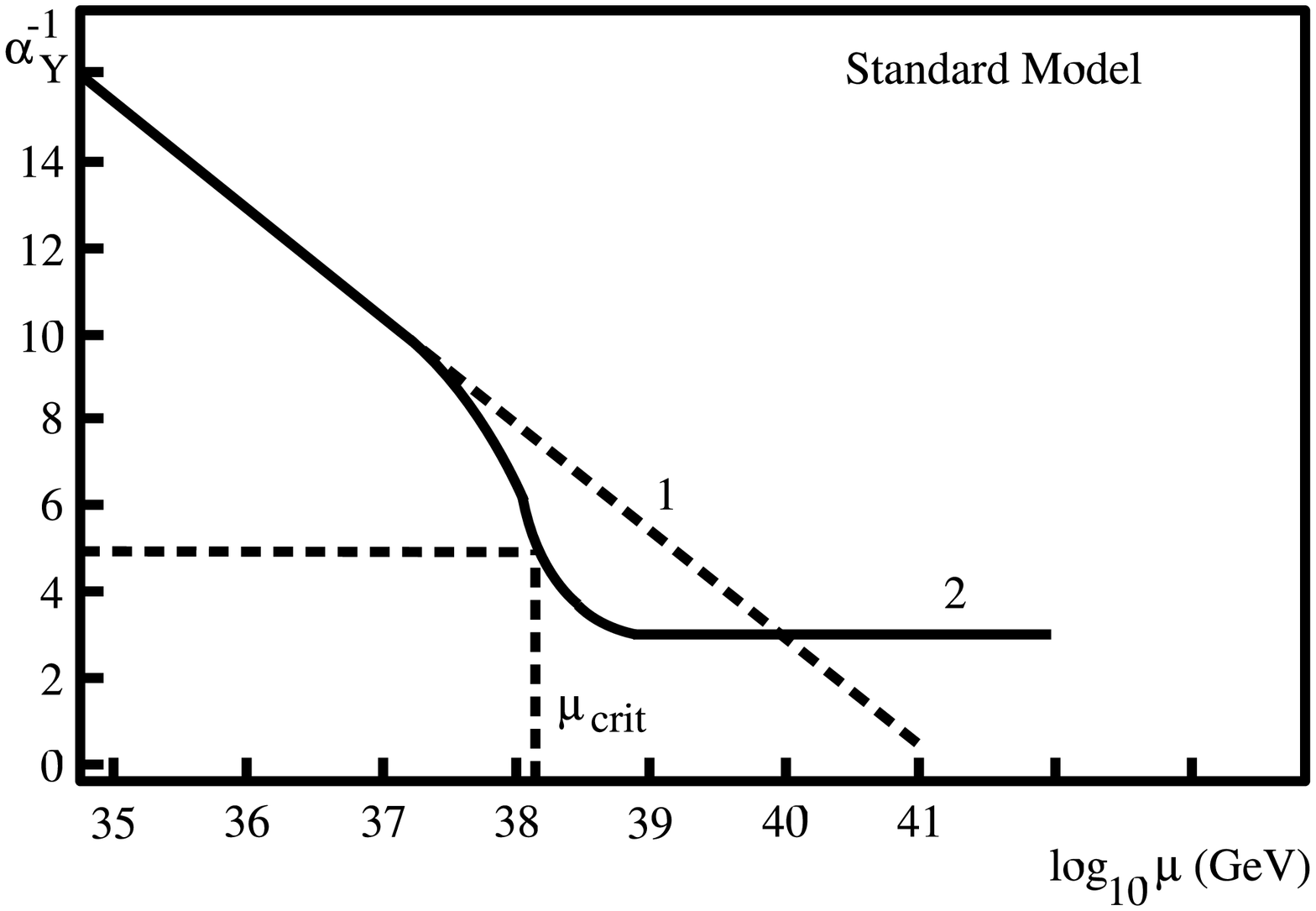}}
\caption{The evolution of the $U(1)_Y$ fine structure constant in the Standard
Model with influence of monopoles at very high energies; $\mu=\mu_{crit}$
is a critical point corresponding to the phase transition
``confinement--deconfinement''.}
\lb{f17a}
\efi

\clearpage\newpage
\bfi
\epsfxsize=14cm
{\epsfbox{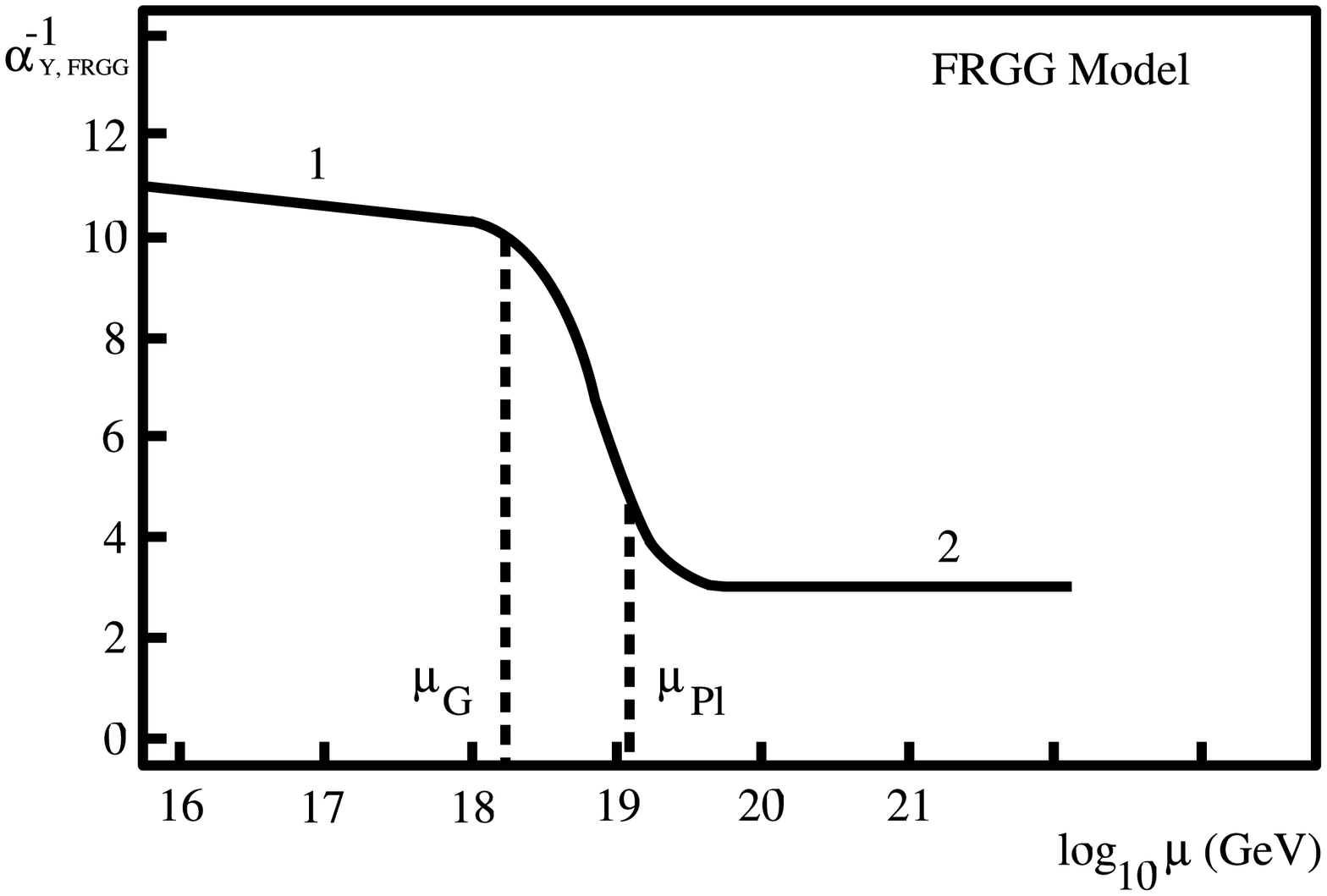}}
\caption{The evolution of the one family fine structure constant in the Family
replicated gauge group model with the phase transition point at
$\mu_{crit}=\mu_{Pl}$, and the FRGG symmetry breaking point at
$\mu=\mu_{\rm G}$.}
\lb{f17b}
\efi

\clearpage\newpage
\bfi
\epsfxsize=14cm
{\epsfbox{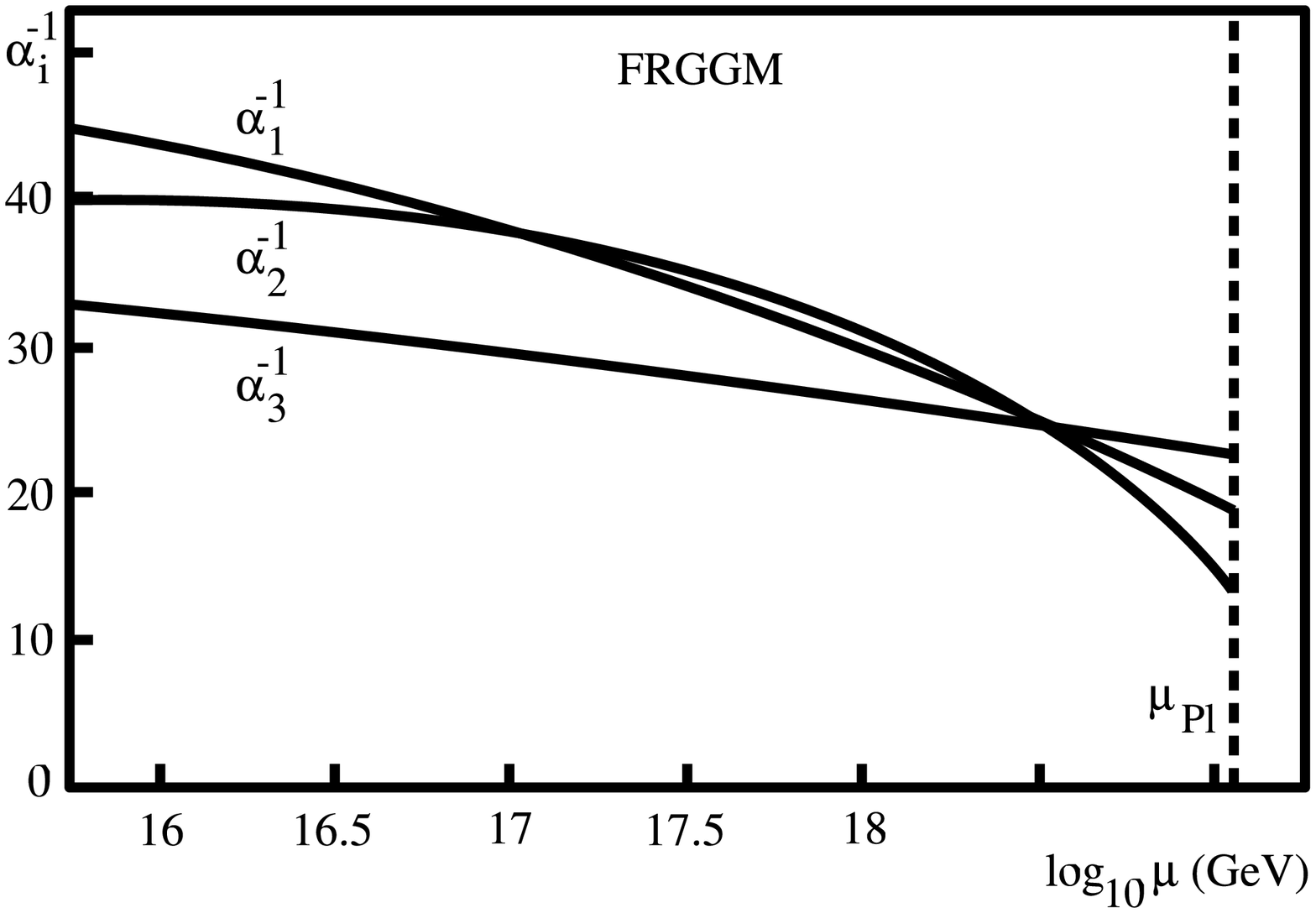}}
\caption{The evolution of fine structure constants
$\alpha^{-1}_{1,\, 2,\, 3}(\mu)$
beyond the Standard model in the Family replicated gauge group model (FRGGM)
with influence of monopoles near the Planck scale.}
\lb{f18}
\efi

\clearpage\newpage
\bfi
\epsfxsize=14cm
{\epsfbox{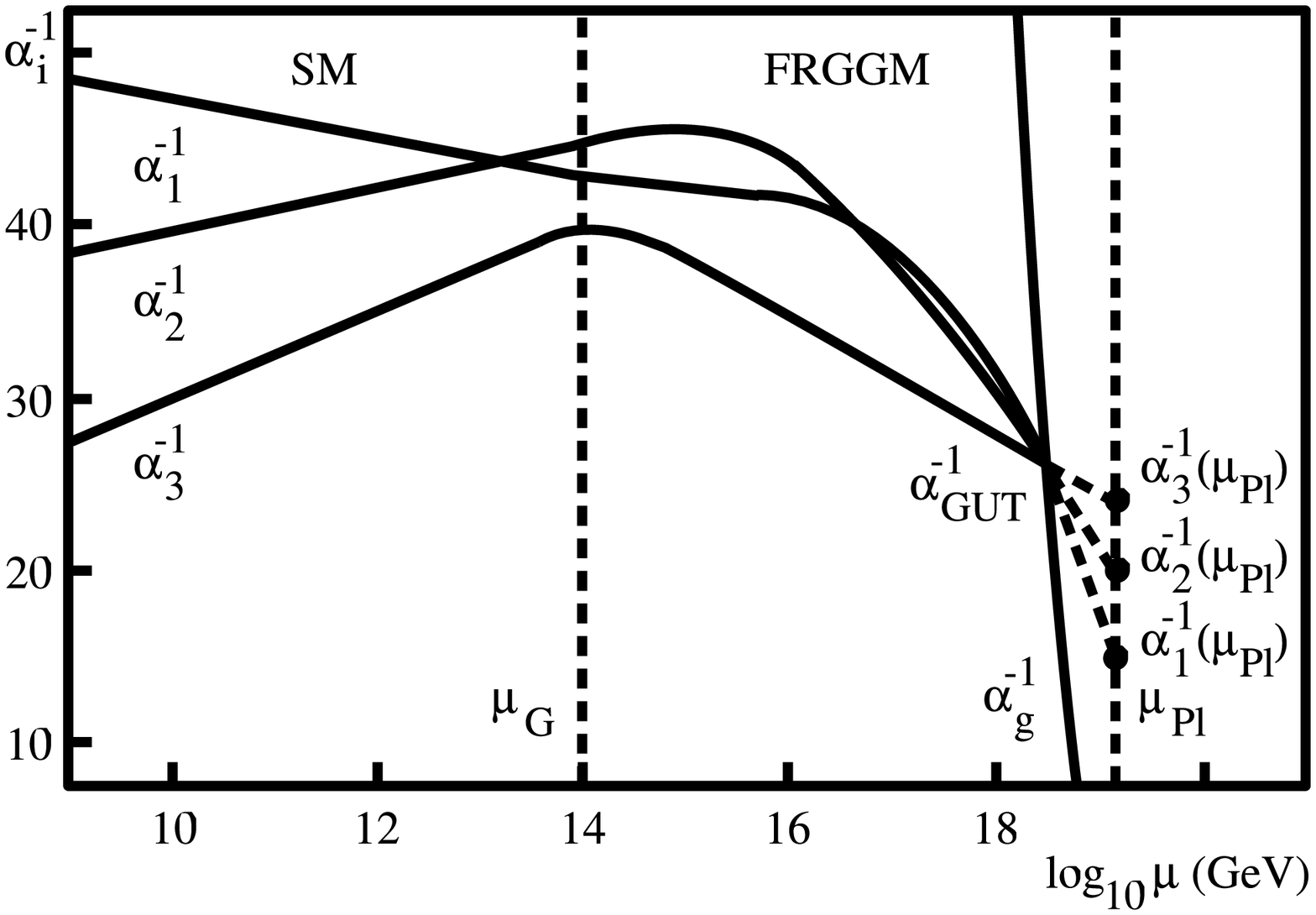}}
\caption{The evolution of $\alpha^{-1}_{1,\, 2,\, 3}(\mu)$
in the Standard model (SM)
and beyond it. The breakdown of FRGG occurs at $\mu_G\sim 10^{14}$ GeV.
It is shown the possibility of the $[SU(5)]^3$ SUSY unification of all gauge
interactions, including gravity, at $\alpha_{\rm GUT}^{-1}\approx 27$ and
$x_{\rm GUT}\approx 18.4$, where $x=\log_{10}(\mu)$ (GeV).}
\lb{f19}
\efi

\clearpage\newpage
\begin{table}
\caption{FRGGM by Froggatt--Nielsen with the gauge group
(SMG)$^3\times U(1)_f$ . Best fit to the conventional experimental
data. All masses are running masses at 1 GeV except the top quark mass
$M_t$ which is the pole mass.}
\bc
\bd
\ba{|c|l|l|}
\hline
&{\mbox {Fitted}}&{\mbox{Experimental}}\\
\hline
m_u & 3.6~{\mbox{MeV}} & 4~{\mbox{MeV}}\\
\hline
m_d&7.0~{\mbox{MeV}}&9~{\mbox{MeV}}\\
\hline
m_e&0.87~{\mbox{MeV}}&0.5~{\mbox{MeV}}\\
\hline
m_c&1.02~{\mbox{GeV}}&1.4~{\mbox{GeV}}\\
\hline
m_s&400~{\mbox{MeV}}&200~{\mbox{MeV}}\\
\hline
m_{\mu}&88~{\mbox{MeV}}&105~{\mbox{MeV}}\\
\hline
M_t&192~{\mbox{GeV}}&180~{\mbox{GeV}}\\
\hline
m_b&8.3~{\mbox{GeV}}&6.3~{\mbox{GeV}}\\
\hline
m_{\tau}&1.27~{\mbox{GeV}}&1.78~{\mbox{GeV}}\\
\hline
V_{us}&0.18&0.22\\
\hline
V_{cb}&0.018&0.041\\
\hline
V_{ub}&0.0039&0.0035\\
\hline
\ea
\ed
\ec
\lb{table1}
\end{table}

\clearpage\newpage
\begin{table}
\caption{FRGGM by Froggatt--Nielsen--Takanishi with the gauge group
(SMG$\times U(1)_{(B-L)})^3$. Best fit to the conventional
experimental data. All masses are running masses at $1~\mbox{\rm
GeV}$ except the top quark mass which is the pole mass.}
\bc
\bd
\ba{|c|c|c|}
\hline 
 & {\rm Fitted} & {\rm Experimental} \\ \hline
m_u & 4.4~\mbox{\rm MeV} & 4~\mbox{\rm MeV} \\ \hline m_d &
4.3~\mbox{\rm MeV} & 9~\mbox{\rm MeV} \\ \hline m_e &
1.6~\mbox{\rm MeV} & 0.5~\mbox{\rm MeV} \\ \hline m_c &
0.64~\mbox{\rm GeV} & 1.4~\mbox{\rm GeV} \\ \hline m_s &
295~\mbox{\rm MeV} & 200~\mbox{\rm MeV} \\ \hline m_{\mu} &
111~\mbox{\rm MeV} & 105~\mbox{\rm MeV} \\ \hline M_t &
202~\mbox{\rm GeV} & 180~\mbox{\rm GeV} \\ \hline m_b &
5.7~\mbox{\rm GeV} & 6.3~\mbox{\rm GeV} \\ \hline m_{\tau} &
1.46~\mbox{\rm GeV} & 1.78~\mbox{\rm GeV} \\ \hline V_{us} & 0.11
& 0.22 \\ \hline V_{cb} & 0.026 & 0.041 \\ \hline V_{ub} &
0.0027
& 0.0035 \\ \hline \Delta m^2_{\odot} & 9.0 \times
10^{-5}~\mbox{\rm eV}^2 & 5.0 \times 10^{-5}~\mbox{\rm eV}^2 \\
\hline \Delta m^2_{\rm atm} & 1.7 \times 10^{-3}~\mbox{\rm eV}^2 &
2.5 \times 10^{-3}~\mbox{\rm eV}^2\\ \hline \tan^2\theta_{\odot}
&0.26 & 0.34\\ \hline \tan^2\theta_{\rm atm}& 0.65 & 1.0\\
\hline
\tan^2\theta_{\rm chooz}  & 2.9 \times 10^{-2} & <2.6 \times
10^{-2}\\
\hline
\ea
\ed
\ec
\label{table2}
\end{table}


\clearpage\newpage

\begin{thebibliography}{999}
\bibitem{1}
C.D.~Froggatt, H.B.~Nielsen, Phys.Lett. {\bf B368}, 96 (1996).
\bibitem{2}
C.D.~Froggatt, L.V.~Laperashvili, H.B.~Nielsen, to appear in
Yad.Fiz. {\bf 68}, $N^o$ 8 (2005); ArXiv: hep-ph/0407102.
\bibitem{3}
C.D.~Froggatt, H.B.~Nielsen, {\it Hierarchy problem and a new bound
state}, in: {\it Proc. to the Euroconference on Symmetries Beyond
the Standard Model}, Slovenia, Portoroz, 2003 (DMFA, Zaloznistvo,
Ljubljana, 2003), p.73; ArXiv: hep-ph/0312218.
\bibitem{4}
C.D.~Froggatt, H.B.~Nielsen, L.V.~Laperashvili, {\it Hierarchy problem
and a bound state of 6 $t$ and 6 $\bar t$}. Invited talk by
H.B.~Nielsen at the {\it Coral Gables Conference on Launching of
Belle Epoque in High--Energy Physics and Cosmology (CG2003)},
 Ft.Lauderdale, Florida, USA, 17--21
Dec., 2003 (see Proceedings); ArXiv:
hep-ph/0406110.
\bibitem{5}
C.D.~Froggatt, L.V.~Laperashvili, H.B.~Nielsen, {\it A new bound state
6t + 6 anti--t and the fundamental--weak scale hierarchy in the
Standard Model}, in: {\it Proceeding to 13th International Seminar on
High--Energy Physics `Quarks 2004'}, Pushkinskie Gory, Russia, 24--30
May, 2004 (World Sci., 2004); ArXiv: hep-ph/0410243.
\bibitem{6}
A.G.~Riess {\itshape et al.}, Astron.J. {\bf 116}, 1009 (1998);
S.~Perlmutter {\itshape et al.}, Astrophys.J. {\bf 517}, 565
(1999); C.~Bennett {\itshape et al.}, ArXiv: astro--ph/0302207;
D.~Spergel {\itshape et al.}, ArXiv: astro--ph/0302209.
\bibitem{7}
D.L.~Bennett, C.D.~Froggatt, H.B.~Nielsen, in: {\it Proceedings of
the 27th International Conference on High Energy Physics}, Glasgow,
Scotland, 1994, Ed. by P.~Bussey, I.~Knowles (IOP Publishing
Ltd, 1995), p.557; {\it Perspectives in Particle Physics '94}, Ed.
by D.~Klabu\u{c}ar, I.~Picek, D.~Tadi\'{c} (World Scientific,
Singapore, 1995), p.255; ArXiv: hep-ph/9504294;
\newline C.D.~Froggatt, H.B.~Nielsen, {\it Influence from the Future},
ArXiv: hep-ph/9607375.
\bibitem{8}
D.L.~Bennett, H.B.~Nielsen, Int.J.Mod.Phys. {\bf A9}, 5155
(1994).
\bibitem{9}
D.L.~Bennett, H.B.~Nielsen, {\it The Multiple Point Principle: realized
vacuum in Nature is maximally degenerate}, in: {\it Proceedings to
the Euroconference on Symmetries Beyond the Standard Model},
Slovenia, Portoroz, 2003 (DMFA, Zaloznistvo, Ljubljana, 2003),
p.235.
\bibitem{10}
C.D.~Froggatt, H.B.~Nielsen, {\it Trying to understand the Standard
Model parameters}. Invited talk by H.B.~Nielsen at {\it the XXXI
ITEP Winter School of Physics}, Moscow, Russia, 18--26 Feb.,
2003; published in Surveys High Energy Phys. {\bf 18}, 55 (2003);
ArXiv: hep-ph/0308144.
\bibitem{11}
D.L.~Bennett, H.B.~Nielsen, I.~Picek, Phys.Lett. {\bf B208}, 275
(1988).
\bibitem{12}
C.D.~Froggatt, H.B.~Nielsen, {\it Origin of Symmetries}, (World
Sci., Singapore, 1991).
\bibitem{13}
L.V.~Laperashvili, Yad.Fiz. {\bf 57}, 501 (1994) [Phys.At.Nucl.
{\bf 57}, 471 (1994)]; Yad.Fiz. {\bf 59}, 172 (1996)
[Phys.Atom.Nucl. {\bf 59}, 162 (1996)].
\bibitem{14}
C.D.~Froggatt, L.V.~Laperashvili, H.B.~Nielsen, Y.~Takanishi,
{\it Family Replicated Gauge Group Models}, in: {\it Proceedings of the
Fifth International Conference `Symmetry in Nonlinear Mathematical
Physics'}, Kiev, Ukraine, 23--29 June, 2003, Ed. by A.G.~Nikitin,
V.M.~Boyko, R.O.~Popovich, I.A.~Yehorchenko (Institute of Mathematics
of NAS of Ukraine, Kiev, 2004), V.50, Part 2, p.737; ArXiv:
hep-ph/0309129.
\bibitem{15}
C.D.~Froggatt, G.~Lowe, H.B.~Nielsen, Phys.Lett. {\bf B311}, 163
(1993); Nucl.Phys. {\bf B414}, 579 (1994); ibid {\bf B420}, 3
(1994); C.D.~Froggatt, H.B.~Nielsen, D.J.~Smith, Phys.Lett. {\bf
B235}, 150 (1996); C.D.~Froggatt, M.~Gibson, H.B.~Nielsen,
D.J.~Smith, Int.J.Mod.Phys. {\bf A13}, 5037 (1998).
\bibitem{15a}
C.D.~Froggatt, L.V.~Laperashvili, H.B.~Nielsen, {\it SUSY or NOT SUSY},
{\it `SUSY98'}, Oxford, 10--17 July, 1998; hepnts1.rl.ac.uk/susy98/.
\bibitem{16}
H.B.~Nielsen, Y.~Takanishi, Nucl.Phys. {\bf B588}, 281 (2000);
ibid, {\bf B604}, 405 (2001); Phys.Lett. {\bf B507}, 241 (2001);
C.D.~Froggatt, H.B.~Nielsen, Y.~Takanishi, Nucl.Phys. {\bf
B631}, 285 (2002).
\bibitem{16a}
H.B.~Nielsen, Y.~Takanishi, Phys.Lett. {\bf B543}, 249 (2002).
\bibitem{17}
L.V.~Laperashvili, H.B.~Nielsen, {\it Anti--grand unification and
critical coupling universality}. A talk given by L.V.L. at {\it the
8th Lomonosov Conference on Elementary Particle Physics}, Moscow,
Russia, 25--29 Aug., 1997; ArXiv: hep-ph/9711388.
\bibitem{18}
L.V.~Laperashvili, H.B.~Nielsen, Mod.Phys.Lett. {\bf A12}, 73
(1997).
\bibitem{19}
L.V.~Laperashvili, {\it Anti--grand unification and the phase
transitions at the Planck scale in gauge theories}, in: {\it
Proceedings of the 4th International Symposium on Frontiers of
Fundamental Physics}, Hyderabad, India, 11--13 Dec., 2000, Ed. by
B.G.~Sidharth (World Scientific, Singapore, 2001); ArXiv:
hep-th/0101230.
\bibitem{20}
L.V.~Laperashvili, H.B.~Nielsen, {\it Multiple Point Principle and
phase transition in gauge theories}, in: {\it Proceedings of the
International Workshop on `What Comes Beyond the Standard Model'},
Bled, Slovenia, 29 June -- 9 July, 1998 (DMFA, Zaloznistvo,
Ljubljana, 1999), p.15; ArXiv: hep-ph/9905357.
\bibitem{21}
L.V.~Laperashvili, H.B.~Nielsen, Int.J.Mod.Phys. {\bf A16}, 2365
(2001); ArXiv: hep-th/0010260.
\bibitem{21aa}
L.V.~Laperashvili, H.B.~Nielsen, D.A.~Ryzhikh, Int.J.Mod.Phys. {\bf A16},
3989 (2001); ArXiv: hep-th/0105275.
\bibitem{21a}
L.V.~Laperashvili, H.B.~Nielsen, D.A.~Ryzhikh, Yad.Fiz. {\bf 65},
377 (2002) [Phys.At.Nucl. {\bf 65}, 353 (2002)]; ArXiv:
hep-th/0109023.
\bibitem{21b}
L.V.~Laperashvili, D.A.~Ryzhikh, H.B.~Nielsen, {\it Multiple Point
Model and phase transition couplings in the two--loop approximation
of dual scalar electrodynamics},in:
{\it Proceedings `Bled
2000--2002, What comes beyond the standard model'} (DMFA, Zaloznistvo,
Ljubljana, 2002), Vol. 2, pp.131--141; ArXiv: hep-ph/0112183.
\bibitem{21c}
L.V.~Laperashvili, D.A.~Ryzhikh, {\it Phase transition in gauge
theories and the Planck scale physics}, preprint ITEP--24--01, Oct.,
2001, 82pp; ArXiv: hep-ph/0212221.
\bibitem{21cc}
L.V.~Laperashvili, D.A.~Ryzhikh, {\it $[SU(5)]^3$ SUSY unification},
hep-preprint dedicated to the 60th jubilee of Holger Bech Nielsen;
ArXiv: hep-th/0112142.
\bibitem{21d}
L.V.~Laperashvili, H.B.~Nielsen, D.A.~Ryzhikh, DESY--02--188, Nov.,
2002. 49pp; Int.J.Mod.Phys. {\bf A18}, 4403 (2003); ArXiv:
hep-th/0211224.
\bibitem{21dd}
L.V.~Laperashvili,
{\it Generalized dual symmetry of non--Abelian theories, monopoles and
dyons}, in preparation, 2005; see also
ArXiv: hep-th/0211227.
\bibitem{22}
C.D.~Froggatt, L.V.~Laperashvili, R.B.~Nevzorov, H.B.~Nielsen,
ICTP Internal Report IC/IR/2003/16, Italy, Miramare--Trieste,
Oct., 2003; Yad.Fiz. {\bf 67}, 601 (2004) [Phys.At.Nucl. {\bf
67}, 582 (2004)]; ArXiv: hep-ph/0310127.
\bibitem{23}
C.D.~Froggatt, L.V.~Laperashvili, R.B.~Nevzorov, H.B.~Nielsen,
{\it No--scale supergravity and the multiple point principle}, in:
{\it Proceedings to the 7th Workshop `What comes beyond the standard
model'}, Bled, Slovenia, 19--31 July, 2004 (DMFA, Zaloznistvo,
Ljubljana, M.~Breskvar et al., Dec., 2004), pp.17--27; ArXiv:
hep-ph/0411273.
\bibitem{24}
C.D.~Froggatt, L.V.~Laperashvili, R.B.~Nevzorov, H.B.~Nielsen,
M.~Sher, {\it Two Higgs doublet model and Multiple Point Principle},
GUTPA--04--12--01, Dec., 2004, 14pp.; in: {\it Proceedings to the 7th
Workshop `What Comes Beyond the Standard Model'}, Bled, Slovenia,
19--31 July, 2004 (DMFA, Zaloznistvo, Ljubljana, M.~Breskvar
et al., Dec 2004), pp.28--39; ArXiv: hep-ph/0412333;
ArXiv: hep-ph/0412333; ArXiv: hep-ph/0412208.
\bibitem{25}
L.V.~Laperashvili, {\it The Multiple Point Principle and Higgs bosons},
in: {\it Proceedings of the International Bogolyubov Conference on
Problems of Theoretical and Mathematical Physics}, Moscow--Dubna,
Russia, 2--6 Sep., 2004; ArXiv: hep-ph/0411177.
\bibitem{26}
C.R.~Das, C.D.~Froggatt, L.V.~Laperashvili, H.B.~Nielsen,
{\it Degenerate vacua, see--saw scale and flipped $SU(5)$}, to be published
in 2005.
\bibitem{28} S.~Coleman, E.~Weinberg, Phys. Rev. {\bf
D7}, 1888 (1973).
\bibitem{29}
M.~Sher, Phys.Rep. {\bf 179}, 274 (1989).
\bibitem{30}
C.G.~Callan, Phys.Rev. {\bf D2}, 1541 (1970).
\bibitem{31}
K.~Symanzik, in: {\it Fundamental Interactions at High Energies},
Ed. A.~Perlmutter (Gordon and Breach, New York, 1970).
\bibitem{32}
C.D.~Froggatt, H.B.~Nielsen, Y.~Takanishi, Phys.Rev. {\bf D64},
113014 (2001).
\bibitem{33}
Particle Data Group, K.~Hagiwara et al., Phys.Rev. {\bf D66},
010001 (2002).
\bibitem{12p}
Yu.M.~Makeenko, M.I.~Polikarpov, {\it X School in Physics at
ITEP},
1983, No.3, pp. 3, 51;\\
Yu.M.~Makeenko, Uspehi in Physics, {\bf 27}, 401 (1984).
\bibitem{1s}
K.~Wilson, Phys.Rev. {\bf D10}, 2445 (1974).
\bibitem{2s}
M.~Creutz, I.~Jacobs, C.~Rebbi, Phys.Rev. {\bf D20}, 1915 (1979).
\bibitem{3s}
B.~Lautrup, M.~Nauenberg, Phys.Lett. {\bf B95}, 63 (1980).
\bibitem{4s}
M.~Creutz, Phys.Rev. {\bf D21}, 2308 (1980); Phys.Rev.Lett. {\bf
45}, 313 (1980).
\bibitem{5s}
B.~Lautrup, M.~Nauenberg, Phys.Rev.Lett. {\bf 45}, 1755 (1980).
\bibitem{6s}
M.~Creutz, Phys.Rev.Lett. {\bf 46}, 1441 (1981).
\bibitem{7s}
G.~Bhanot, M.~Creutz, Phys.Rev. {\bf D24}, 3212 (1981).
\bibitem{8s}
G.~Bhanot, Phys.Lett. B{\bf 108}, 337 (1982).
\bibitem{8ss}
C.P.~Bachas, R.F.~Dashen, Nucl.Phys. {\bf B210}, 583 (1982).
\bibitem{9ss}
A.A.~Migdal, {\it Problems and perspectives of gauge theories}, in:
E.~Seiler,{\it The phase structure of finite temperature lattice gauge
theories}, Munich, Max Planck Institute, 1986 (translated in Russian
language).
\bibitem{9s}
G.~Bhanot, Nucl.Phys. {\bf B205}, 168 (1982); Phys.Rev. {\bf D24},
461 (1981); Nucl.Phys. {\bf B378}, 633 (1992).
\bibitem{10s}
J.~Jersak, T.~Neuhaus, P.M.~Zerwas, Phys.Lett. {\bf B133}, 103
(1983); Nucl.Phys. {\bf B251}, 299 (1985).
\bibitem{10ss}
N.~Arkani--Hamed, Invited talk at the {\it Conference on Hierarchy
Problems in Four and More Dimensions}, ICTP, Italy, Trieste, 1--4 Oct.,
2003.
\bibitem{10sx}
G.~Volovik, JETP Lett. {\bf 79}, 101 (2004).
\bibitem{10as}
G.S.~Bali, {\it Overview from Lattice QCD}, plenary talk presented at
{\it Nuclear and Particle physics with CEBAF at Jefferson Lab},
Dubrovnik, 3--10 Nov., 1998; ArXiv: hep-lat/9901023.
\bibitem{11s}
T.~Suzuki, Nucl.Phys.Proc.Suppl. {\bf 30}, 176 (1993);\\
R.W.Haymaker, Phys.Rep. {\bf 315}, 153 (1999).
\bibitem{12s}
M.N.~Chernodub, M.I.~Polikarpov, in {\it Confinement, Duality and
Non--perturbative Aspects of QCD},
p.387, Ed. by Pierre van Baal, Plenum Press, 1998; ArXiv: hep-th/9710205;
M.N.~Chernodub, F.V.~Gubarev, M.I.~Polikarpov, A.I.~Veselov,
Prog.Ther.Phys.Suppl. {\bf 131}, 309 (1998); ArXiv: hep-lat/9802036;
M.N.~Chernodub, F.V.~Gubarev, M.I.~Polikarpov, V.I.~Zakharov,
{\it Magnetic monopoles,
alive}, ArXiv: hep-th/0007135; {\it Towards Abelian--like formulation of
the
dual gluodynamics}, ArXiv: hep-th/0010265.
\bibitem{[22]}
T.~Banks, R.~Myerson, J.~Kogut. Nucl.Phys. {\bf B129}, 493 (1977).
\bibitem{[23]}
J.~Villain. J.de Phys. {\bf 36}, 581 (1975).
\bibitem{[24]}
J.L.~Cardy. Nucl.Phys. {\bf B170}, 369 (1980).
\bibitem{[25]}
J.M.~Luck. Nucl.Phys. {\bf B210}, 111 (1982).
\bibitem{13p}
G.~Parisi, R.~Petronzio, F.~Rapuano. Phys.Lett. {\bf B128}, 418 (1983);\\
E.~Marinari, M.~Guagnelli, M.P.~Lombardo, G.~Parisi, G.~Salina,
Proceedings {\it Lattice '91'}, Tsukuba 1991, p.278--280;
Nucl.Phys.Proc.Suppl. {\bf B26}, 278 (1992).
\bibitem{14p}
M.~L\"uscher, K.~Symanzik, P.~Weisz, Nucl.Phys. {\bf B173}, 365 (1980).\\
C.~Surlykke, {\it On monopole suppression in lattice QED}, preprint NBI,
1994.
\bibitem{14pa}
A.~Goldhaber, Hsiang--nan Li, R.R.~Parwani. Phys.Rev. {\bf D51}, 919 (1995).
\bibitem{14pb}
H.B.~Nielsen, P.~Olesen. Nucl.Phys. {\bf B160}, 380 (1979).
\bibitem{15p}
Yu.A.~Simonov, Yad.Fiz. {\bf 58}, 113 (1995); A.M.~Badalian,
Yu.A.~Simonov, Yad.Fiz. {\bf 60}, 714 (1997); A.M.~Badalian,
D.S.~Kuzmenko, {\it Freezing of QCD coupling alpha(s) affects the short
distance static potential}, ArXiv: hep-ph/0104097.
\bibitem{13s}
T.~Suzuki, Progr.Theor.Phys. {\bf 80}, 929 (1988); S.~Maedan,
T.~Suzuki, Progr.Theor.Phys. {\bf 81}, 229 (1989).
\bibitem{22y}
D.R.T.~Jones, Nucl.Phys. {\bf B75}, 531 (1974); Phys.Rev. {\bf
D25}, 581 (1982).
\bibitem{23y}
M.~Fischler, C.T.~Hill, Nucl.Phys. {\bf B193}, 53 (1981).
\bibitem{24y}
I.~Jack, H.~Osborn, J.Phys. {\bf A16}, 1101 (1983).
\bibitem{25y}
M.E.~Machacek, M.T.~Vaughn, Nucl.Phys. {\bf B222}, 83 (1983);
ibid, {\bf B249}, 70 (1985).
\bibitem{26y}
H.~Arason, D.J.~Castano, B.~Kesthelyi, S.~Mikaelian, E.J.~Piard,
P.~Ramond, B.D.~Wright, Phys.Rev. {\bf D46}, 3945 (1992).
\bibitem{27y}
C.~Ford, D.R.T.~Jones, P.W.~Stephenson, M.B.~Einhorn, Nucl.Phys. {\bf B395},
17 (1993); C.~Ford, I.~Jack, D.R.T.~Jones, Nucl.Phys. {\bf B387},
373 (1992); Erratum--ibid, {\bf B504}, 551, (1997);
ArXiv: hep-ph/0111190.
\bibitem{28y}
O.V.~Tarasov, A.A.~Vladimirov, A.Yu.~Zharkov, Phys.Lett. {\bf B93},
429 (1980);\\
S.~Larin, T.~Ritberg, J.~Vermaseren, Phys. Lett. {\bf B400}, 379
(1997);\\
M.~Czakon, Nucl.Phys. {\bf B710}, 485 (2005).
\bibitem{29y}
J.~Jersak, T.~Neuhaus, H.~Pfeiffer, Phys.Rev. {\bf D60}, 054502
(1999).
\bibitem{30y}
A.A.~Abrikosov, Soviet JETP, {\bf 32}, 1442 (1957).
\bibitem{31y}
H.B.~Nielsen, P.~Olesen, Nucl.Phys. {\bf B61}, 45 (1973).
\bibitem{32y}
E.T.~Akhmedov, M.N.~Chernodub, M.I.~Polikarpov, M.A.~Zubkov,
Phys.Rev. {\bf D53}, 2087 (1996).
\bibitem{24p}
G.~'t Hooft, Nucl.Phys. {\bf B190}, 455 (1981).
\bibitem{25p}
Yu.A.~Simonov, Uspehi in Physics {\bf 39}, 313 (1996).
\bibitem{25pa}
M.N.~Chernodub, F.V.~Gubarev, JETP Lett. {\bf 62}, 100 (1995);\\
M.N.~Chernodub, F.V.~Gubarev, M.I.~Polikarpov, JETP Lett. {\bf 69}, 169
(1999);\\
M.N.~Chernodub, Phys.Rev. {\bf D69}, 094504 (2004).
\bibitem{25pb}
G.S.~Bali, C.~Schlighter, K.~Shilling, Progr.Theor.Phys.Suppl. {\bf 131},
645 (1998).
\bibitem{25pc}
L.D.~Faddeev, A.J.~Niemi, Phys.Lett. {\bf B525}, 195 (2002).
\bibitem{10na}
L.~Nottale, Int.J.Mod.Phys. {\bf A7}, 4899 (1992); L.~Nottale,
{\it Fractal Space--Time and Microphysics}, Singapore: World
Scientific, 1993; L.~Nottale, in: {\it `Frontiers of Fundamental
Physics', Proceedings of Forth International Symposium}, Hyderabad,
India, 11--13 Dec., 2000; ed. B.G.~Sidharth 
(World Scientific, Singapore, 2001).
\bibitem{10nb}
B.G.~Sidharth, Int.J.Mod.Phys. {\bf A13}, 2599 (1998).
\bibitem{32a}
H.P.~Nilles, Phys.Rep. {\bf 110}, 1 (1984).
\bibitem{33a}
P.~Langacker, N.~Polonsky, Phys.Rev. {\bf D47}, 4028 (1993); ibid,
{\bf D49}, 1454 (1994); ibid, {\bf D52}, 3081 (1995).
\bibitem{39}
M.B.~Green, J.~Schwarz, Phys.Lett. {\bf B149}, 117 (1984).
\bibitem{41}
A.G.~Agnese, R.~Festa, Phys.Lett. {\bf A227}, 165 (1997).
\bibitem{41r}
J.~Schwinger, Phys.Rev. {\bf 144}, 1087 (1966); {\bf 151}, 1048;
1055 (1966); {\bf 173}, 1536 (1968); Science {\bf 165}, 757
(1969); {\bf 166}, 690 (1969).
\bibitem{42r}
W.~Deans, Nucl.Phys. {\bf B197}, 307 (1982).
\bibitem{43r}
C.~Panagiotakopoulos, J.Phys. {\bf A16}, 133 (1983).
\bibitem{43ar}
P.A.M.~Dirac, Proc.R.Soc. (London) {\bf A33}, 60 (1931).
\bibitem{44r}
G.~Calucci, R.~Iengo, Nucl.Phys. {\bf B223}, 501 (1983).
\bibitem{45r}
G.J.~Goebel, M.T.~Thomaz, Phys.Rev. {\bf D30}, 823 (1984).
\bibitem{46r}
S.~Coleman, in: {\it The Unity of the Fundamental Interactions,
Eriche Lectures}, 1982, ed. A.Zichichi (Plenum, 1983), p.21.
\bibitem{47r}
L.V.~Laperashvili, H.B.~Nielsen, Mod.Phys.Lett. {\bf
A14}, 2797 (1999).
\bibitem{48r}
M.~Blagojevic, P.~Senjanovic, Phys.Rep. {\bf 157}, 234 (1988).
\bibitem{49r}
D.Zwanziger, Phys.Rev. {\bf D3}, 343 (1971).
\bibitem{50r}
R.A.~Brandt, F.~Neri, D.~Zwanziger, Phys.Rev. {\bf D19}, 1153 (1979).
\bibitem{51r}
F.V.~Gubarev, M.I.~Polikarpov, V.I.~Zakharov, Phys.Lett. {\bf B438},
147 (1998).
\bibitem{53r}
L.P.~Gamberg, K.A.~Milton, Phys.Rev. {\bf D61}, 075013 (2000).
\bibitem{40}
L.V.~Laperashvili, H.B.~Nielsen, {\it The problem of monopoles in the
Standard and Family replicated models}. A talk given by L.V.L. at
the {\it 11th Lomonosov Conference on Elementary Particle Physics},
Moscow, Russia, 21--27 Aug., 2003; ArXiv: hep-th/0311261.
\bibitem{42}
N.N.~Bogolyubov, D.V.~Shirkov, JETP, {\bf 30}, 77 (1956).
\bibitem{22q}
S.G.~Matinyan, G.K.~Savvidy, Nucl.Phys. {\bf B134}, 539 (1978).
\bibitem{23q}
P.A.~Kovalenko, L.V.~Laperashvili, Phys. of Atom.Nucl. {\bf 62},
1729 (1999).
\bibitem{23qq}
P.A.~Kovalenko, L.V.~Laperashvili, {\it The effective QCD Lagrangian
and renormalization group approach}, ITEP--PH--11--97, Aug 1997, 5pp.
{\it Talk given at 8th Lomonosov Conference on Elementary Particle
Physics}, Moscow, Russia, 25--29 Aug., 1997; ArXiv: hep-ph/9711390.
\bibitem{40ao}
D.I.~Olive, Phys.Rep. {\bf 49}, 165 (1979).
\bibitem{40bo}
L.~Michel, L.~O'Raifeartaigh, K.C.~Wali, Phys.Rev. {\bf D15}, 3641
(1977).
\bibitem{44}
S.~Rajpoot, Nucl.Phys.Proc.Suppl. {\bf A51}, 50 (1996).
\bibitem{45}
S.~Dimopoulos, G.~Dvali, R.~Rattazzi, G.F.~Guidice, Nucl.Phys.
{\bf B510}, 12 (1998).
\bibitem{46}
F.S.~Ling, P.~Ramond, Phys.Rev. {\bf D67}, 115010 (2003).

\end{thebibliography}
\end{document}